\documentclass[twocolumn,superscriptaddress,showpacs,nofootinbib,preprintnumbers,secnumarabic,amssymb, nobibnotes, aps, prd]{revtex4-2}
\usepackage[utf8]{inputenc}
\usepackage{graphicx}
\usepackage{latexsym,amsmath,amssymb,amsthm,lmodern,float,url}
\usepackage{natbib}
\usepackage{color}
\usepackage{microtype}
\usepackage{import}
\usepackage{bbold}
\usepackage[plain]{fancyref}
\usepackage{varioref}
\usepackage{slashed}
\usepackage{multirow}
\usepackage{tikz}
\usepackage{scrextend}
\usepackage{braket}
\usetikzlibrary{shapes}
\usetikzlibrary{positioning}
\usepackage[normalem]{ulem}

\usepackage{qcircuit}

\usepackage[colorlinks=true,backref=false, linktocpage=true,
citecolor=blue,urlcolor=blue,linkcolor=blue,pdfpagemode=UseOutlines]{hyperref}
\hypersetup{%
  bookmarksnumbered=true,
  pdftitle = {},
  pdfsubject = {},
  pdfauthor = {},
  pdfkeywords = {}
}

\let\Re\undefined

\DeclareMathOperator{\Tr}{Tr}
\DeclareMathOperator{\Re}{Re}

\DeclareMathOperator{\tr}{Tr}
\DeclareMathOperator{\re}{Re}

\newcommand{\bi}{\mathbb{BI}}
\newcommand{\bo}{\mathbb{BO}}
\newcommand{\btt}{\mathbb{BT}}

\newcommand{\bfj}{\mathbf{j}}
\newcommand{\bfk}{\mathbf{k}}
\newcommand{\bfu}{\mathbf{u}}
\newcommand{\bft}{\mathbf{t}}

\definecolor{blugrn}{RGB}{0,158,115}

\begin{document}
\preprint{FERMILAB-PUB-23-753-SQMS-T}
\title{Primitive Quantum Gates for an \texorpdfstring{$SU(2)$}{SU(2)} Discrete Subgroup: Binary Octahedral}
\author{Erik J. Gustafson}
\affiliation{Superconducting and Quantum Materials System Center (SQMS), Batavia, Illinois, 60510, USA.}
\affiliation{Fermi National Accelerator Laboratory, Batavia,  Illinois, 60510, USA}
\affiliation{Quantum Artificial Intelligence Laboratory (QuAIL),
NASA Ames Research Center, Moffett Field, CA, 94035, USA}
\affiliation{USRA Research Institute for Advanced Computer Science (RIACS), Mountain View, CA, 94043, USA}

\author{Henry Lamm}
\email{hlamm@fnal.gov}
\affiliation{Superconducting and Quantum Materials System Center (SQMS), Batavia, Illinois, 60510, USA.}
\affiliation{Fermi National Accelerator Laboratory, Batavia, Illinois, 60510, USA}
\author{Felicity Lovelace}
\email{fl16@uic.edu}
\affiliation{Department of Physics, University of Illinois at Chicago, Chicago, Illinois 60607, USA}

\date{\today}

\begin{abstract}
We construct a primitive gate set for the digital quantum simulation of the 48-element binary octahedral ($\mathbb{BO}$) group. This nonabelian discrete group better approximates $SU(2)$ lattice gauge theory than previous work on the binary tetrahedral group at the cost of one additional qubit -- for a total of six -- per gauge link. The necessary primitives are the inversion gate, the group multiplication gate, the trace gate, and the $\mathbb{BO}$ Fourier transform.
\end{abstract}

\maketitle

\section{Introduction}
The possibilities for quantum utility in lattice gauge theories (LGT) are legion~\cite{Klco:2021lap,Banuls:2019bmf,Bauer:2022hpo,DiMeglio:2023nsa}. Perhaps foremost, it provides an elegant solution to the sign problem which results in exponential scaling of classical computing resources~\cite{Gattringer:2016kco} which precludes large-scale simulations of dynamics, at finite fermion density, and in the presence of topological terms. In order to study these fundamental physics topics, a number of quantum subroutines are required.  

The first task is preparing strongly-coupled states of interest including ground states~\cite{Kuhn:2014rha,Kokail:2018eiw,Chakraborty:2020uhf,Yamamoto:2021vxp,Desai:2021oiy,Farrell:2023fgd,Kane:2023jdo}, thermal states~\cite{2010PhRvL.105q0405B,PhysRevLett.108.080402,Lamm:2018siq,Klco:2019xro,Harmalkar:2020mpd,motta2020determining,deJong:2021wsd,Xie:2022jgj,Davoudi:2022uzo,Ball:2022dxy,Saroni:2023uob}, and colliding particles~\cite{Jordan:2011ne,Jordan:2011ci,Garcia-Alvarez:2014uda,Jordan:2014tma,Jordan:2017lea,Moosavian:2017tkv,brandao2019finite,Gustafson:2019mpk,Gustafson:2019vsd,Gustafson:2020yfe,Kreshchuk:2023btr}. For applications to dynamics, the time-evolution operator $U(t)=e^{-iHt}$ must be approximated and many different choices exist: trotterization~\cite{PhysRevX.11.011020,Davoudi:2022xmb}, random compilation~\cite{PhysRevLett.123.070503,Shaw:2020udc}, Taylor series~\cite{PhysRevLett.114.090502}, qubitization~\cite{Low2019hamiltonian}, quantum walks~\cite{berry2009black}, signal processing~\cite{PhysRevLett.118.010501}, linear combination of unitaries~\cite{childs2012hamiltonian,Shaw:2020udc}, and variational approaches~\cite{cirstoiu2020variational,gibbs2021longtime,Yao:2021ddt,Nagano:2023uaq}; each with their own tradeoffs. Important alongside state preparation and evolution is the need to develop efficient techniques~\cite{Roggero:2018hrn,Roggero:2019srp,Kanasugi:2023wxu,Gustafson:2023ayr} and formulations~\cite{Lamm:2019bik,Bauer:2019qxa,Echevarria:2020wct,Xu:2021tey,Bauer:2021gup,Cohen:2021imf,Barata:2022wim,Czajka:2022plx,Farrell:2022vyh,Bedaque:2022ftd,Ikeda:2023zil,Ciavarella:2020vqm} for measuring physical observables. Necessary for acheiving this, one may further use algorithmic improvements such as error mitigation \& correction~\cite{Huffman:2021gsi,Klco:2021jxl,Charles:2023zbl,Gustafson:2022xlj,rajput2021quantum,Gustafson:2023swx,Halimeh:2019svu,Lamm:2020jwv,Tran:2020azk,Kasper:2020owz,Halimeh:2020ecg,VanDamme:2020rur,Nguyen:2021hyk,Halimeh:2021vzf,ARahman:2022tkr,Yeter-Aydeniz:2022vuy,Gustafson:2023swx}, improved Hamiltonians~\cite{Carena:2022kpg,Ciavarella:2023mfc} and quantum smearing~\cite{Gustafson:2022hjf} to reduce errors from quantum noise and theoretical approximations. Beyond direct simulations, quantum computers could also accelerate classical lattice gauge theory simulations by reducing autocorrelation~\cite{Temme:2009wa,Clemente:2020lpr,Yamamoto:2022jes,Ballini:2023ljs} and optimizing interpolating operators~\cite{Avkhadiev:2019niu,Avkhadiev:2022ttx}.

Across this cornucopia, there exist a set of fundamental group theoretic operations that LGT requires~\cite{Lamm:2019bik}. Via this identification of primitive subroutines, the problem of formulating quantum algorithms for LGT can be divided into deriving said group-dependent primitives~\cite{Alam:2021uuq,Gustafson:2022xdt,Zache:2023dko} and group-independent algorithmic design~\cite{Carena:2022kpg,Gustafson:2022hjf,Zache:2023cfj,Zache:2023dko}.

With this notion in mind, one can turn to digitizing the infinite-dimensional Hilbert space of the gauge bosons. Many proposals exist that prioritizing theoretical and algorithmic facets of the problem differently~\cite{Zohar:2012ay,Zohar:2012xf,Zohar:2013zla,Zohar:2014qma,Zohar:2015hwa,Zohar:2016iic,Klco:2019evd,Ciavarella:2021nmj,Bender:2018rdp,Liu:2020eoa,Hackett:2018cel,Alexandru:2019nsa,Yamamoto:2020eqi,Haase:2020kaj,Armon:2021uqr,PhysRevD.99.114507,Bazavov:2015kka,Zhang:2018ufj,Unmuth-Yockey:2018ugm,Unmuth-Yockey:2018xak,Kreshchuk:2020dla,Kreshchuk:2020aiq,Raychowdhury:2018osk,Raychowdhury:2019iki,Davoudi:2020yln,Wiese:2014rla,Luo:2019vmi,Brower:2020huh,Mathis:2020fuo,Singh:2019jog,Singh:2019uwd,Buser:2020uzs,Bhattacharya:2020gpm,Barata:2020jtq,Kreshchuk:2020kcz,Ji:2020kjk,Bauer:2021gek,Gustafson:2021qbt,Hartung:2022hoz,Grabowska:2022uos,Murairi:2022zdg}.  For some digitized theories, there may be no nontrivial continuum limit~\cite{Hasenfratz:2001iz,Caracciolo:2001jd,Hasenfratz:2000hd,PhysRevE.57.111,PhysRevE.94.022134,car_article,Singh:2019jog,Singh:2019uwd,Bhattacharya:2020gpm,Zhou:2021qpm,Caspar:2022llo}. Furthermore, the efficacy of a gauge digitization can be dimension-dependent~\cite{Davoudi:2020yln,Zohar:2021nyc,Alam:2021uuq,Gustafson:2022xdt}.  While all digitizations consider the relative quantum memory costs, the consequences of digitization on gate costs are more limited.  Despite this, recent work has made abundantly clear that the number of expensive fixed-point arithmetic required can vary by orders of magnitude~\cite{Kan:2021xfc,Davoudi:2022xmb}.

One promising digitization that uses comparatively few qubits and avoids fixed-point arithmetic is the discrete subgroup approximation~\cite{Bender:2018rdp,Hackett:2018cel,Alexandru:2019nsa,Yamamoto:2020eqi,Ji:2020kjk,Haase:2020kaj,Carena:2021ltu,Armon:2021uqr,Gonzalez-Cuadra:2022hxt,Charles:2023zbl}. This method was explored in the 70's and 80's to reduce memory and runtime of Euclidean LGT on classical computers. The replacement of $U(1)$ by $\mathbb{Z}_N$ was considered first~\cite{Creutz:1979zg,Creutz:1982dn} and eventually extended to to the crystal-like subgroups of $SU(N)$~\cite{Bhanot:1981xp,Petcher:1980cq,Bhanot:1981pj,Hackett:2018cel,Alexandru:2019nsa,Ji:2020kjk,Ji:2022qvr,Alexandru:2021jpm,Carena:2022hpz}. Some studies were even performed with dynamical fermions~\cite{Weingarten:1980hx,Weingarten:1981jy}. Theoretical work has established that the discrete subgroup approximation corresponds to continuous groups broken by a Higgs mechanism~\cite{Kogut:1980qb,romers2007discrete,Fradkin:1978dv,Harlow:2018tng,Horn:1979fy}. 

LGT calculations are performed at fixed lattice spacing $a=a(\beta)$ which for asymptotically free theories approaches zero as $\beta\rightarrow\infty$. Finite $a$ leads to discrepancies from the continuum results, but provided one simulates in the \emph{scaling regime} below $a_s(\beta_s)$, these errors are well-behaved.  On the lattice, the breakdown of the discrete subgroup approximation manifests as a \emph{freezeout} $a_f$ (or coupling $\beta_f$) where the gauge links become "frozen" to the identity. Despite this, the approximation error for Euclidean calculations can be tolerable provided $a_s\gtrsim a_f$ ($\beta_s\lesssim\beta_f$)~\cite{Alexandru:2019nsa,Alexandru:2021jpm}. Further, a connection between the couplings and lattice spacings of Minkowski and Euclidean lattice field theories has been shown~\cite{Carena:2021ltu,Clemente:2022cka,Carena:2022hpz}, which suggests similarly controllable digitization error on large-scale quantum simulations and a way for determining viable approximations. 

The freezing transitions are known in $3+1d$ when the Wilson action is used. Given the known connection between the Wilson action and the Kogut-Susskind Hamiltonian $H_{KS}$~\cite{Creutz:1984mg}, this provides insight into the viable groups for quantum simuations. Approximating $U(1)$ by $\mathbb{Z}_{n>5}$ satisfies $\beta_f>\beta_s$, with $\beta_f\propto 1-\cos^{-1}(2\pi/n)$. In the case of nonabelian gauge groups, there are limited number of crystal-like subgroups. $SU(2)$, with $\beta_s=2.2$ has three: the 24-element binary tetrahedral $\mathbb{BT}$ ($\beta_f=2.24(8)$), the 48-element binary octahedral $\mathbb{BO}$ ($\beta_f=3.26(8)$), and the 120-element binary icosahedral $\bi$ ($\beta_f=5.82(8)$)~\cite{Alexandru:2019nsa}. Thus, while $\mathbb{BT}$ require only 5 qubits, due to its low $\beta_f$ it is unlikely $H_{KS}$ can be used for quantum simulation but a modified or improved Hamiltonians $H_{I}$~\cite{Carena:2022kpg,Ciavarella:2023mfc} could prove sufficient~\cite{Alexandru:2019nsa,Ji:2020kjk,Ji:2022qvr,Alexandru:2021jpm}.

In this work, we consider the smallest crystal-like subgroup of a $SU(2)$ with $a_f>a_s$ -- $\mathbb{BO}$ which requires 6 qubits per register. A number of smaller nonabelian groups have been considered previously. Quantum simulations of the $2N$-element dihedral groups, $D_N$, while not crystal-like, have been extensively studied~\cite{Bender:2018rdp,Lamm:2019bik,Alam:2021uuq,Fromm:2022vaj}. The 8-element $\mathbb{Q}_8$ subgroup of $SU(2)$ has also been investigated~\cite{Gonzalez-Cuadra:2022hxt}. In \cite{Gustafson:2022xdt}, quantum circuits for $\btt$ were constructed and resource estimates were obtained using the $H_I$ of~\cite{Carena:2022kpg}.

\begin{figure}[!th]
\centering
    \includegraphics[width=0.9\linewidth]{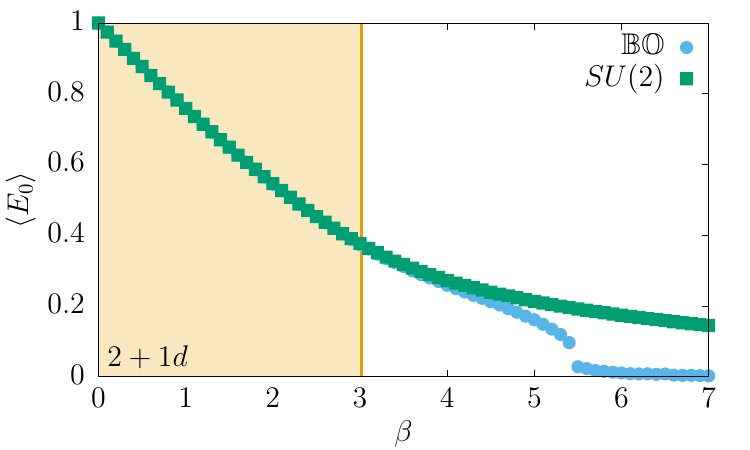}
    \includegraphics[width=0.9\linewidth]{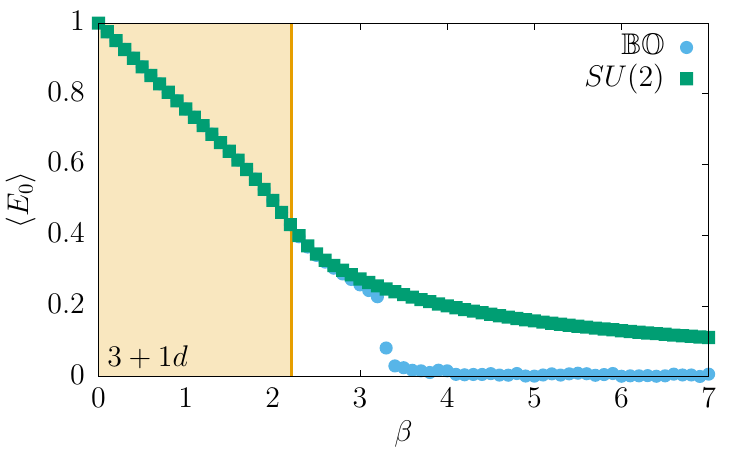}
    \caption{\label{fig:freezing}Euclidean calculations of lattice energy density $\langle E_0\rangle$ of $\bo$ as measured by the expectation value of the plaquette as a function of Wilson coupling $\beta$  on $8^d$ lattices for (top) $2+1d$ (bottom) $3+1d$. The shaded region indicates $\beta\leq \beta_s$.} 
\end{figure}

In the interest of studying near-term quantum simulations, we should consider $2+1d$ theories in addition to $3+1d$ ones. Using classical lattice simulations, we determined $\beta_f>\beta_s$ in both space-times for the Wilson action (See Fig.~\ref{fig:freezing}). Thus quantum simulations with $\bo$ can be performed with the Kogut-Susskind Hamiltonian~\cite{PhysRevD.11.395}, although using an improved Hamiltonian can reduce either qubits or lattice spacing errors~\cite{Carena:2022kpg}.

In this paper, the four necessary primitive quantum gates (inversion, multiplication, trace, and Fourier) for quantum  simulation of $\bo$ theories on qubit-based computers are constructed. In Sec.~\ref{sec:group}, important group theory for $\bo$ is summarized and the qubit encoding is presented. A brief review of entangling gates used is found in Sec.~\ref{sec:qgates}. Sec.~\ref{sec:gates} provides an overview of the primitive gates. This is followed by
explicit quantum circuits for each gate for $\bo$: 
the inversion gate in Sec.~\ref{sec:inverse}, the multiplication gate in Sec.~\ref{sec:multiplication}, the trace gate in Sec.~\ref{sec:trace}, and the Fourier transform gate in Sec.~\ref{sec:fourier}. Using these gates, Sec.~\ref{sec:resources} presents a resource estimates for simulating $3+1d$ $SU(2)$.  We conclude and discuss future work in Sec.~\ref{sec:conclusions}.

\section{\label{sec:group}Properties of \texorpdfstring{$\bo$}{BO}}
The simulation of LGT requires defining a register where one can store the state of a bosonic link variable which we call a $G-$register. Toconstruct the $\bo-$register in term of integers, it is necessary to map the 48 elements of $\bo$ to the integers $[0,47]$. A clean way to obtain this is to write every element of $\mathbb{BO}$ as an ordered product of five generators with exponents written in terms of the binary variables $x_i$ with $i=[1,6]$:

\begin{equation}
\label{eq:pres}
    g = (-1)^{x_1} \mathbf{j}^{x_2} \mathbf{k}^{x_3} \mathbf{u}^{2 x_4 + x_5} \mathbf{t}^{x_6},
\end{equation}
with 
\begin{equation}
    \mathbf{u} = -\frac{1}{2} (\mathbb{1}+ \mathbf{i} + \mathbf{j} + \mathbf{k})\,\text{  and  }    \mathbf{t}=\frac{1}{\sqrt{2}}(\mathbb{1}+\mathbf{i})
\end{equation}
and $\mathbf{i}$, $\mathbf{j}$, $\mathbf{k}$ are the unit quaternions which in the 2d irreducible representation (irrep) correspond to Pauli matrices. With the construction of Eq.~(\ref{eq:pres}), the $\bo-$register is given by a binary qubit encoding with the ordering $|x_6x_5x_4x_3x_2x_1\rangle$.  While there exist $2^6$ possible state in a 6 qubit register, we only consider the 48 states where $x_4 + x_5 \leq 1$ represent the group elements.  The states where $x_4 = x_5 = 1$ correspond to \emph{forbidden states}. In this work we will use a short hand $\ket{N}$ where $N$ is the decimal representation of the binary $x_6x_5x_4x_3x_2x_1$.
For example, 
\begin{equation}
\frac{1}{\sqrt{2}}\begin{pmatrix}
\text{-}i&\text{-}1\\1&i
\end{pmatrix}=(-1)^1\bfj^0\bfk^1\bfu^{2\times 0+0}\bft^1\rightarrow|100101\rangle=|37\rangle
\end{equation}

The $\mathbf{i}$, $\mathbf{j}$, and $\mathbf{k}$ generators anti-commute with each other. Additional useful relations are:
\begin{equation}
\begin{split}
\label{eq:genrel}
&\mathbf{i}^2=\mathbf{j}^2=\mathbf{k}^2=-\mathbb{1},~ \mathbf{u}^3=\mathbb{1},~\mathbf{t}^2=\mathbf{i}\\
&\mathbf{i} \mathbf{j} = \mathbf{k}, ~ \mathbf{j} \mathbf{k} = \mathbf{i}, ~ \mathbf{k} \mathbf{i} = \mathbf{j},\\
&\mathbf{i}\mathbf{u}=\mathbf{u}\mathbf{k},~\mathbf{j}\mathbf{u}=\mathbf{u}\mathbf{i},~\mathbf{k}\mathbf{u}=\mathbf{u}\mathbf{j},\\
&\mathbf{i}\mathbf{t}=\mathbf{t}\mathbf{i},~-\mathbf{j}\mathbf{t}=\mathbf{t}\mathbf{k},~\mathbf{k}\mathbf{t}=\mathbf{t}\mathbf{j},
\end{split}
\end{equation}

\begin{table}[b]
    \caption{Character table of $\mathbb{BO}$ from \cite{Grimus:2011fk} and an enumeration of the elements in the given class.}
    \label{tab:charbt}
    \centering
    \begin{tabular}{c||c|c|c|c|c|c|c|c}
Size & 1 & 1 & 12 & 6 & 8 & 8 & 6 & 6\\
Ord. & 1 & 2 & 4 & 4 & 6 & 3 & 8 & 8 \\
\hline\hline               
$\rho_1$ & 1 & 1 & 1 & 1 & 1 & 1 & 1 & 1\\
$\rho_2$ & 1 & 1 & $-1$ & 1 & 1 & 1 & $-1$ & $-1$\\
$\rho_3$ & 2 & $2$ & 0 &2 & $-1$ & $-1$ & 0 & 0\\
$\rho_4$ & 2 & $-2$ & 0 & 0& 1 & $-1$ & $\sqrt{2}$ & $-\sqrt{2}$ \\
$\rho_5$ & 2 & $-2$ & 0 & 0& 1 & $-1$ & $-\sqrt{2}$ & $\sqrt{2}$ \\
$\rho_6$ & 3 & 3 & $-1$ & $-1$ & 0 & 0 & 1 & 1\\
$\rho_7$ & 3 & 3 & 1 & $-1$ & 0 & 0 & $-1$ & $-1$\\
$\rho_8$ & 4 & $-4$ & 0 & 0 & $-1$ & 1 & 0 & 0\\
\hline 
$\ket{g}$ & \tiny $\ket{0}$ &\tiny  $\ket{1}$ & \tiny $\ket{34}$-$\ket{37}$& \tiny $\ket{2}$-$\ket{7}$ & \tiny $\ket{9}$,$\ket{11}$ & \tiny $\ket{8}$,$\ket{10}$ & \tiny $\ket{32}$,$\ket{39}$&\tiny $\ket{33}$,$\ket{38}$ \\
\tiny \phantom{x} & \phantom{x} & \phantom{x} & \tiny $\ket{44}$-$\ket{49}$& \phantom{x}&\tiny  $\ket{13}$,$\ket{15}$ & \tiny $\ket{12}$,$\ket{14}$ & \tiny $\ket{41}$,$\ket{43}$&\tiny $\ket{40}$,$\ket{42}$ \\
\tiny \phantom{x} & \phantom{x} & \phantom{x} &\tiny  $\ket{52}$,$\ket{53}$& \phantom{x} &\tiny  $\ket{17}$,$\ket{18}$ & \tiny $\ket{16}$,$\ket{19}$&\tiny  $\ket{50}$,$\ket{54}$&\tiny $\ket{51}$,$\ket{55}$ \\
\tiny \phantom{x} & \phantom{x} & \phantom{x} & \phantom{x} &  & \tiny $\ket{20}$,$\ket{22}$ & \tiny $\ket{21}$,$\ket{23}$ & \phantom{x}& \\
    \end{tabular}
\end{table}

The character table (Table~\ref{tab:charbt}) lists important group properties; the different irreps can be identified by the value of their character acting on each element. An irrep's dimension is the value of the character of $\mathbb{1}$.  There are three $1d$ irreps, three $2d$ irreps (one real and two complex), and one $3d$ irrep.  To derive the Fourier transform, it is necessary to know a matrix presentation of each irrep. Based on our qubit mapping, given a presentation of $-1,\mathbf{i},\mathbf{j},$ and $\mathbf{l}$ we can construct any element of the group from Eq.~(\ref{eq:pres}). With the $n$-th root of unity $\omega_n=e^{2\pi i/n}$, the matrix presentations of our generators in each irrep are found in Tab.~\ref{tab:irreps}.

\begin{table*}
    \caption{Matrix representations of the generators used for digitization of $\bo$}
    \label{tab:irreps}
    \centering
    \begin{tabular}{c|ccccc}
         g&  -1&  $\bfj$&  $\bfk$&  $\bfu$& $\bft$\\
         \hline\hline
         $\rho_1$ &  1&  1&  1&  1& 1\\
         $\rho_2$&  1&  1&  1&  1& -1\\
         $\rho_3$&  $\begin{pmatrix}
             1 &0\\0 &1
         \end{pmatrix}$ & $\begin{pmatrix}
             1 &0\\0 &1
         \end{pmatrix}$ & $\begin{pmatrix}
             1 &0\\0 &1
         \end{pmatrix}$ & $\begin{pmatrix}
             \omega_3^2 &0\\0 &\omega_3
         \end{pmatrix}$ & $\begin{pmatrix}
             0 &1\\1 &0
         \end{pmatrix}$\\
         $\rho_4$& $\begin{pmatrix}
     -1 & 0 \\  0 & -1
    \end{pmatrix}$ & $\begin{pmatrix}
        0 & 1 \\ -1 & 0
    \end{pmatrix}$ & $\begin{pmatrix}
        i & 0 \\ 0 & -i
    \end{pmatrix}$ & $\frac{1}{2}\begin{pmatrix}
        -1-i & -1+i \\ 1+i & -1+i
    \end{pmatrix}$
    & $\frac{1}{\sqrt{2}}\begin{pmatrix}
        1 & - i \\ -i & 1
    \end{pmatrix}$ \\
         $\rho_5$& $\begin{pmatrix}
     -1 & 0 \\
     0 & -1
    \end{pmatrix}$  & $\begin{pmatrix}
       0 & 1\\ -1& 0 
    \end{pmatrix}$ & $\begin{pmatrix}
        i & 0 \\ 0 & -i
    \end{pmatrix}$  & $\frac{1}{2}\begin{pmatrix}
       -1-i & -1+i \\ 1+i & -1+i
    \end{pmatrix}$ 
    & $\frac{1}{\sqrt{2}}\begin{pmatrix}
        -1 &  i \\ i & -1
    \end{pmatrix}$ \\

    
         $\rho_6$&  $\begin{pmatrix}
             1 & 0 & 0 \\
             0 & 1 & 0 \\
             0 & 0 & 1 \\
         \end{pmatrix}$  & $\begin{pmatrix}
             -1 & 0 & 0 \\
             0 & 1 & 0 \\
             0 & 0 & -1 \\
         \end{pmatrix}$ & $\begin{pmatrix}
             1 & 0 & 0 \\
             0 & -1 & 0 \\
             0 & 0 & -1 \\
         \end{pmatrix}$ & $\begin{pmatrix}
             0 & 1 & 0 \\
             0 & 0 & 1 \\
             1 & 0 & 0 \\
         \end{pmatrix}$ & $\begin{pmatrix}
             0 & -1 & 0 \\
             1 & 0 & 0 \\
             0 & 0 & -1 \\
         \end{pmatrix}$\\
         
    $\rho_7$& $\begin{pmatrix}
             1 & 0 & 0 \\
             0 & 1 & 0 \\
             0 & 0 & 1 \\
         \end{pmatrix}$ & $\begin{pmatrix}
             -1 & 0 & 0 \\ 0 & 1 & 0 \\ 0 & 0 & -1
         \end{pmatrix}$ & $\begin{pmatrix}
             1 & 0 & 0 \\ 0 & -1 & 0 \\ 0 & 0 & -1
         \end{pmatrix}$ & $\begin{pmatrix}
             0 & 1 & 0 \\ 0 & 0 & 1 \\ 1 & 0 & 0
         \end{pmatrix}$ & $\begin{pmatrix}
             0 & 1 & 0 \\ -1 & 0 & 0 \\ 0 & 0 & 1
         \end{pmatrix}$\\     
         $\rho_8$&  $\begin{pmatrix}
             -1 & 0 & 0 & 0 \\
             0 & -1 & 0 & 0\\
             0 & 0 & -1 & 0\\
             0 & 0 & 0 & -1\\
         \end{pmatrix}$ & $\begin{pmatrix}
             0 & -i & 0 & 0 \\
             -i & 0 & 0 & 0\\
             0 & 0 & -i & 0\\
             0 & 0 & 0 & i\\
         \end{pmatrix}$ & $\begin{pmatrix}
             i & 0 & 0 & 0 \\
             0 & -i & 0 & 0\\
             0 & 0 & 0 & -i\\
             0 & 0 & -i & 0\\
         \end{pmatrix}$ & $\frac{\omega_3}{2}\begin{pmatrix}
    (-1-i)\omega_3 & (1+i)\omega_3 & 0 & 0 \\
    (-1+i)\omega_3 & (-1+i)\omega_3 & 0 & 0 \\
    0 & 0 & -1 + i & 1 + i \\
    0 & 0 & -1 + i& -1 - i
\end{pmatrix}$ & $\begin{pmatrix}
    0 & 0 & 0 & -1\\ 0 & 0 & 1 & 0 \\ 1 & 0 & 0 & 0 \\ 0 & 1 & 0 & 0
\end{pmatrix}$\\
    \end{tabular}
\end{table*}

\section{Qubit Gates}
\label{sec:qgates}

To construct our primitive gates, we chose a universal, albeit redundant, basic qubit quantum gate set.  We use the Pauli gates $p=X,Y,Z$ and their arbitrary rotation generalizations $R_p(\theta)=e^{i\theta p/2}$.  When decomposing onto fault-tolerant devices, how these are decomposed in terms of the $T=\text{diag}(1,e^{i\pi/4})$ gate becomes relevant to resource estimations.

We also use the SWAP gate
        \begin{equation*}
            \text{SWAP} ~\ket{a}\otimes\ket{b} = \ket{b}\otimes\ket{a},
        \end{equation*}
and CNOT gate
        \begin{equation*}
            \text{CNOT} \ket{a}\otimes\ket{b} = \ket{a}\otimes\ket{b\oplus a},
        \end{equation*}
We further use the multiqubit C$^n$NOT -- of which C$^2$NOT is called the Toffoli gate -- and CSWAP (Fredkin) gates. The C$^n$NOT gate consists of one target qubit and $n$ control qubits. For example, the Toffoli in terms of modular arithmetic is
        \begin{equation*}
            \text{C}^2\text{NOT} \ket{a}\otimes\ket{b}\otimes\ket{c}=\ket{a}\otimes\ket{b}\otimes\ket{c \oplus ab}.
        \end{equation*}
The CSWAP gate swaps two qubit states if the control is in the $\ket{1}$ state:
        \begin{equation*}
        \begin{split}
            \text{CSWAP}\ket{a}\otimes\ket{b}\otimes\ket{c}=&\ket{a}\otimes\ket{b(1\oplus a)\oplus ac}\\&\otimes\ket{c(1\oplus a)\oplus a b}.\\
            \end{split}
        \end{equation*}
        
\section{\label{sec:gates}Overview of Primitive Gates}

One can define any quantum circuit for gauge theories via a set of primitive gates, of which one choice is: inversion $\mathfrak U_{-1}$, multiplication $\mathfrak U_{\times}$, trace $\mathfrak U_{\rm Tr}$, and Fourier transform $\mathfrak U_{F}$~\cite{Lamm:2019bik}. The inversion gate, $\mathfrak U_{-1}$, takes a $G$-register to its inverse:
\begin{equation}
\mathfrak U_{-1} \left|g\right> = \left|g^{-1}\right>\text.
\end{equation}
$\mathfrak U_{\times}$ takes a target $G-$register and changes it to the left product controlled by a second $G-$register:
\begin{equation}
    \mathfrak U_{\times} \ket{g}\ket{h} = \ket{g} \ket{gh}.
\end{equation}
Left multiplication is sufficient for a minimal set as right multiplication can be obtained from two applications of $\mathfrak U_{-1}$ and $\mathfrak U_{\times}$, albeit resource costs can be further reduced by an explicit construction~\cite{Carena:2022kpg}.

Traces of group elements generally define the lattice Hamiltonian.  We can implement the evolution with respect to these terms via:
\begin{equation}
\mathfrak U_{\Tr}(\theta) \left|g\right> = e^{i \theta \Re\Tr g} \left|g\right>.
\label{eqn:trace-gate}
\end{equation}

\begin{figure}
    \centering
    \includegraphics[width=0.8\linewidth]{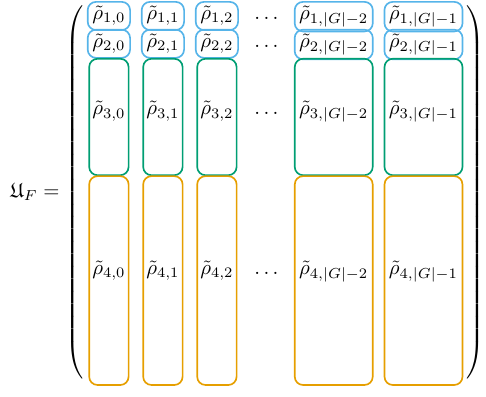}
    \caption{Example $\mathfrak U_{F}$ from Eq.~(\ref{eqn:Fourier-group}) using $\tilde\rho_{i,j}=\sqrt{d_\rho/|G|}\rho_{i,j}$ where $\rho_{i,j}=\rho_i(g_j)$. This example has four irreps with $d_1=d_2<d_3<d_4$.  $\mathfrak U_{F}$ is square since $\sum_\rho d_{\rho}^2=|G|$}
    \label{fig:qft_cartoon}
\end{figure}

The final gate of this set is $\mathfrak U_F$. The Fourier transform, $\hat f$, of a function $f$ over a finite $G$ is
\begin{eqnarray}
\hat{f}(\rho) = \sum_{g \in G}\sqrt{\frac{d_{\rho}}{|G|}}  f(g) \rho(g),
\label{eqn:Fourier-group}
\end{eqnarray}
where $\vert G \vert$ is the size of the group, $d_{\rho}$ is the dimensionality of the irreducible representation (irrep) $\rho$.
The inverse transform is given by
\begin{eqnarray}
\label{eq:dft}
f(g) =  \sum_{\rho \in \hat{G}} \sqrt{\frac{d_{\rho}}{|G|}} \tr{(\hat{f}(\rho) \rho(g^{-1}))},
\end{eqnarray}
where the dual $\hat{G}$ is the set of irreducible representations (irrep) of $G$. A paradigm of this unitary matrix is show in Fig.~\ref{fig:qft_cartoon} which can then be transformed into a gate. $\mathfrak U_{F}$ then acts on a single $G$-register with some amplitudes $f(g)$ which rotate it into the irrep basis:
\begin{equation}
\label{eq:uft}
\mathfrak U_F \sum_{g \in G} f(g)\left|g \right>
=
\sum_{\rho \in \hat G} \hat f(\rho)_{ij} \left|\rho,i,j\right>.
\end{equation}

\section{\label{sec:inverse}Inversion Gate}

Consider a $\bo-$register storing the group element given by $g=(-1)^{x_1} \mathbf{j}^{x_2} \mathbf{k}^{x_3} \mathbf{u}^{2 x_4 + x_5} \mathbf{t}^{x_6}$. The effect of the inversion gate on this register is to transform it to
\begin{equation}
    |g\rangle=|x_1 x_2 x_3 x_4 x_5 x_6\rangle\rightarrow |g^{-1}\rangle=|y_1 y_2 y_3 y_4 y_5 y_6\rangle.
\end{equation}
Where $y_i$ must be determined. With Eq.~(\ref{eq:pres}), $g^{-1}$ is 
\begin{equation}
\label{eq:inverseexpression}
    g^{-1} = \bft^{8 - x_6} \bfu^{3 - (2x_4 + x_5)}\bfk^{x_3} \bfj^{x_2}(-1)^{x_1 + x_2 + x_3} 
\end{equation}
A systematic way to build the $\mathfrak{U}_{-1}$ of $\bo$ is to embed the inverse gates of subgroups of $\bo$. We start with the expression of Eq.~(\ref{eq:inverseexpression}), and begin by reordering the $\mathbb{Q}_8$ subgroup so that the element is of the form:
\begin{align}
\label{eq:q8inverseofbo}
\bfk^{x_3} \bfj^{x_2}(-1)^{x_1 + x_2 + x_3} &= 
(-1)^{a_1} \bfj^{a_2}\bfk^{a_3},
\end{align}
where $a_1 = x_1 + x_2 + x_3 + x_2 x_3$ and $a_2 = x_2$, $a_3 = x_3$.
The next step is commuting thru $\bfu$ to obtain
\begin{equation}
\label{eq:btinverseofbo}
\bfu^{3 - (2x_4 + x_5)}(-1)^{a_1} \bfj^{a_2}\bfk^{a_3}=(-1)^{b_1} \bfj^{b_2}\bfk^{b_3}\bfu^{2b_4 + b_5},
\end{equation}
which corresponds to the $\btt$ inverse gate where
\begin{equation}
\begin{split}
\label{eq:btinvrel}
    b_2 & = a_2 (1 + x_4) + a_3 (x_4 + x_5)\\
    b_3 & = a_3 (1 + x_5) + a_2 (x_4 + x_5)\\
    b_4 &= x_5\\
    b_5 &= x_4
\end{split}
\end{equation}
while $b_1=a_1$ is unchanged. Finally, commuting thru $\bft$,
\begin{equation}
    \bft^{8 - x_6}(-1)^{b_1} \bfj^{b_2}\bfk^{b_3}\bfu^{2b_4 + b_5}=(-1)^{y_1} \bfj^{y_2}\bfk^{y_3}\bfu^{2y_4 + y_5}\bft^{y_6}.
\end{equation}
Propagating $\bft^{x_6}$ through the $\mathbb{Q}_8$ portion yields $
    g^{-1} = (-1)^{c_1}\mathbf{j}^{c_2}\mathbf{k}^{c_3}\mathbf{t}^{x_6}\mathbf{u}^{2b_4 + b_5}$ where,
\begin{align}
    c_1 &= b_1 + x_6(1-b_3)(1-b_2)\notag\\
    c_2 &= (1-b_3)x_6 + (1- x_6)b_2\notag\\
    c_3 &= (1-b_2)x_6 + (1- x_6)b_3\notag.
\end{align}
Finally propagating through $\bft^{x_6}$ through $\bfu^{2b_4 + b_5}$ yields 
\begin{align}
    y_1 & = c_1 + x_6(b_4(1 - c_2) + c_3b_5)\notag\\
    y_2 & = c_2 + b_4 x_6 \notag\\
    y_3 & = c_3 + (b_4 + b_5) x_6\notag\\
    y_4 & = x_6 b_5 + (1 - x_6) b_4\notag\\
    y_5 & = x_6 b_4 + (1 - x_6) b_5
\end{align}
with $y_6=x_6$. Together, this yields for $|g\rangle$ to $|g^{-1}\rangle$:
\begin{align}
y_1 =& x_{1} + x_{2} x_{3} x_{6} + x_{2} x_{3} + x_{2} x_{5} x_{6} + x_{2} x_{6} + x_{2} \notag\\
&+ x_{3} x_{4} x_{6} + x_{3} x_{6} + x_{3} + x_{4} x_{6} + x_{6} \notag\\
y_2 =& x_{2} x_{4} + x_{2} x_{5} x_{6} + x_{2} x_{6} + x_{2} + x_{3} x_{4} x_{6} + x_{3} x_{4}\notag\\
&+ x_{3} x_{5} + x_{3} x_{6} + x_{5} x_{6} + x_{6} \notag\\
y_3 = & (x_2 + x_3)(x_5 + x_6) + x_6(x_4 (x_3 + 1) + x_5(x_2 + 1))\notag\\
& + x_2 x_4 + x_3 + x_6\notag\\
y_4 = &(x_{4} + x_{5})x_{6} + x_{5} \notag\\
y_5 = &(x_{4} + x_5) x_{6} + x_{4}
\end{align}
where $y_6=x_6$ is unaffected.

\begin{figure}
    \centering
    \includegraphics[width=\linewidth]{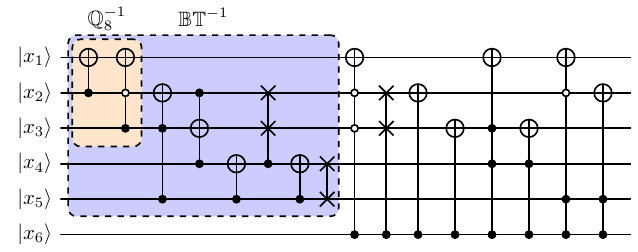}
    \caption{Quantum circuit for $\mathfrak{U}_{-1}$ for $\mathbb{BO}$.  The subgates correspond to  $\mathfrak{U}_{-1}$ for $\mathbb{Q}_8$ ($\btt$) in orange (blue)}
    \label{fig:boinvcirc}
\end{figure}

\section{\label{sec:multiplication}Multiplication Gate}
\begin{figure*}
    \centering
    \includegraphics[width=1\linewidth]{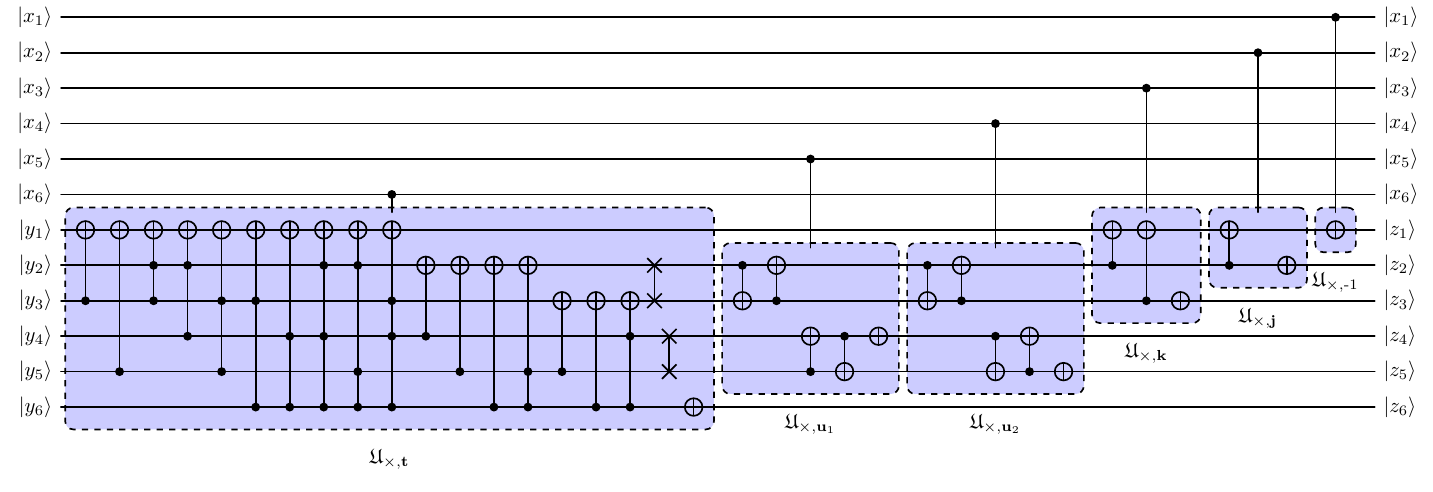}
    \caption{The decomposition of the multiplication gate, $\mathfrak{U}_{\times}$, into a product of multiplication by each generator.}
    \label{fig:multgate}
\end{figure*}

Given two $\bo-$registers $\ket{g}=\ket{\prod x_i}$ and $\ket{h}=\ket{\prod y_i}$, 
we want $\ket{gh}=\ket{\prod z_i}$. Here, we again decompose
\begin{equation}
    \mathfrak U_{\times}=\mathfrak U_{\times,\text{-}1}\mathfrak U_{\times,\bfj}\mathfrak U_{\times,\bfk}\mathfrak U_{\times,\bfu_2}\mathfrak U_{\times,\bfu_1}\mathfrak U_{\times,\bft}
\end{equation} 
into gates controlled by a single qubits. For each step, we use temporary variables indexed by other letters, e.g. $a_i$. The first gate, $\mathfrak U_{\times,\bft}$, controlled by $x_6$, sets:
\begin{equation}
\begin{split}
   a_1 &= y_2y_4y_6 + y_3y_4y_6 + y_2y_5y_6 + y_2y_3 + y_2y_4\\
   &\phantom{xx}+ y_3y_5+ y_3y_6 + y_4y_6 + y_1 + y_3 + y_5 \\
    a_2 &= y_4y_6 + y_3 + y_5 + y_6\\
    a_3 &= y_5y_6 + y_2 + y_4 + y_5 + y_6\\
    a_4 &= y_5\\
    a_5 &= y_4\\
    z_6 &= y_6 + 1
\end{split}
\end{equation}
Then one acts with $\mathfrak U_{\times,\bfu_1}$ which is controlled by $x_5$:
\begin{equation}
\begin{split}
    b_2 &= a_{2} + a_{3}\\
    b_3 &= a_{2}\\
    b_4 &= a_{5}\\
    b_5 &= a_{4} + a_{5} + 1
\end{split}
\end{equation}
where $a_1$ and $z_6$ do not change. Next, applying $\mathfrak U_{\times,\bfu_2}$ controlled by $x_4$,
\begin{equation}
\begin{split}
    c_2 &= b_3\\
    c_3 &= b_{2} + b_{3}\\
    z_4 &= b_{4} + b_{5} + 1\\
    z_5 &= b_4
\end{split}
\end{equation}
with $a_1$ and $z_6$ unchanged. $\mathfrak U_{\times,\bfk}$, controlled by $x_3$, sets
\begin{equation}
\begin{split}
    d_1 &=a_{1} + c_{2} + c_{3}\\
    z_3 &= c_{3} + 1
\end{split}
\end{equation}
where $c_2,z_4,z_5$, and $z_6$ are unchanged. Then, $\mathfrak U_{\times,\bfj}$ controlled by $x_2$ is:
\begin{equation}
\begin{split}
    e_1 &= d_{1} + c_{2}\\
    z_2 &= c_{2} + 1
\end{split}
\end{equation}
with $z_3$ thru $z_6$ unchanged. Finally, $\mathfrak U_{\times,\text{-}1}$ needs $z_1=e_1+1$ controlled by $x_1$ i.e. a CNOT. The full $\mathfrak{U}_{\times}$ is in Fig.~\ref{fig:multgate}.

\section{\label{sec:trace}Trace gate}
For simulating LGT, the Hamiltonian requires gauge-invariant operators.  Without matter, all terms can be constructed from traces of $g$ -- and thus $\mathfrak U_{\rm Tr}$ -- in the fundamental irrep, $\rho_4$.  The Tab.~\ref{tab:charbt} provides us with $\re\tr(g_i)$. In previous works~\cite{Lamm:2019bik,Alam:2021uuq,Gustafson:2022xdt}, $\mathfrak{U}_{\rm Tr}$ was derived from a Hamiltonian, $H_{\rm Tr}$ with $\re\tr( g_i)$ as eigenvalues. For our $\bo$-register, $H_{\rm Tr}$ would require 20 Pauli strings 
which results in $\mathfrak U_{\rm Tr}=e^{i\theta H_{\rm Tr}}$ decomposing into at least 20 $R_Z(\theta)$ gates.  In a fault-tolerant calculation, $R_Z$ gates require synthesis from T gates, and thus can be unduly expensive.  

\begin{figure*}
    \centering
    \includegraphics[width=1\linewidth]{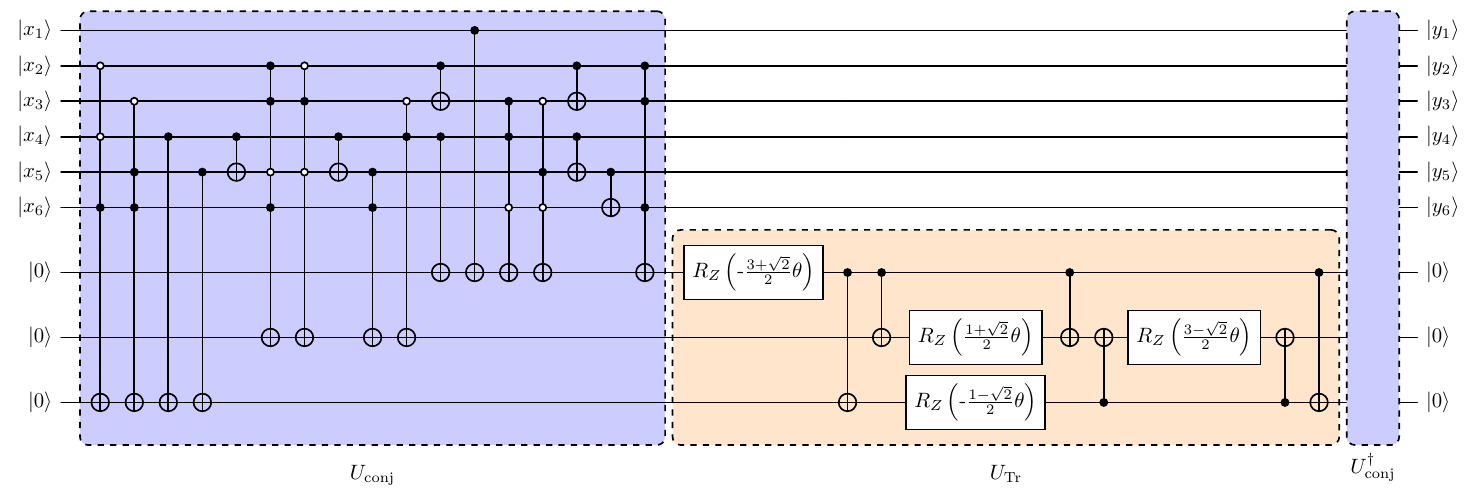}
    \caption{A qubit implementation of $\mathfrak U_{\rm Tr}$.}
    \label{fig:squish+trace}
\end{figure*}

Here, we explore a different implementation of $\mathfrak U_{\rm Tr}$ that is advantageous for discrete groups where $\re\tr (g)$ are limited by the number of conjugacy classes. In this case, $\mathfrak U_{\rm Tr}$ could be decomposed into two gates: a gate $U_{conj}$ to map the 48 $\ket{g}$ to the 8 conjugacy classes $\ket{c}$, and a gate $U_{\rm Tr}$ which computes the traces for each conjugacy class. Together, we obtain a circuit for $\mathfrak U_{\tr}=U_{\rm{conj}} U_{\rm{Tr}}(\theta) U_{\rm{conj}}^{\dagger}$ which should have fewer $R_Z(\theta)$ at the cost of additional C$^N$NOT gates which have $\epsilon$-independent T gate costs. Taking the assignment of Table~\ref{tab:traces} for the traces we derive
\begin{equation*}
    (-1)^{x_1} \bfj^{x_2} \bfk^{x_3} \bfu^{2 x_4 + x_5} \bft^{x_6} \mapsto (v_1,v_2,v_3)
\end{equation*}
where
\begin{align}
    v_1 &= (x_6 + 1)(x_1 + (x_4 + x_5)(1 + x_2)(1 + x_3)) \notag\\&+ x_6(x_5x_1+x_4(1+x_1)+(1+x_4+x_5)(x_2x_3 + x_1))\notag\\
    v_2 &= (x_6 + 1)((1 + x_2)(1+x_3)(1+x_4+x_5)) \notag\\&+ x_6(x_5x_2 + x_4(1+x_3)+(1+x_4+x_5)(1+x_2+x_3))\notag\\
    v_3 &= (x_6 + 1)(x_4 + x_5) + x_6(x_5x_2 + x_4(1+x_3) \notag\\&+ (1 + x_4 + x_5)(1 + x_2 + x_3)).
\end{align}
 A qubit-based circuit for $\mathfrak U_{\rm Tr}$ is shown in Fig.~\ref{fig:squish+trace}. With this, we will estimate the reduction in fault-tolerant resource gate costs in Sec.~\ref{sec:resources} compared to the $H_{\tr}$ method.

\begin{table}
\caption{Trace elements and associated bit strings for the conjugacy classes stored in the ancilla.  Note, $v_1$ is a sign bit.}
\label{tab:traces}
\begin{tabular}{c|cccccccc}
\hline\hline
$\re\tr(g)$ & 0 & 0 & 2 & -$2$ & 1 & -$1$ & $\sqrt{2}$ & -$\sqrt{2}$ \\\hline
$v1$& 0 & 1 & 0 & 1 & 0 & 1 &0 & 1\\
$v2$& 0 & 0 & 1 & 1 & 0 & 0 &1 & 1\\
$v3$& 0 & 0 & 0 & 0 & 1 & 1 &1 & 1
\\\hline\hline
\end{tabular}
\end{table}
\section{\label{sec:fourier}Fourier Transform}
The standard $n$-qubit quantum Fourier transform (QFT)~\cite{nielsen_chuang_2010} corresponds to the quantum version of the fast Fourier transform of $\mathbb{Z}_{2^n}$. 
Quantum Fourier transforms over some nonabelian groups exist~\cite{hoyer1997efficient,beals1997quantum,puschel1999fast,moore2006generic,Alam:2021uuq}. Alas, for the crystal-like subgroups, efficient QFT circuits are currently unknown~\cite{childs2010quantum}. In general, no clear algorithmic way to construct the QFT exists.  Thus, we instead construct a suboptimal $\mathfrak U_{F}$ from Eq.~(\ref{eqn:Fourier-group}) from the irreps. 

Since $\bo$ has 48 elements, on a qubit device $\mathfrak U_{F}$ must be embedded into a larger $2^d\times2^d$ unitary. The columns index qubit bitstrings from $\ket{0}$ to $\ket{63}$ where the physical states are given by the subset of $\ket{g}$ in Table~\ref{tab:charbt}.  We then index the irreducible representation $\rho_i$ sequentially from $i=1$ to $i=8$, and construct the matrix, including appropriate padding for the forbidden states. This matrix was then passed to the \textsc{Qiskit} v0.43.1 transpiler, and an optimized version of $\mathfrak U_{F}$ needed 1839 CNOTs, 166 $R_X$, 1996 $R_Y$, and 3401 $R_Z$ gates\footnote{In \cite{Gustafson:2022xdt}, the \textsc{Qiskit} v0.37.2 was used, yielding for the $\btt$  $\mathfrak U_{F}$ of 1025 CNOTs, 1109 $R_Y$, and 2139 $R_Z$ gates.  With the v0.43.1 transpiler, we obtain 442 CNOTs, 40 $R_X$, 494 $R_Y$, and 835 $R_Z$.}; the Fourier gate is the most expensive qubit primitive and dominates simulation costs.

\section{\label{sec:resources}Resource Estimates}
Clearly quantum practicality with $\bo$ will require error corrections. Since the Eastin-Knill theorem restricts QEC codes from having a universal transversal sets of gates~\cite{PhysRevLett.102.110502}, compromises must be made.  Typically, the Clifford gates are transversal \cite{Chuang:1996hw,1996PhRvA..54.1098C,PhysRevLett.77.793,1996RSPSA.452.2551S,PhysRevA.54.4741} while the T gate is not. Thus T gate counts are often used in fault-tolerant resource analysis~\cite{Chuang:1996hw,1997RuMaS..52.1191K}. Recently, novel universal sets have been proposed with transversal $\btt,\bo,\bi$ gates~\cite{Kubischta:2023nlb,Denys:2023syu,Jain:2023deu,Denys:2022iyj} which deserve investigation for use with LGT.

\begin{table}[]
    \centering
    \caption{Number of physical T gates and clean ancilla required to implement logical gates for (top) basic gates taken from~\cite{Chuang:1996hw} (bottom) primitive gates for $\bo$.}
    \label{tab:tgatecost}
    \begin{tabular}{ccc}
    \hline\hline
         Gate & T gates & Clean ancilla\\
         \hline
         C$^2$NOT & 7 & 0\\
         C$^3$NOT & 21 & 1\\
         C$^4$NOT & 35 & 2\\
         CSWAP & 7 & 0\\
         $R_Z$ & 1.15 $\log_2(1/\epsilon)$ & 0\\
         \hline
         $\mathcal{U}_{-1}$ & 112 & 1\\
         $\mathcal{U}_{\times}$ & 392 & 4\\
         $\mathcal{U}_{Tr}$ & $350$\footnote{With 6 additional ancilla, this can be reduced to $188$}$ + 4.6 \log_2(1 / \epsilon)$& 2\\
         $\mathcal{U}_{FT}$ & $11370.1\log_2(1/\epsilon)$ & 0\\\hline\hline
    \end{tabular}
\end{table}

The Toffoli gate requires 7 T gates~\cite{Chuang:1996hw} and one method for constructing the C$^n$NOT gates is with $2\lceil\log_2{n}\rceil-1$ Toffoli gates and $n-2$ dirty ancilla qubits\footnote{A \emph{clean} ancilla is in state $\ket{0}$. \emph{Dirty} ancillae have unknown states.} which can be reused later~\cite{2019arXiv190401671B,Chuang:1996hw,PhysRevA.52.3457}. We arrive at the cost for the $R_Z$ gates via \cite{PhysRevLett.114.080502} where these gates can be approximated to precision $\epsilon$ with on average $1.15 \log_2(1/\epsilon))$ T gates (and at worst $-9 + 4 \log_2(1/\epsilon)$~\cite{10.5555/2685188.2685198}). Further, $R_Y$ and $R_X$ can be replaced by at most 3 $R_Z$. With these, we can construct gate estimates for $\bo$ (See Tab.~\ref{tab:tgatecost}). We note that for $\epsilon\lesssim 10^{-6}$, the 20 $R_Z(\theta)$ required for $e^{i\theta H_{\tr}}$ is more expensive than the $\mathfrak U_{\tr}$ constructed using $U_{\rm conj}$.

\begin{table}[ht]
   \caption{Number of primitive gates per link per $\delta t$ neglecting boundary effects as a function of $d$ for $H_{KS}$ and $H_{I}$.}
    \label{tab:primcost}
    \begin{tabular}{c|c|c}
    \hline\hline
    Gate & $N[H_{KS}]$&$N[H_{I}]$\\
    \hline
    $\mathfrak U_F$ & 2  &4\\
    $\mathfrak U_{\rm Tr}$&$\frac{1}{2}(d-1)$ &$\frac{3}{2}(d-1)$\\
    $\mathfrak U_{-1}$& $3(d-1)$ & $2+11(d-1)$\\
    $\mathfrak U_{\times}$& $6(d-1)$ &$4+26(d-1)$\\
    \hline\hline
    \end{tabular}
\end{table}
Primitive gate costs for implementing $H_{KS}$~\cite{PhysRevD.11.395} and $H_I$~\cite{Carena:2022kpg}, per link per Trotter step $\delta t$ are shown in Tab.~\ref{tab:primcost}.  Using these, we can determine the total T gate count $N^{H}_T=C^H_T\times d L^d N_t$ for a $d$ spatial lattice simulated for a time $t=N_t\delta t$. We find that for $H_{KS}$
\begin{equation}
    C^{KS}_{T}=2863( d-1)+(22737.8+ 2.3d)\log_2\frac{1}{\epsilon}.
\end{equation}
With this, the total synthesis error $\epsilon_T$ can be estimated as the sum of $\epsilon$ from each $R_Z$.  In the case of $H_{KS}$ this is 
\begin{equation}
    \epsilon_T=2 (9886 + d)d L^d N_t\times \epsilon.
\end{equation}
If one looks to reduce lattice spacing errors for a fixed number of qubits, one can use $H_I$ which would require
\begin{equation}
    C^{I}_{T}=11949d-10157+(45473.3+ 6.9d)\log_2\frac{1}{\epsilon}.
\end{equation}
where the total synthesis error is
\begin{equation}
    \epsilon_T = 2 (19771 + 3 d) d L^d N_t \times\epsilon
\end{equation}
Following \cite{Cohen:2021imf,Kan:2021xfc,Gustafson:2022xdt}, we will make resource estimates based on our primitive gates for the calculation of the shear viscosity $\eta$ on a $L^3=10^3$ lattice evolved for $N_t=50$, and total synthesis error of $\epsilon_T=10^{-8}$. Considering only the time evolution and neglecting state preparation (which can be substantial~\cite{Xie:2022jgj,Davoudi:2022uzo,Avkhadiev:2019niu,Gustafson:2022hjf,peruzzo2014variational,Gustafson:2019mpk,Gustafson:2019vsd,Harmalkar:2020mpd,Gustafson:2020yfe,Jordan:2017lea,PhysRevLett.108.080402,brandao2019finite,Clemente:2020lpr,motta2020determining,deJong:2021wsd,Gustafson:2023ayr,Kane:2023jdo}), Kan and Nam estimated $3\times10^{19}$ T gates would be required for an pure-gauge $SU(2)$ simulation of $H_{KS}$. This estimate used a truncated electric-field digitization which requires substantial fixed-point arithmetic -- greatly inflating the T gate cost. Using $\btt$, we presently estimate that $1.1\times10^{11}$ T gates would be sufficient\footnote{This supersedes~\cite{Gustafson:2022xdt} due to the improved transpiler, proper consideration of $\epsilon_T$, and better cost estimates of $R_X,R_Y$.} but only $H_I$ could be used. Here, the $\bo$ group would require $4.1\times 10^{11}$ T gates if $H_{I}$ were used -- albeit with smaller lattice spacing errors than Kan and Nam since they used $H_{KS}$. Thus $\bo$ reduces the gate costs of~\cite{Kan:2021xfc} by $10^8$ for fixed $L$ by avoiding quantum fixed-point arithmetic while allowing for smaller lattice spacings that $\btt$. If $H_{KS}$ were used the T gate cost is reduced to $2.0\times 10^{11}$.  Alternatively, if we take the heuristic of~\cite{Carena:2022kpg} that for $a=\mathcal{O}(0.1\text{ fm})$ using $H_{I}$ could reduce $L$ by half for fixed lattice spacing errors with only $4.9 \times 10^{10}$ T gates.  Similar to $\btt$, $\mathfrak{U}_{F}$ dominates the simulations -- 99\% and 98\% of the total cost of simulation $H_{KS}$ and $H_I$ respectively.

\section{Conclusions}
\label{sec:conclusions}
In this paper, we constructed the necessary primitive gates for the simulation of $\bo$ -- a crystal-like subgroup of $SU(2)$ -- gauge theories and quantum resource estimates were made for the simulation of pure $SU(2)$ shear viscosity.  Compared to previous fault-tolerant qubit estimates of electric basis truncations, we require $10^{8}$ fewer T gates by avoiding quantum fixed point arithmetic via the discrete group approximation. Further, reducing digitization error compared to $\btt$ increase the total cost by a factor of $\sim4$ perhaps suggesting a $|G|^2$ scaling. 

Looking forward, primitive gates should be constructed for $\bi$ and to the subgroups of $SU(3)$.  At the cost of more qubits, $\bi$ would allow smaller digitization error and lattice spacings. To further reduce the qubit-based simulation gate costs for all discrete subgroup approximations, the formalism for deriving the quantum Fourier transform for crystal-like groups remains of paramount interest.

\begin{acknowledgements}
The authors thank M. S. Alam, R. Van de Water and M. Wagman for insightful comments and support over the course of this work. EG was supported by the NASA Academic Mission Services, Contract No. NNA16BD14C. HL and FL are supported by the Department of Energy through the Fermilab QuantiSED program in the area of ``Intersections of QIS and Theoretical Particle Physics". This material is based on work supported by the U.S. Department of Energy, Office of Science, National Quantum Information Science Research Centers, Superconducting Quantum Materials and Systems Center (SQMS) under contract number DE-AC02-07CH11359 (EG). Fermilab is operated by Fermi Research Alliance, LLC under contract number DE-AC02-07CH11359 with the United States Department of Energy. 

\end{acknowledgements}
\bibliography{ref}

\begin{thebibliography}{187}%
\makeatletter
\providecommand \@ifxundefined [1]{%
 \@ifx{#1\undefined}
}%
\providecommand \@ifnum [1]{%
 \ifnum #1\expandafter \@firstoftwo
 \else \expandafter \@secondoftwo
 \fi
}%
\providecommand \@ifx [1]{%
 \ifx #1\expandafter \@firstoftwo
 \else \expandafter \@secondoftwo
 \fi
}%
\providecommand \natexlab [1]{#1}%
\providecommand \enquote  [1]{``#1''}%
\providecommand \bibnamefont  [1]{#1}%
\providecommand \bibfnamefont [1]{#1}%
\providecommand \citenamefont [1]{#1}%
\providecommand \href@noop [0]{\@secondoftwo}%
\providecommand \href [0]{\begingroup \@sanitize@url \@href}%
\providecommand \@href[1]{\@@startlink{#1}\@@href}%
\providecommand \@@href[1]{\endgroup#1\@@endlink}%
\providecommand \@sanitize@url [0]{\catcode `\\12\catcode `\$12\catcode
  `\&12\catcode `\#12\catcode `\^12\catcode `\_12\catcode `\%12\relax}%
\providecommand \@@startlink[1]{}%
\providecommand \@@endlink[0]{}%
\providecommand \url  [0]{\begingroup\@sanitize@url \@url }%
\providecommand \@url [1]{\endgroup\@href {#1}{\urlprefix }}%
\providecommand \urlprefix  [0]{URL }%
\providecommand \Eprint [0]{\href }%
\providecommand \doibase [0]{https://doi.org/}%
\providecommand \selectlanguage [0]{\@gobble}%
\providecommand \bibinfo  [0]{\@secondoftwo}%
\providecommand \bibfield  [0]{\@secondoftwo}%
\providecommand \translation [1]{[#1]}%
\providecommand \BibitemOpen [0]{}%
\providecommand \bibitemStop [0]{}%
\providecommand \bibitemNoStop [0]{.\EOS\space}%
\providecommand \EOS [0]{\spacefactor3000\relax}%
\providecommand \BibitemShut  [1]{\csname bibitem#1\endcsname}%
\let\auto@bib@innerbib\@empty
\bibitem [{\citenamefont {Klco}\ \emph {et~al.}(2022)\citenamefont {Klco},
  \citenamefont {Roggero},\ and\ \citenamefont {Savage}}]{Klco:2021lap}%
  \BibitemOpen
  \bibfield  {author} {\bibinfo {author} {\bibfnamefont {N.}~\bibnamefont
  {Klco}}, \bibinfo {author} {\bibfnamefont {A.}~\bibnamefont {Roggero}},\ and\
  \bibinfo {author} {\bibfnamefont {M.~J.}\ \bibnamefont {Savage}},\ }\bibfield
   {title} {\bibinfo {title} {{Standard model physics and the digital quantum
  revolution: thoughts about the interface}},\ }\href
  {https://doi.org/10.1088/1361-6633/ac58a4} {\bibfield  {journal} {\bibinfo
  {journal} {Rept. Prog. Phys.}\ }\textbf {\bibinfo {volume} {85}},\ \bibinfo
  {pages} {064301} (\bibinfo {year} {2022})}\BibitemShut {NoStop}%
\bibitem [{\citenamefont {Ba\~nuls}\ \emph {et~al.}(2020)\citenamefont
  {Ba\~nuls} \emph {et~al.}}]{Banuls:2019bmf}%
  \BibitemOpen
  \bibfield  {author} {\bibinfo {author} {\bibfnamefont {M.~C.}\ \bibnamefont
  {Ba\~nuls}} \emph {et~al.},\ }\bibfield  {title} {\bibinfo {title}
  {{Simulating Lattice Gauge Theories within Quantum Technologies}},\ }\href
  {https://doi.org/10.1140/epjd/e2020-100571-8} {\bibfield  {journal} {\bibinfo
   {journal} {Eur. Phys. J. D}\ }\textbf {\bibinfo {volume} {74}},\ \bibinfo
  {pages} {165} (\bibinfo {year} {2020})},\ \Eprint
  {https://arxiv.org/abs/1911.00003} {arXiv:1911.00003 [quant-ph]} \BibitemShut
  {NoStop}%
\bibitem [{\citenamefont {Bauer}\ \emph {et~al.}(2023)\citenamefont {Bauer}
  \emph {et~al.}}]{Bauer:2022hpo}%
  \BibitemOpen
  \bibfield  {author} {\bibinfo {author} {\bibfnamefont {C.~W.}\ \bibnamefont
  {Bauer}} \emph {et~al.},\ }\bibfield  {title} {\bibinfo {title} {{Quantum
  Simulation for High-Energy Physics}},\ }\href
  {https://doi.org/10.1103/PRXQuantum.4.027001} {\bibfield  {journal} {\bibinfo
   {journal} {PRX Quantum}\ }\textbf {\bibinfo {volume} {4}},\ \bibinfo {pages}
  {027001} (\bibinfo {year} {2023})},\ \Eprint
  {https://arxiv.org/abs/2204.03381} {arXiv:2204.03381 [quant-ph]} \BibitemShut
  {NoStop}%
\bibitem [{\citenamefont {Di~Meglio}\ \emph {et~al.}(2023)\citenamefont
  {Di~Meglio} \emph {et~al.}}]{DiMeglio:2023nsa}%
  \BibitemOpen
  \bibfield  {author} {\bibinfo {author} {\bibfnamefont {A.}~\bibnamefont
  {Di~Meglio}} \emph {et~al.},\ }\href@noop {} {\bibinfo {title} {{Quantum
  Computing for High-Energy Physics: State of the Art and Challenges. Summary
  of the QC4HEP Working Group}}} (\bibinfo {year} {2023}),\ \Eprint
  {https://arxiv.org/abs/2307.03236} {arXiv:2307.03236 [quant-ph]} \BibitemShut
  {NoStop}%
\bibitem [{\citenamefont {Gattringer}\ and\ \citenamefont
  {Langfeld}(2016)}]{Gattringer:2016kco}%
  \BibitemOpen
  \bibfield  {author} {\bibinfo {author} {\bibfnamefont {C.}~\bibnamefont
  {Gattringer}}\ and\ \bibinfo {author} {\bibfnamefont {K.}~\bibnamefont
  {Langfeld}},\ }\bibfield  {title} {\bibinfo {title} {{Approaches to the sign
  problem in lattice field theory}},\ }\href
  {https://doi.org/10.1142/S0217751X16430077} {\bibfield  {journal} {\bibinfo
  {journal} {Int. J. Mod. Phys. A}\ }\textbf {\bibinfo {volume} {31}},\
  \bibinfo {pages} {1643007} (\bibinfo {year} {2016})},\ \Eprint
  {https://arxiv.org/abs/1603.09517} {arXiv:1603.09517 [hep-lat]} \BibitemShut
  {NoStop}%
\bibitem [{\citenamefont {Kühn}\ \emph {et~al.}(2014)\citenamefont {Kühn},
  \citenamefont {Cirac},\ and\ \citenamefont {Ba\~nuls}}]{Kuhn:2014rha}%
  \BibitemOpen
  \bibfield  {author} {\bibinfo {author} {\bibfnamefont {S.}~\bibnamefont
  {Kühn}}, \bibinfo {author} {\bibfnamefont {J.~I.}\ \bibnamefont {Cirac}},\
  and\ \bibinfo {author} {\bibfnamefont {M.-C.}\ \bibnamefont {Ba\~nuls}},\
  }\bibfield  {title} {\bibinfo {title} {{Quantum simulation of the Schwinger
  model: A study of feasibility}},\ }\href
  {https://doi.org/10.1103/PhysRevA.90.042305} {\bibfield  {journal} {\bibinfo
  {journal} {Phys. Rev. A}\ }\textbf {\bibinfo {volume} {90}},\ \bibinfo
  {pages} {042305} (\bibinfo {year} {2014})},\ \Eprint
  {https://arxiv.org/abs/1407.4995} {arXiv:1407.4995 [quant-ph]} \BibitemShut
  {NoStop}%
\bibitem [{\citenamefont {Kokail}\ \emph {et~al.}(2018)\citenamefont {Kokail}
  \emph {et~al.}}]{Kokail:2018eiw}%
  \BibitemOpen
  \bibfield  {author} {\bibinfo {author} {\bibfnamefont {C.}~\bibnamefont
  {Kokail}} \emph {et~al.},\ }\href@noop {} {\bibinfo {title} {{Self-Verifying
  Variational Quantum Simulation of the Lattice Schwinger Model}}} (\bibinfo
  {year} {2018}),\ \Eprint {https://arxiv.org/abs/1810.03421} {arXiv:1810.03421
  [quant-ph]} \BibitemShut {NoStop}%
\bibitem [{\citenamefont {Chakraborty}\ \emph {et~al.}(2022)\citenamefont
  {Chakraborty}, \citenamefont {Honda}, \citenamefont {Izubuchi}, \citenamefont
  {Kikuchi},\ and\ \citenamefont {Tomiya}}]{Chakraborty:2020uhf}%
  \BibitemOpen
  \bibfield  {author} {\bibinfo {author} {\bibfnamefont {B.}~\bibnamefont
  {Chakraborty}}, \bibinfo {author} {\bibfnamefont {M.}~\bibnamefont {Honda}},
  \bibinfo {author} {\bibfnamefont {T.}~\bibnamefont {Izubuchi}}, \bibinfo
  {author} {\bibfnamefont {Y.}~\bibnamefont {Kikuchi}},\ and\ \bibinfo {author}
  {\bibfnamefont {A.}~\bibnamefont {Tomiya}},\ }\bibfield  {title} {\bibinfo
  {title} {{Classically emulated digital quantum simulation of the Schwinger
  model with a topological term via adiabatic state preparation}},\ }\href
  {https://doi.org/10.1103/PhysRevD.105.094503} {\bibfield  {journal} {\bibinfo
   {journal} {Phys. Rev. D}\ }\textbf {\bibinfo {volume} {105}},\ \bibinfo
  {pages} {094503} (\bibinfo {year} {2022})},\ \Eprint
  {https://arxiv.org/abs/2001.00485} {arXiv:2001.00485 [hep-lat]} \BibitemShut
  {NoStop}%
\bibitem [{\citenamefont {Yamamoto}(2021{\natexlab{a}})}]{Yamamoto:2021vxp}%
  \BibitemOpen
  \bibfield  {author} {\bibinfo {author} {\bibfnamefont {A.}~\bibnamefont
  {Yamamoto}},\ }\bibfield  {title} {\bibinfo {title} {{Quantum variational
  approach to lattice gauge theory at nonzero density}},\ }\href
  {https://doi.org/10.1103/PhysRevD.104.014506} {\bibfield  {journal} {\bibinfo
   {journal} {Phys. Rev. D}\ }\textbf {\bibinfo {volume} {104}},\ \bibinfo
  {pages} {014506} (\bibinfo {year} {2021}{\natexlab{a}})},\ \Eprint
  {https://arxiv.org/abs/2104.10669} {arXiv:2104.10669 [hep-lat]} \BibitemShut
  {NoStop}%
\bibitem [{\citenamefont {Desai}\ \emph {et~al.}(2021)\citenamefont {Desai},
  \citenamefont {Feng}, \citenamefont {Hassan}, \citenamefont {Kodumagulla},\
  and\ \citenamefont {McGuigan}}]{Desai:2021oiy}%
  \BibitemOpen
  \bibfield  {author} {\bibinfo {author} {\bibfnamefont {R.}~\bibnamefont
  {Desai}}, \bibinfo {author} {\bibfnamefont {Y.}~\bibnamefont {Feng}},
  \bibinfo {author} {\bibfnamefont {M.}~\bibnamefont {Hassan}}, \bibinfo
  {author} {\bibfnamefont {A.}~\bibnamefont {Kodumagulla}},\ and\ \bibinfo
  {author} {\bibfnamefont {M.}~\bibnamefont {McGuigan}},\ }\href@noop {}
  {\bibinfo {title} {{Z3 gauge theory coupled to fermions and quantum
  computing}}} (\bibinfo {year} {2021}),\ \Eprint
  {https://arxiv.org/abs/2106.00549} {arXiv:2106.00549 [quant-ph]} \BibitemShut
  {NoStop}%
\bibitem [{\citenamefont {Farrell}\ \emph
  {et~al.}(2023{\natexlab{a}})\citenamefont {Farrell}, \citenamefont {Illa},
  \citenamefont {Ciavarella},\ and\ \citenamefont {Savage}}]{Farrell:2023fgd}%
  \BibitemOpen
  \bibfield  {author} {\bibinfo {author} {\bibfnamefont {R.~C.}\ \bibnamefont
  {Farrell}}, \bibinfo {author} {\bibfnamefont {M.}~\bibnamefont {Illa}},
  \bibinfo {author} {\bibfnamefont {A.~N.}\ \bibnamefont {Ciavarella}},\ and\
  \bibinfo {author} {\bibfnamefont {M.~J.}\ \bibnamefont {Savage}},\
  }\href@noop {} {\bibinfo {title} {{Scalable Circuits for Preparing Ground
  States on Digital Quantum Computers: The Schwinger Model Vacuum on 100
  Qubits}}} (\bibinfo {year} {2023}{\natexlab{a}}),\ \Eprint
  {https://arxiv.org/abs/2308.04481} {arXiv:2308.04481 [quant-ph]} \BibitemShut
  {NoStop}%
\bibitem [{\citenamefont {Kane}\ \emph {et~al.}(2023)\citenamefont {Kane},
  \citenamefont {Gomes},\ and\ \citenamefont {Kreshchuk}}]{Kane:2023jdo}%
  \BibitemOpen
  \bibfield  {author} {\bibinfo {author} {\bibfnamefont {C.~F.}\ \bibnamefont
  {Kane}}, \bibinfo {author} {\bibfnamefont {N.}~\bibnamefont {Gomes}},\ and\
  \bibinfo {author} {\bibfnamefont {M.}~\bibnamefont {Kreshchuk}},\ }\href@noop
  {} {\bibinfo {title} {{Nearly-optimal state preparation for quantum
  simulations of lattice gauge theories}}} (\bibinfo {year} {2023}),\ \Eprint
  {https://arxiv.org/abs/2310.13757} {arXiv:2310.13757 [quant-ph]} \BibitemShut
  {NoStop}%
\bibitem [{\citenamefont {Bilgin}\ and\ \citenamefont
  {Boixo}(2010)}]{2010PhRvL.105q0405B}%
  \BibitemOpen
  \bibfield  {author} {\bibinfo {author} {\bibfnamefont {E.}~\bibnamefont
  {Bilgin}}\ and\ \bibinfo {author} {\bibfnamefont {S.}~\bibnamefont {Boixo}},\
  }\bibfield  {title} {\bibinfo {title} {Preparing thermal states of quantum
  systems by dimension reduction},\ }\href
  {https://doi.org/10.1103/PhysRevLett.105.170405} {\bibfield  {journal}
  {\bibinfo  {journal} {Phys. Rev. Lett.}\ }\textbf {\bibinfo {volume} {105}},\
  \bibinfo {pages} {170405} (\bibinfo {year} {2010})}\BibitemShut {NoStop}%
\bibitem [{\citenamefont {Riera}\ \emph {et~al.}(2012)\citenamefont {Riera},
  \citenamefont {Gogolin},\ and\ \citenamefont
  {Eisert}}]{PhysRevLett.108.080402}%
  \BibitemOpen
  \bibfield  {author} {\bibinfo {author} {\bibfnamefont {A.}~\bibnamefont
  {Riera}}, \bibinfo {author} {\bibfnamefont {C.}~\bibnamefont {Gogolin}},\
  and\ \bibinfo {author} {\bibfnamefont {J.}~\bibnamefont {Eisert}},\
  }\bibfield  {title} {\bibinfo {title} {Thermalization in nature and on a
  quantum computer},\ }\href {https://doi.org/10.1103/PhysRevLett.108.080402}
  {\bibfield  {journal} {\bibinfo  {journal} {Phys. Rev. Lett.}\ }\textbf
  {\bibinfo {volume} {108}},\ \bibinfo {pages} {080402} (\bibinfo {year}
  {2012})}\BibitemShut {NoStop}%
\bibitem [{\citenamefont {Lamm}\ and\ \citenamefont
  {Lawrence}(2018)}]{Lamm:2018siq}%
  \BibitemOpen
  \bibfield  {author} {\bibinfo {author} {\bibfnamefont {H.}~\bibnamefont
  {Lamm}}\ and\ \bibinfo {author} {\bibfnamefont {S.}~\bibnamefont
  {Lawrence}},\ }\bibfield  {title} {\bibinfo {title} {{Simulation of
  Nonequilibrium Dynamics on a Quantum Computer}},\ }\href
  {https://doi.org/10.1103/PhysRevLett.121.170501} {\bibfield  {journal}
  {\bibinfo  {journal} {Phys. Rev. Lett.}\ }\textbf {\bibinfo {volume} {121}},\
  \bibinfo {pages} {170501} (\bibinfo {year} {2018})},\ \Eprint
  {https://arxiv.org/abs/1806.06649} {arXiv:1806.06649 [quant-ph]} \BibitemShut
  {NoStop}%
\bibitem [{\citenamefont {Klco}\ and\ \citenamefont
  {Savage}(2019)}]{Klco:2019xro}%
  \BibitemOpen
  \bibfield  {author} {\bibinfo {author} {\bibfnamefont {N.}~\bibnamefont
  {Klco}}\ and\ \bibinfo {author} {\bibfnamefont {M.~J.}\ \bibnamefont
  {Savage}},\ }\href@noop {} {\bibinfo {title} {{Minimally-Entangled State
  Preparation of Localized Wavefunctions on Quantum Computers}}} (\bibinfo
  {year} {2019}),\ \Eprint {https://arxiv.org/abs/1904.10440} {arXiv:1904.10440
  [quant-ph]} \BibitemShut {NoStop}%
\bibitem [{\citenamefont {Harmalkar}\ \emph {et~al.}(2020)\citenamefont
  {Harmalkar}, \citenamefont {Lamm},\ and\ \citenamefont
  {Lawrence}}]{Harmalkar:2020mpd}%
  \BibitemOpen
  \bibfield  {author} {\bibinfo {author} {\bibfnamefont {S.}~\bibnamefont
  {Harmalkar}}, \bibinfo {author} {\bibfnamefont {H.}~\bibnamefont {Lamm}},\
  and\ \bibinfo {author} {\bibfnamefont {S.}~\bibnamefont {Lawrence}} (\bibinfo
  {collaboration} {NuQS}),\ }\href@noop {} {\bibinfo {title} {{Quantum
  Simulation of Field Theories Without State Preparation}}} (\bibinfo {year}
  {2020}),\ \Eprint {https://arxiv.org/abs/2001.11490} {arXiv:2001.11490
  [hep-lat]} \BibitemShut {NoStop}%
\bibitem [{\citenamefont {Motta}\ \emph {et~al.}(2020)\citenamefont {Motta},
  \citenamefont {Sun}, \citenamefont {Tan}, \citenamefont {O’Rourke},
  \citenamefont {Ye}, \citenamefont {Minnich}, \citenamefont {Brandao},\ and\
  \citenamefont {Chan}}]{motta2020determining}%
  \BibitemOpen
  \bibfield  {author} {\bibinfo {author} {\bibfnamefont {M.}~\bibnamefont
  {Motta}}, \bibinfo {author} {\bibfnamefont {C.}~\bibnamefont {Sun}}, \bibinfo
  {author} {\bibfnamefont {A.~T.}\ \bibnamefont {Tan}}, \bibinfo {author}
  {\bibfnamefont {M.~J.}\ \bibnamefont {O’Rourke}}, \bibinfo {author}
  {\bibfnamefont {E.}~\bibnamefont {Ye}}, \bibinfo {author} {\bibfnamefont
  {A.~J.}\ \bibnamefont {Minnich}}, \bibinfo {author} {\bibfnamefont {F.~G.}\
  \bibnamefont {Brandao}},\ and\ \bibinfo {author} {\bibfnamefont {G.~K.-L.}\
  \bibnamefont {Chan}},\ }\bibfield  {title} {\bibinfo {title} {Determining
  eigenstates and thermal states on a quantum computer using quantum imaginary
  time evolution},\ }\href@noop {} {\bibfield  {journal} {\bibinfo  {journal}
  {Nature Physics}\ }\textbf {\bibinfo {volume} {16}},\ \bibinfo {pages} {205}
  (\bibinfo {year} {2020})}\BibitemShut {NoStop}%
\bibitem [{\citenamefont {de~Jong}\ \emph {et~al.}(2021)\citenamefont
  {de~Jong}, \citenamefont {Lee}, \citenamefont {Mulligan}, \citenamefont
  {P\l{}osko\'n}, \citenamefont {Ringer},\ and\ \citenamefont
  {Yao}}]{deJong:2021wsd}%
  \BibitemOpen
  \bibfield  {author} {\bibinfo {author} {\bibfnamefont {W.~A.}\ \bibnamefont
  {de~Jong}}, \bibinfo {author} {\bibfnamefont {K.}~\bibnamefont {Lee}},
  \bibinfo {author} {\bibfnamefont {J.}~\bibnamefont {Mulligan}}, \bibinfo
  {author} {\bibfnamefont {M.}~\bibnamefont {P\l{}osko\'n}}, \bibinfo {author}
  {\bibfnamefont {F.}~\bibnamefont {Ringer}},\ and\ \bibinfo {author}
  {\bibfnamefont {X.}~\bibnamefont {Yao}},\ }\href@noop {} {\bibinfo {title}
  {{Quantum simulation of non-equilibrium dynamics and thermalization in the
  Schwinger model}}} (\bibinfo {year} {2021}),\ \Eprint
  {https://arxiv.org/abs/2106.08394} {arXiv:2106.08394 [quant-ph]} \BibitemShut
  {NoStop}%
\bibitem [{\citenamefont {Xie}\ \emph {et~al.}(2022)\citenamefont {Xie},
  \citenamefont {Guo}, \citenamefont {Xing}, \citenamefont {Xue}, \citenamefont
  {Zhang},\ and\ \citenamefont {Zhu}}]{Xie:2022jgj}%
  \BibitemOpen
  \bibfield  {author} {\bibinfo {author} {\bibfnamefont {X.-D.}\ \bibnamefont
  {Xie}}, \bibinfo {author} {\bibfnamefont {X.}~\bibnamefont {Guo}}, \bibinfo
  {author} {\bibfnamefont {H.}~\bibnamefont {Xing}}, \bibinfo {author}
  {\bibfnamefont {Z.-Y.}\ \bibnamefont {Xue}}, \bibinfo {author} {\bibfnamefont
  {D.-B.}\ \bibnamefont {Zhang}},\ and\ \bibinfo {author} {\bibfnamefont
  {S.-L.}\ \bibnamefont {Zhu}} (\bibinfo {collaboration} {QuNu}),\ }\bibfield
  {title} {\bibinfo {title} {{Variational thermal quantum simulation of the
  lattice Schwinger model}},\ }\href
  {https://doi.org/10.1103/PhysRevD.106.054509} {\bibfield  {journal} {\bibinfo
   {journal} {Phys. Rev. D}\ }\textbf {\bibinfo {volume} {106}},\ \bibinfo
  {pages} {054509} (\bibinfo {year} {2022})},\ \Eprint
  {https://arxiv.org/abs/2205.12767} {arXiv:2205.12767 [quant-ph]} \BibitemShut
  {NoStop}%
\bibitem [{\citenamefont {Davoudi}\ \emph {et~al.}(2023)\citenamefont
  {Davoudi}, \citenamefont {Mueller},\ and\ \citenamefont
  {Powers}}]{Davoudi:2022uzo}%
  \BibitemOpen
  \bibfield  {author} {\bibinfo {author} {\bibfnamefont {Z.}~\bibnamefont
  {Davoudi}}, \bibinfo {author} {\bibfnamefont {N.}~\bibnamefont {Mueller}},\
  and\ \bibinfo {author} {\bibfnamefont {C.}~\bibnamefont {Powers}},\
  }\bibfield  {title} {\bibinfo {title} {{Towards Quantum Computing Phase
  Diagrams of Gauge Theories with Thermal Pure Quantum States}},\ }\href
  {https://doi.org/10.1103/PhysRevLett.131.081901} {\bibfield  {journal}
  {\bibinfo  {journal} {Phys. Rev. Lett.}\ }\textbf {\bibinfo {volume} {131}},\
  \bibinfo {pages} {081901} (\bibinfo {year} {2023})},\ \Eprint
  {https://arxiv.org/abs/2208.13112} {arXiv:2208.13112 [hep-lat]} \BibitemShut
  {NoStop}%
\bibitem [{\citenamefont {Ball}\ and\ \citenamefont
  {Cohen}(2023)}]{Ball:2022dxy}%
  \BibitemOpen
  \bibfield  {author} {\bibinfo {author} {\bibfnamefont {C.}~\bibnamefont
  {Ball}}\ and\ \bibinfo {author} {\bibfnamefont {T.~D.}\ \bibnamefont
  {Cohen}},\ }\bibfield  {title} {\bibinfo {title} {{Boltzmann distributions on
  a quantum computer via active cooling}},\ }\href
  {https://doi.org/10.1016/j.nuclphysa.2023.122708} {\bibfield  {journal}
  {\bibinfo  {journal} {Nucl. Phys. A}\ }\textbf {\bibinfo {volume} {1038}},\
  \bibinfo {pages} {122708} (\bibinfo {year} {2023})},\ \Eprint
  {https://arxiv.org/abs/2212.06730} {arXiv:2212.06730 [quant-ph]} \BibitemShut
  {NoStop}%
\bibitem [{\citenamefont {Saroni}\ \emph {et~al.}(2023)\citenamefont {Saroni},
  \citenamefont {Lamm}, \citenamefont {Orth},\ and\ \citenamefont
  {Iadecola}}]{Saroni:2023uob}%
  \BibitemOpen
  \bibfield  {author} {\bibinfo {author} {\bibfnamefont {J.}~\bibnamefont
  {Saroni}}, \bibinfo {author} {\bibfnamefont {H.}~\bibnamefont {Lamm}},
  \bibinfo {author} {\bibfnamefont {P.~P.}\ \bibnamefont {Orth}},\ and\
  \bibinfo {author} {\bibfnamefont {T.}~\bibnamefont {Iadecola}},\ }\bibfield
  {title} {\bibinfo {title} {{Reconstructing thermal quantum quench dynamics
  from pure states}},\ }\href {https://doi.org/10.1103/PhysRevB.108.134301}
  {\bibfield  {journal} {\bibinfo  {journal} {Phys. Rev. B}\ }\textbf {\bibinfo
  {volume} {108}},\ \bibinfo {pages} {134301} (\bibinfo {year} {2023})},\
  \Eprint {https://arxiv.org/abs/2307.14508} {arXiv:2307.14508 [quant-ph]}
  \BibitemShut {NoStop}%
\bibitem [{\citenamefont {Jordan}\ \emph {et~al.}(2012)\citenamefont {Jordan},
  \citenamefont {Lee},\ and\ \citenamefont {Preskill}}]{Jordan:2011ne}%
  \BibitemOpen
  \bibfield  {author} {\bibinfo {author} {\bibfnamefont {S.~P.}\ \bibnamefont
  {Jordan}}, \bibinfo {author} {\bibfnamefont {K.~S.~M.}\ \bibnamefont {Lee}},\
  and\ \bibinfo {author} {\bibfnamefont {J.}~\bibnamefont {Preskill}},\
  }\bibfield  {title} {\bibinfo {title} {{Quantum Algorithms for Quantum Field
  Theories}},\ }\href {https://doi.org/10.1126/science.1217069} {\bibfield
  {journal} {\bibinfo  {journal} {Science}\ }\textbf {\bibinfo {volume}
  {336}},\ \bibinfo {pages} {1130} (\bibinfo {year} {2012})},\ \Eprint
  {https://arxiv.org/abs/1111.3633} {arXiv:1111.3633 [quant-ph]} \BibitemShut
  {NoStop}%
\bibitem [{\citenamefont {Jordan}\ \emph
  {et~al.}(2014{\natexlab{a}})\citenamefont {Jordan}, \citenamefont {Lee},\
  and\ \citenamefont {Preskill}}]{Jordan:2011ci}%
  \BibitemOpen
  \bibfield  {author} {\bibinfo {author} {\bibfnamefont {S.~P.}\ \bibnamefont
  {Jordan}}, \bibinfo {author} {\bibfnamefont {K.~S.~M.}\ \bibnamefont {Lee}},\
  and\ \bibinfo {author} {\bibfnamefont {J.}~\bibnamefont {Preskill}},\
  }\bibfield  {title} {\bibinfo {title} {{Quantum Computation of Scattering in
  Scalar Quantum Field Theories}},\ }\href@noop {} {\bibfield  {journal}
  {\bibinfo  {journal} {Quant. Inf. Comput.}\ }\textbf {\bibinfo {volume}
  {14}},\ \bibinfo {pages} {1014} (\bibinfo {year} {2014}{\natexlab{a}})},\
  \Eprint {https://arxiv.org/abs/1112.4833} {arXiv:1112.4833 [hep-th]}
  \BibitemShut {NoStop}%
\bibitem [{\citenamefont {García-Álvarez}\ \emph {et~al.}(2015)\citenamefont
  {García-Álvarez}, \citenamefont {Casanova}, \citenamefont {Mezzacapo},
  \citenamefont {Egusquiza}, \citenamefont {Lamata}, \citenamefont {Romero},\
  and\ \citenamefont {Solano}}]{Garcia-Alvarez:2014uda}%
  \BibitemOpen
  \bibfield  {author} {\bibinfo {author} {\bibfnamefont {L.}~\bibnamefont
  {García-Álvarez}}, \bibinfo {author} {\bibfnamefont {J.}~\bibnamefont
  {Casanova}}, \bibinfo {author} {\bibfnamefont {A.}~\bibnamefont {Mezzacapo}},
  \bibinfo {author} {\bibfnamefont {I.~L.}\ \bibnamefont {Egusquiza}}, \bibinfo
  {author} {\bibfnamefont {L.}~\bibnamefont {Lamata}}, \bibinfo {author}
  {\bibfnamefont {G.}~\bibnamefont {Romero}},\ and\ \bibinfo {author}
  {\bibfnamefont {E.}~\bibnamefont {Solano}},\ }\bibfield  {title} {\bibinfo
  {title} {{Fermion-Fermion Scattering in Quantum Field Theory with
  Superconducting Circuits}},\ }\href
  {https://doi.org/10.1103/PhysRevLett.114.070502} {\bibfield  {journal}
  {\bibinfo  {journal} {Phys. Rev. Lett.}\ }\textbf {\bibinfo {volume} {114}},\
  \bibinfo {pages} {070502} (\bibinfo {year} {2015})},\ \Eprint
  {https://arxiv.org/abs/1404.2868} {arXiv:1404.2868 [quant-ph]} \BibitemShut
  {NoStop}%
\bibitem [{\citenamefont {Jordan}\ \emph
  {et~al.}(2014{\natexlab{b}})\citenamefont {Jordan}, \citenamefont {Lee},\
  and\ \citenamefont {Preskill}}]{Jordan:2014tma}%
  \BibitemOpen
  \bibfield  {author} {\bibinfo {author} {\bibfnamefont {S.~P.}\ \bibnamefont
  {Jordan}}, \bibinfo {author} {\bibfnamefont {K.~S.~M.}\ \bibnamefont {Lee}},\
  and\ \bibinfo {author} {\bibfnamefont {J.}~\bibnamefont {Preskill}},\
  }\href@noop {} {\bibinfo {title} {{Quantum Algorithms for Fermionic Quantum
  Field Theories}}} (\bibinfo {year} {2014}{\natexlab{b}}),\ \Eprint
  {https://arxiv.org/abs/1404.7115} {arXiv:1404.7115 [hep-th]} \BibitemShut
  {NoStop}%
\bibitem [{\citenamefont {Jordan}\ \emph {et~al.}(2018)\citenamefont {Jordan},
  \citenamefont {Krovi}, \citenamefont {Lee},\ and\ \citenamefont
  {Preskill}}]{Jordan:2017lea}%
  \BibitemOpen
  \bibfield  {author} {\bibinfo {author} {\bibfnamefont {S.~P.}\ \bibnamefont
  {Jordan}}, \bibinfo {author} {\bibfnamefont {H.}~\bibnamefont {Krovi}},
  \bibinfo {author} {\bibfnamefont {K.~S.}\ \bibnamefont {Lee}},\ and\ \bibinfo
  {author} {\bibfnamefont {J.}~\bibnamefont {Preskill}},\ }\bibfield  {title}
  {\bibinfo {title} {{BQP-completeness of Scattering in Scalar Quantum Field
  Theory}},\ }\href {https://doi.org/10.22331/q-2018-01-08-44} {\bibfield
  {journal} {\bibinfo  {journal} {Quantum}\ }\textbf {\bibinfo {volume} {2}},\
  \bibinfo {pages} {44} (\bibinfo {year} {2018})},\ \Eprint
  {https://arxiv.org/abs/1703.00454} {arXiv:1703.00454 [quant-ph]} \BibitemShut
  {NoStop}%
\bibitem [{\citenamefont {Hamed~Moosavian}\ and\ \citenamefont
  {Jordan}(2018)}]{Moosavian:2017tkv}%
  \BibitemOpen
  \bibfield  {author} {\bibinfo {author} {\bibfnamefont {A.}~\bibnamefont
  {Hamed~Moosavian}}\ and\ \bibinfo {author} {\bibfnamefont {S.}~\bibnamefont
  {Jordan}},\ }\bibfield  {title} {\bibinfo {title} {{Faster Quantum Algorithm
  to simulate Fermionic Quantum Field Theory}},\ }\href
  {https://doi.org/10.1103/PhysRevA.98.012332} {\bibfield  {journal} {\bibinfo
  {journal} {Phys. Rev.}\ }\textbf {\bibinfo {volume} {A98}},\ \bibinfo {pages}
  {012332} (\bibinfo {year} {2018})},\ \Eprint
  {https://arxiv.org/abs/1711.04006} {arXiv:1711.04006 [quant-ph]} \BibitemShut
  {NoStop}%
\bibitem [{\citenamefont {Brand{\~a}o}\ and\ \citenamefont
  {Kastoryano}(2019)}]{brandao2019finite}%
  \BibitemOpen
  \bibfield  {author} {\bibinfo {author} {\bibfnamefont {F.~G.}\ \bibnamefont
  {Brand{\~a}o}}\ and\ \bibinfo {author} {\bibfnamefont {M.~J.}\ \bibnamefont
  {Kastoryano}},\ }\bibfield  {title} {\bibinfo {title} {Finite correlation
  length implies efficient preparation of quantum thermal states},\ }\href@noop
  {} {\bibfield  {journal} {\bibinfo  {journal} {Communications in Mathematical
  Physics}\ }\textbf {\bibinfo {volume} {365}},\ \bibinfo {pages} {1} (\bibinfo
  {year} {2019})}\BibitemShut {NoStop}%
\bibitem [{\citenamefont {Gustafson}\ \emph
  {et~al.}(2019{\natexlab{a}})\citenamefont {Gustafson}, \citenamefont
  {Meurice},\ and\ \citenamefont {Unmuth-Yockey}}]{Gustafson:2019mpk}%
  \BibitemOpen
  \bibfield  {author} {\bibinfo {author} {\bibfnamefont {E.}~\bibnamefont
  {Gustafson}}, \bibinfo {author} {\bibfnamefont {Y.}~\bibnamefont {Meurice}},\
  and\ \bibinfo {author} {\bibfnamefont {J.}~\bibnamefont {Unmuth-Yockey}},\
  }\href@noop {} {\bibinfo {title} {{Quantum simulation of scattering in the
  quantum Ising model}}} (\bibinfo {year} {2019}{\natexlab{a}}),\ \Eprint
  {https://arxiv.org/abs/1901.05944} {arXiv:1901.05944 [hep-lat]} \BibitemShut
  {NoStop}%
\bibitem [{\citenamefont {Gustafson}\ \emph
  {et~al.}(2019{\natexlab{b}})\citenamefont {Gustafson}, \citenamefont
  {Dreher}, \citenamefont {Hang},\ and\ \citenamefont
  {Meurice}}]{Gustafson:2019vsd}%
  \BibitemOpen
  \bibfield  {author} {\bibinfo {author} {\bibfnamefont {E.}~\bibnamefont
  {Gustafson}}, \bibinfo {author} {\bibfnamefont {P.}~\bibnamefont {Dreher}},
  \bibinfo {author} {\bibfnamefont {Z.}~\bibnamefont {Hang}},\ and\ \bibinfo
  {author} {\bibfnamefont {Y.}~\bibnamefont {Meurice}},\ }\href@noop {}
  {\bibinfo {title} {{Benchmarking quantum computers for real-time evolution of
  a $(1+1)$ field theory with error mitigation}}} (\bibinfo {year}
  {2019}{\natexlab{b}}),\ \Eprint {https://arxiv.org/abs/1910.09478}
  {arXiv:1910.09478 [hep-lat]} \BibitemShut {NoStop}%
\bibitem [{\citenamefont {Gustafson}\ and\ \citenamefont
  {Lamm}(2021)}]{Gustafson:2020yfe}%
  \BibitemOpen
  \bibfield  {author} {\bibinfo {author} {\bibfnamefont {E.~J.}\ \bibnamefont
  {Gustafson}}\ and\ \bibinfo {author} {\bibfnamefont {H.}~\bibnamefont
  {Lamm}},\ }\bibfield  {title} {\bibinfo {title} {{Toward quantum simulations
  of $\mathbb{Z}_2$ gauge theory without state preparation}},\ }\href
  {https://doi.org/10.1103/PhysRevD.103.054507} {\bibfield  {journal} {\bibinfo
   {journal} {Phys. Rev. D}\ }\textbf {\bibinfo {volume} {103}},\ \bibinfo
  {pages} {054507} (\bibinfo {year} {2021})},\ \Eprint
  {https://arxiv.org/abs/2011.11677} {arXiv:2011.11677 [hep-lat]} \BibitemShut
  {NoStop}%
\bibitem [{\citenamefont {Kreshchuk}\ \emph {et~al.}(2023)\citenamefont
  {Kreshchuk}, \citenamefont {Vary},\ and\ \citenamefont
  {Love}}]{Kreshchuk:2023btr}%
  \BibitemOpen
  \bibfield  {author} {\bibinfo {author} {\bibfnamefont {M.}~\bibnamefont
  {Kreshchuk}}, \bibinfo {author} {\bibfnamefont {J.~P.}\ \bibnamefont
  {Vary}},\ and\ \bibinfo {author} {\bibfnamefont {P.~J.}\ \bibnamefont
  {Love}},\ }\href@noop {} {\bibinfo {title} {{Simulating Scattering of
  Composite Particles}}} (\bibinfo {year} {2023}),\ \Eprint
  {https://arxiv.org/abs/2310.13742} {arXiv:2310.13742 [quant-ph]} \BibitemShut
  {NoStop}%
\bibitem [{\citenamefont {Childs}\ \emph {et~al.}(2021)\citenamefont {Childs},
  \citenamefont {Su}, \citenamefont {Tran}, \citenamefont {Wiebe},\ and\
  \citenamefont {Zhu}}]{PhysRevX.11.011020}%
  \BibitemOpen
  \bibfield  {author} {\bibinfo {author} {\bibfnamefont {A.~M.}\ \bibnamefont
  {Childs}}, \bibinfo {author} {\bibfnamefont {Y.}~\bibnamefont {Su}}, \bibinfo
  {author} {\bibfnamefont {M.~C.}\ \bibnamefont {Tran}}, \bibinfo {author}
  {\bibfnamefont {N.}~\bibnamefont {Wiebe}},\ and\ \bibinfo {author}
  {\bibfnamefont {S.}~\bibnamefont {Zhu}},\ }\bibfield  {title} {\bibinfo
  {title} {Theory of {T}rotter error with commutator scaling},\ }\href
  {https://doi.org/10.1103/PhysRevX.11.011020} {\bibfield  {journal} {\bibinfo
  {journal} {Phys. Rev. X}\ }\textbf {\bibinfo {volume} {11}},\ \bibinfo
  {pages} {011020} (\bibinfo {year} {2021})}\BibitemShut {NoStop}%
\bibitem [{\citenamefont {Davoudi}\ \emph {et~al.}(2022)\citenamefont
  {Davoudi}, \citenamefont {Shaw},\ and\ \citenamefont
  {Stryker}}]{Davoudi:2022xmb}%
  \BibitemOpen
  \bibfield  {author} {\bibinfo {author} {\bibfnamefont {Z.}~\bibnamefont
  {Davoudi}}, \bibinfo {author} {\bibfnamefont {A.~F.}\ \bibnamefont {Shaw}},\
  and\ \bibinfo {author} {\bibfnamefont {J.~R.}\ \bibnamefont {Stryker}},\
  }\href@noop {} {\bibinfo {title} {{General quantum algorithms for Hamiltonian
  simulation with applications to a non-Abelian lattice gauge theory}}}
  (\bibinfo {year} {2022}),\ \Eprint {https://arxiv.org/abs/2212.14030}
  {arXiv:2212.14030 [hep-lat]} \BibitemShut {NoStop}%
\bibitem [{\citenamefont {Campbell}(2019)}]{PhysRevLett.123.070503}%
  \BibitemOpen
  \bibfield  {author} {\bibinfo {author} {\bibfnamefont {E.}~\bibnamefont
  {Campbell}},\ }\bibfield  {title} {\bibinfo {title} {Random compiler for fast
  {H}amiltonian simulation},\ }\href
  {https://doi.org/10.1103/PhysRevLett.123.070503} {\bibfield  {journal}
  {\bibinfo  {journal} {Phys. Rev. Lett.}\ }\textbf {\bibinfo {volume} {123}},\
  \bibinfo {pages} {070503} (\bibinfo {year} {2019})}\BibitemShut {NoStop}%
\bibitem [{\citenamefont {Shaw}\ \emph {et~al.}(2020)\citenamefont {Shaw},
  \citenamefont {Lougovski}, \citenamefont {Stryker},\ and\ \citenamefont
  {Wiebe}}]{Shaw:2020udc}%
  \BibitemOpen
  \bibfield  {author} {\bibinfo {author} {\bibfnamefont {A.~F.}\ \bibnamefont
  {Shaw}}, \bibinfo {author} {\bibfnamefont {P.}~\bibnamefont {Lougovski}},
  \bibinfo {author} {\bibfnamefont {J.~R.}\ \bibnamefont {Stryker}},\ and\
  \bibinfo {author} {\bibfnamefont {N.}~\bibnamefont {Wiebe}},\ }\bibfield
  {title} {\bibinfo {title} {{Quantum Algorithms for Simulating the Lattice
  Schwinger Model}},\ }\href {https://doi.org/10.22331/q-2020-08-10-306}
  {\bibfield  {journal} {\bibinfo  {journal} {Quantum}\ }\textbf {\bibinfo
  {volume} {4}},\ \bibinfo {pages} {306} (\bibinfo {year} {2020})},\ \Eprint
  {https://arxiv.org/abs/2002.11146} {arXiv:2002.11146 [quant-ph]} \BibitemShut
  {NoStop}%
\bibitem [{\citenamefont {Berry}\ \emph {et~al.}(2015)\citenamefont {Berry},
  \citenamefont {Childs}, \citenamefont {Cleve}, \citenamefont {Kothari},\ and\
  \citenamefont {Somma}}]{PhysRevLett.114.090502}%
  \BibitemOpen
  \bibfield  {author} {\bibinfo {author} {\bibfnamefont {D.~W.}\ \bibnamefont
  {Berry}}, \bibinfo {author} {\bibfnamefont {A.~M.}\ \bibnamefont {Childs}},
  \bibinfo {author} {\bibfnamefont {R.}~\bibnamefont {Cleve}}, \bibinfo
  {author} {\bibfnamefont {R.}~\bibnamefont {Kothari}},\ and\ \bibinfo {author}
  {\bibfnamefont {R.~D.}\ \bibnamefont {Somma}},\ }\bibfield  {title} {\bibinfo
  {title} {Simulating {H}amiltonian dynamics with a truncated {T}aylor
  series},\ }\href {https://doi.org/10.1103/PhysRevLett.114.090502} {\bibfield
  {journal} {\bibinfo  {journal} {Phys. Rev. Lett.}\ }\textbf {\bibinfo
  {volume} {114}},\ \bibinfo {pages} {090502} (\bibinfo {year}
  {2015})}\BibitemShut {NoStop}%
\bibitem [{\citenamefont {Low}\ and\ \citenamefont
  {Chuang}(2019)}]{Low2019hamiltonian}%
  \BibitemOpen
  \bibfield  {author} {\bibinfo {author} {\bibfnamefont {G.~H.}\ \bibnamefont
  {Low}}\ and\ \bibinfo {author} {\bibfnamefont {I.~L.}\ \bibnamefont
  {Chuang}},\ }\bibfield  {title} {\bibinfo {title} {Hamiltonian {S}imulation
  by {Q}ubitization},\ }\href {https://doi.org/10.22331/q-2019-07-12-163}
  {\bibfield  {journal} {\bibinfo  {journal} {{Quantum}}\ }\textbf {\bibinfo
  {volume} {3}},\ \bibinfo {pages} {163} (\bibinfo {year} {2019})}\BibitemShut
  {NoStop}%
\bibitem [{\citenamefont {Berry}\ and\ \citenamefont
  {Childs}(2012)}]{berry2009black}%
  \BibitemOpen
  \bibfield  {author} {\bibinfo {author} {\bibfnamefont {D.~W.}\ \bibnamefont
  {Berry}}\ and\ \bibinfo {author} {\bibfnamefont {A.~M.}\ \bibnamefont
  {Childs}},\ }\bibfield  {title} {\bibinfo {title} {Black-box {H}amiltonian
  simulation and unitary implementation},\ }\href@noop {} {\bibfield  {journal}
  {\bibinfo  {journal} {Quantum Information \& Computation}\ }\textbf {\bibinfo
  {volume} {12}} (\bibinfo {year} {2012})}\BibitemShut {NoStop}%
\bibitem [{\citenamefont {Low}\ and\ \citenamefont
  {Chuang}(2017)}]{PhysRevLett.118.010501}%
  \BibitemOpen
  \bibfield  {author} {\bibinfo {author} {\bibfnamefont {G.~H.}\ \bibnamefont
  {Low}}\ and\ \bibinfo {author} {\bibfnamefont {I.~L.}\ \bibnamefont
  {Chuang}},\ }\bibfield  {title} {\bibinfo {title} {Optimal {H}amiltonian
  simulation by quantum signal processing},\ }\href
  {https://doi.org/10.1103/PhysRevLett.118.010501} {\bibfield  {journal}
  {\bibinfo  {journal} {Phys. Rev. Lett.}\ }\textbf {\bibinfo {volume} {118}},\
  \bibinfo {pages} {010501} (\bibinfo {year} {2017})}\BibitemShut {NoStop}%
\bibitem [{\citenamefont {Childs}\ and\ \citenamefont
  {Wiebe}(2012)}]{childs2012hamiltonian}%
  \BibitemOpen
  \bibfield  {author} {\bibinfo {author} {\bibfnamefont {A.~M.}\ \bibnamefont
  {Childs}}\ and\ \bibinfo {author} {\bibfnamefont {N.}~\bibnamefont {Wiebe}},\
  }\href@noop {} {\bibinfo {title} {Hamiltonian simulation using linear
  combinations of unitary operations}} (\bibinfo {year} {2012})\BibitemShut
  {NoStop}%
\bibitem [{\citenamefont {Cirstoiu}\ \emph {et~al.}(2020)\citenamefont
  {Cirstoiu}, \citenamefont {Holmes}, \citenamefont {Iosue}, \citenamefont
  {Cincio}, \citenamefont {Coles},\ and\ \citenamefont
  {Sornborger}}]{cirstoiu2020variational}%
  \BibitemOpen
  \bibfield  {author} {\bibinfo {author} {\bibfnamefont {C.}~\bibnamefont
  {Cirstoiu}}, \bibinfo {author} {\bibfnamefont {Z.}~\bibnamefont {Holmes}},
  \bibinfo {author} {\bibfnamefont {J.}~\bibnamefont {Iosue}}, \bibinfo
  {author} {\bibfnamefont {L.}~\bibnamefont {Cincio}}, \bibinfo {author}
  {\bibfnamefont {P.~J.}\ \bibnamefont {Coles}},\ and\ \bibinfo {author}
  {\bibfnamefont {A.}~\bibnamefont {Sornborger}},\ }\bibfield  {title}
  {\bibinfo {title} {Variational fast forwarding for quantum simulation beyond
  the coherence time},\ }\href@noop {} {\bibfield  {journal} {\bibinfo
  {journal} {npj Quantum Information}\ }\textbf {\bibinfo {volume} {6}},\
  \bibinfo {pages} {1} (\bibinfo {year} {2020})}\BibitemShut {NoStop}%
\bibitem [{\citenamefont {Gibbs}\ \emph {et~al.}(2021)\citenamefont {Gibbs},
  \citenamefont {Gili}, \citenamefont {Holmes}, \citenamefont {Commeau},
  \citenamefont {Arrasmith}, \citenamefont {Cincio}, \citenamefont {Coles},\
  and\ \citenamefont {Sornborger}}]{gibbs2021longtime}%
  \BibitemOpen
  \bibfield  {author} {\bibinfo {author} {\bibfnamefont {J.}~\bibnamefont
  {Gibbs}}, \bibinfo {author} {\bibfnamefont {K.}~\bibnamefont {Gili}},
  \bibinfo {author} {\bibfnamefont {Z.}~\bibnamefont {Holmes}}, \bibinfo
  {author} {\bibfnamefont {B.}~\bibnamefont {Commeau}}, \bibinfo {author}
  {\bibfnamefont {A.}~\bibnamefont {Arrasmith}}, \bibinfo {author}
  {\bibfnamefont {L.}~\bibnamefont {Cincio}}, \bibinfo {author} {\bibfnamefont
  {P.~J.}\ \bibnamefont {Coles}},\ and\ \bibinfo {author} {\bibfnamefont
  {A.}~\bibnamefont {Sornborger}},\ }\href@noop {} {\bibinfo {title} {Long-time
  simulations with high fidelity on quantum hardware}} (\bibinfo {year}
  {2021}),\ \Eprint {https://arxiv.org/abs/2102.04313} {arXiv:2102.04313
  [quant-ph]} \BibitemShut {NoStop}%
\bibitem [{\citenamefont {Yao}\ \emph {et~al.}(2021)\citenamefont {Yao},
  \citenamefont {Gomes}, \citenamefont {Zhang}, \citenamefont {Wang},
  \citenamefont {Ho}, \citenamefont {Iadecola},\ and\ \citenamefont
  {Orth}}]{Yao:2021ddt}%
  \BibitemOpen
  \bibfield  {author} {\bibinfo {author} {\bibfnamefont {Y.-X.}\ \bibnamefont
  {Yao}}, \bibinfo {author} {\bibfnamefont {N.}~\bibnamefont {Gomes}}, \bibinfo
  {author} {\bibfnamefont {F.}~\bibnamefont {Zhang}}, \bibinfo {author}
  {\bibfnamefont {C.-Z.}\ \bibnamefont {Wang}}, \bibinfo {author}
  {\bibfnamefont {K.-M.}\ \bibnamefont {Ho}}, \bibinfo {author} {\bibfnamefont
  {T.}~\bibnamefont {Iadecola}},\ and\ \bibinfo {author} {\bibfnamefont
  {P.~P.}\ \bibnamefont {Orth}},\ }\bibfield  {title} {\bibinfo {title}
  {{Adaptive Variational Quantum Dynamics Simulations}},\ }\href
  {https://doi.org/10.1103/PRXQuantum.2.030307} {\bibfield  {journal} {\bibinfo
   {journal} {PRX Quantum}\ }\textbf {\bibinfo {volume} {2}},\ \bibinfo {pages}
  {030307} (\bibinfo {year} {2021})}\BibitemShut {NoStop}%
\bibitem [{\citenamefont {Nagano}\ \emph {et~al.}(2023)\citenamefont {Nagano},
  \citenamefont {Bapat},\ and\ \citenamefont {Bauer}}]{Nagano:2023uaq}%
  \BibitemOpen
  \bibfield  {author} {\bibinfo {author} {\bibfnamefont {L.}~\bibnamefont
  {Nagano}}, \bibinfo {author} {\bibfnamefont {A.}~\bibnamefont {Bapat}},\ and\
  \bibinfo {author} {\bibfnamefont {C.~W.}\ \bibnamefont {Bauer}},\ }\bibfield
  {title} {\bibinfo {title} {{Quench dynamics of the Schwinger model via
  variational quantum algorithms}},\ }\href
  {https://doi.org/10.1103/PhysRevD.108.034501} {\bibfield  {journal} {\bibinfo
   {journal} {Phys. Rev. D}\ }\textbf {\bibinfo {volume} {108}},\ \bibinfo
  {pages} {034501} (\bibinfo {year} {2023})}\BibitemShut {NoStop}%
\bibitem [{\citenamefont {Roggero}\ and\ \citenamefont
  {Carlson}(2018)}]{Roggero:2018hrn}%
  \BibitemOpen
  \bibfield  {author} {\bibinfo {author} {\bibfnamefont {A.}~\bibnamefont
  {Roggero}}\ and\ \bibinfo {author} {\bibfnamefont {J.}~\bibnamefont
  {Carlson}},\ }\href@noop {} {\bibinfo {title} {{Linear Response on a Quantum
  Computer}}} (\bibinfo {year} {2018}),\ \Eprint
  {https://arxiv.org/abs/1804.01505} {arXiv:1804.01505 [quant-ph]} \BibitemShut
  {NoStop}%
\bibitem [{\citenamefont {Roggero}\ and\ \citenamefont
  {Baroni}(2019)}]{Roggero:2019srp}%
  \BibitemOpen
  \bibfield  {author} {\bibinfo {author} {\bibfnamefont {A.}~\bibnamefont
  {Roggero}}\ and\ \bibinfo {author} {\bibfnamefont {A.}~\bibnamefont
  {Baroni}},\ }\href@noop {} {\bibinfo {title} {{Short-depth circuits for
  efficient expectation value estimation}}} (\bibinfo {year} {2019}),\ \Eprint
  {https://arxiv.org/abs/1905.08383} {arXiv:1905.08383 [quant-ph]} \BibitemShut
  {NoStop}%
\bibitem [{\citenamefont {Kanasugi}\ \emph {et~al.}(2023)\citenamefont
  {Kanasugi}, \citenamefont {Tsutsui}, \citenamefont {Nakagawa}, \citenamefont
  {Maruyama}, \citenamefont {Oshima},\ and\ \citenamefont
  {Sato}}]{Kanasugi:2023wxu}%
  \BibitemOpen
  \bibfield  {author} {\bibinfo {author} {\bibfnamefont {S.}~\bibnamefont
  {Kanasugi}}, \bibinfo {author} {\bibfnamefont {S.}~\bibnamefont {Tsutsui}},
  \bibinfo {author} {\bibfnamefont {Y.~O.}\ \bibnamefont {Nakagawa}}, \bibinfo
  {author} {\bibfnamefont {K.}~\bibnamefont {Maruyama}}, \bibinfo {author}
  {\bibfnamefont {H.}~\bibnamefont {Oshima}},\ and\ \bibinfo {author}
  {\bibfnamefont {S.}~\bibnamefont {Sato}},\ }\bibfield  {title} {\bibinfo
  {title} {{Computation of Green's function by local variational quantum
  compilation}},\ }\href {https://doi.org/10.1103/PhysRevResearch.5.033070}
  {\bibfield  {journal} {\bibinfo  {journal} {Phys. Rev. Res.}\ }\textbf
  {\bibinfo {volume} {5}},\ \bibinfo {pages} {033070} (\bibinfo {year}
  {2023})},\ \Eprint {https://arxiv.org/abs/2303.15667} {arXiv:2303.15667
  [quant-ph]} \BibitemShut {NoStop}%
\bibitem [{\citenamefont {Gustafson}\ \emph {et~al.}(2023)\citenamefont
  {Gustafson}, \citenamefont {Lamm},\ and\ \citenamefont
  {Unmuth-Yockey}}]{Gustafson:2023ayr}%
  \BibitemOpen
  \bibfield  {author} {\bibinfo {author} {\bibfnamefont {E.~J.}\ \bibnamefont
  {Gustafson}}, \bibinfo {author} {\bibfnamefont {H.}~\bibnamefont {Lamm}},\
  and\ \bibinfo {author} {\bibfnamefont {J.}~\bibnamefont {Unmuth-Yockey}},\
  }\bibfield  {title} {\bibinfo {title} {{Quantum mean estimation for lattice
  field theory}},\ }\href {https://doi.org/10.1103/PhysRevD.107.114511}
  {\bibfield  {journal} {\bibinfo  {journal} {Phys. Rev. D}\ }\textbf {\bibinfo
  {volume} {107}},\ \bibinfo {pages} {114511} (\bibinfo {year} {2023})},\
  \Eprint {https://arxiv.org/abs/2303.00094} {arXiv:2303.00094 [hep-lat]}
  \BibitemShut {NoStop}%
\bibitem [{\citenamefont {Lamm}\ \emph {et~al.}(2019)\citenamefont {Lamm},
  \citenamefont {Lawrence},\ and\ \citenamefont {Yamauchi}}]{Lamm:2019bik}%
  \BibitemOpen
  \bibfield  {author} {\bibinfo {author} {\bibfnamefont {H.}~\bibnamefont
  {Lamm}}, \bibinfo {author} {\bibfnamefont {S.}~\bibnamefont {Lawrence}},\
  and\ \bibinfo {author} {\bibfnamefont {Y.}~\bibnamefont {Yamauchi}} (\bibinfo
  {collaboration} {NuQS}),\ }\bibfield  {title} {\bibinfo {title} {{General
  Methods for Digital Quantum Simulation of Gauge Theories}},\ }\href
  {https://doi.org/10.1103/PhysRevD.100.034518} {\bibfield  {journal} {\bibinfo
   {journal} {Phys. Rev.}\ }\textbf {\bibinfo {volume} {D100}},\ \bibinfo
  {pages} {034518} (\bibinfo {year} {2019})},\ \Eprint
  {https://arxiv.org/abs/1903.08807} {arXiv:1903.08807 [hep-lat]} \BibitemShut
  {NoStop}%
\bibitem [{\citenamefont {Bauer}\ \emph
  {et~al.}(2021{\natexlab{a}})\citenamefont {Bauer}, \citenamefont {de~Jong},
  \citenamefont {Nachman},\ and\ \citenamefont {Provasoli}}]{Bauer:2019qxa}%
  \BibitemOpen
  \bibfield  {author} {\bibinfo {author} {\bibfnamefont {C.~W.}\ \bibnamefont
  {Bauer}}, \bibinfo {author} {\bibfnamefont {W.~A.}\ \bibnamefont {de~Jong}},
  \bibinfo {author} {\bibfnamefont {B.}~\bibnamefont {Nachman}},\ and\ \bibinfo
  {author} {\bibfnamefont {D.}~\bibnamefont {Provasoli}},\ }\bibfield  {title}
  {\bibinfo {title} {{Quantum Algorithm for High Energy Physics Simulations}},\
  }\href {https://doi.org/10.1103/PhysRevLett.126.062001} {\bibfield  {journal}
  {\bibinfo  {journal} {Phys. Rev. Lett.}\ }\textbf {\bibinfo {volume} {126}},\
  \bibinfo {pages} {062001} (\bibinfo {year} {2021}{\natexlab{a}})},\ \Eprint
  {https://arxiv.org/abs/1904.03196} {arXiv:1904.03196 [hep-ph]} \BibitemShut
  {NoStop}%
\bibitem [{\citenamefont {Echevarria}\ \emph {et~al.}(2020)\citenamefont
  {Echevarria}, \citenamefont {Egusquiza}, \citenamefont {Rico},\ and\
  \citenamefont {Schnell}}]{Echevarria:2020wct}%
  \BibitemOpen
  \bibfield  {author} {\bibinfo {author} {\bibfnamefont {M.}~\bibnamefont
  {Echevarria}}, \bibinfo {author} {\bibfnamefont {I.}~\bibnamefont
  {Egusquiza}}, \bibinfo {author} {\bibfnamefont {E.}~\bibnamefont {Rico}},\
  and\ \bibinfo {author} {\bibfnamefont {G.}~\bibnamefont {Schnell}},\
  }\href@noop {} {\bibinfo {title} {{Quantum Simulation of Light-Front Parton
  Correlators}}} (\bibinfo {year} {2020}),\ \Eprint
  {https://arxiv.org/abs/2011.01275} {arXiv:2011.01275 [quant-ph]} \BibitemShut
  {NoStop}%
\bibitem [{\citenamefont {Xu}\ and\ \citenamefont {Xue}(2022)}]{Xu:2021tey}%
  \BibitemOpen
  \bibfield  {author} {\bibinfo {author} {\bibfnamefont {B.}~\bibnamefont
  {Xu}}\ and\ \bibinfo {author} {\bibfnamefont {W.}~\bibnamefont {Xue}},\
  }\bibfield  {title} {\bibinfo {title} {{(3+1)-dimensional Schwinger pair
  production with quantum computers}},\ }\href
  {https://doi.org/10.1103/PhysRevD.106.116007} {\bibfield  {journal} {\bibinfo
   {journal} {Phys. Rev. D}\ }\textbf {\bibinfo {volume} {106}},\ \bibinfo
  {pages} {116007} (\bibinfo {year} {2022})},\ \Eprint
  {https://arxiv.org/abs/2112.06863} {arXiv:2112.06863 [quant-ph]} \BibitemShut
  {NoStop}%
\bibitem [{\citenamefont {Bauer}\ \emph
  {et~al.}(2021{\natexlab{b}})\citenamefont {Bauer}, \citenamefont {Freytsis},\
  and\ \citenamefont {Nachman}}]{Bauer:2021gup}%
  \BibitemOpen
  \bibfield  {author} {\bibinfo {author} {\bibfnamefont {C.~W.}\ \bibnamefont
  {Bauer}}, \bibinfo {author} {\bibfnamefont {M.}~\bibnamefont {Freytsis}},\
  and\ \bibinfo {author} {\bibfnamefont {B.}~\bibnamefont {Nachman}},\
  }\bibfield  {title} {\bibinfo {title} {{Simulating Collider Physics on
  Quantum Computers Using Effective Field Theories}},\ }\href
  {https://doi.org/10.1103/PhysRevLett.127.212001} {\bibfield  {journal}
  {\bibinfo  {journal} {Phys. Rev. Lett.}\ }\textbf {\bibinfo {volume} {127}},\
  \bibinfo {pages} {212001} (\bibinfo {year} {2021}{\natexlab{b}})},\ \Eprint
  {https://arxiv.org/abs/2102.05044} {arXiv:2102.05044 [hep-ph]} \BibitemShut
  {NoStop}%
\bibitem [{\citenamefont {Cohen}\ \emph {et~al.}(2021)\citenamefont {Cohen},
  \citenamefont {Lamm}, \citenamefont {Lawrence},\ and\ \citenamefont
  {Yamauchi}}]{Cohen:2021imf}%
  \BibitemOpen
  \bibfield  {author} {\bibinfo {author} {\bibfnamefont {T.~D.}\ \bibnamefont
  {Cohen}}, \bibinfo {author} {\bibfnamefont {H.}~\bibnamefont {Lamm}},
  \bibinfo {author} {\bibfnamefont {S.}~\bibnamefont {Lawrence}},\ and\
  \bibinfo {author} {\bibfnamefont {Y.}~\bibnamefont {Yamauchi}} (\bibinfo
  {collaboration} {NuQS}),\ }\bibfield  {title} {\bibinfo {title} {{Quantum
  algorithms for transport coefficients in gauge theories}},\ }\href
  {https://doi.org/10.1103/PhysRevD.104.094514} {\bibfield  {journal} {\bibinfo
   {journal} {Phys. Rev. D}\ }\textbf {\bibinfo {volume} {104}},\ \bibinfo
  {pages} {094514} (\bibinfo {year} {2021})},\ \Eprint
  {https://arxiv.org/abs/2104.02024} {arXiv:2104.02024 [hep-lat]} \BibitemShut
  {NoStop}%
\bibitem [{\citenamefont {Barata}\ \emph {et~al.}(2022)\citenamefont {Barata},
  \citenamefont {Du}, \citenamefont {Li}, \citenamefont {Qian},\ and\
  \citenamefont {Salgado}}]{Barata:2022wim}%
  \BibitemOpen
  \bibfield  {author} {\bibinfo {author} {\bibfnamefont {J.~a.}\ \bibnamefont
  {Barata}}, \bibinfo {author} {\bibfnamefont {X.}~\bibnamefont {Du}}, \bibinfo
  {author} {\bibfnamefont {M.}~\bibnamefont {Li}}, \bibinfo {author}
  {\bibfnamefont {W.}~\bibnamefont {Qian}},\ and\ \bibinfo {author}
  {\bibfnamefont {C.~A.}\ \bibnamefont {Salgado}},\ }\bibfield  {title}
  {\bibinfo {title} {{Medium induced jet broadening in a quantum computer}},\
  }\href {https://doi.org/10.1103/PhysRevD.106.074013} {\bibfield  {journal}
  {\bibinfo  {journal} {Phys. Rev. D}\ }\textbf {\bibinfo {volume} {106}},\
  \bibinfo {pages} {074013} (\bibinfo {year} {2022})}\BibitemShut {NoStop}%
\bibitem [{\citenamefont {Czajka}\ \emph {et~al.}(2022)\citenamefont {Czajka},
  \citenamefont {Kang}, \citenamefont {Tee},\ and\ \citenamefont
  {Zhao}}]{Czajka:2022plx}%
  \BibitemOpen
  \bibfield  {author} {\bibinfo {author} {\bibfnamefont {A.~M.}\ \bibnamefont
  {Czajka}}, \bibinfo {author} {\bibfnamefont {Z.-B.}\ \bibnamefont {Kang}},
  \bibinfo {author} {\bibfnamefont {Y.}~\bibnamefont {Tee}},\ and\ \bibinfo
  {author} {\bibfnamefont {F.}~\bibnamefont {Zhao}},\ }\href@noop {} {\bibinfo
  {title} {{Studying chirality imbalance with quantum algorithms}}} (\bibinfo
  {year} {2022}),\ \Eprint {https://arxiv.org/abs/2210.03062} {arXiv:2210.03062
  [hep-ph]} \BibitemShut {NoStop}%
\bibitem [{\citenamefont {Farrell}\ \emph
  {et~al.}(2023{\natexlab{b}})\citenamefont {Farrell}, \citenamefont
  {Chernyshev}, \citenamefont {Powell}, \citenamefont {Zemlevskiy},
  \citenamefont {Illa},\ and\ \citenamefont {Savage}}]{Farrell:2022vyh}%
  \BibitemOpen
  \bibfield  {author} {\bibinfo {author} {\bibfnamefont {R.~C.}\ \bibnamefont
  {Farrell}}, \bibinfo {author} {\bibfnamefont {I.~A.}\ \bibnamefont
  {Chernyshev}}, \bibinfo {author} {\bibfnamefont {S.~J.~M.}\ \bibnamefont
  {Powell}}, \bibinfo {author} {\bibfnamefont {N.~A.}\ \bibnamefont
  {Zemlevskiy}}, \bibinfo {author} {\bibfnamefont {M.}~\bibnamefont {Illa}},\
  and\ \bibinfo {author} {\bibfnamefont {M.~J.}\ \bibnamefont {Savage}},\
  }\bibfield  {title} {\bibinfo {title} {{Preparations for quantum simulations
  of quantum chromodynamics in 1+1 dimensions. II. Single-baryon
  \ensuremath{\beta}-decay in real time}},\ }\href
  {https://doi.org/10.1103/PhysRevD.107.054513} {\bibfield  {journal} {\bibinfo
   {journal} {Phys. Rev. D}\ }\textbf {\bibinfo {volume} {107}},\ \bibinfo
  {pages} {054513} (\bibinfo {year} {2023}{\natexlab{b}})},\ \Eprint
  {https://arxiv.org/abs/2209.10781} {arXiv:2209.10781 [quant-ph]} \BibitemShut
  {NoStop}%
\bibitem [{\citenamefont {Bedaque}\ \emph {et~al.}(2022)\citenamefont
  {Bedaque}, \citenamefont {Khadka}, \citenamefont {Rupak},\ and\ \citenamefont
  {Yusf}}]{Bedaque:2022ftd}%
  \BibitemOpen
  \bibfield  {author} {\bibinfo {author} {\bibfnamefont {P.~F.}\ \bibnamefont
  {Bedaque}}, \bibinfo {author} {\bibfnamefont {R.}~\bibnamefont {Khadka}},
  \bibinfo {author} {\bibfnamefont {G.}~\bibnamefont {Rupak}},\ and\ \bibinfo
  {author} {\bibfnamefont {M.}~\bibnamefont {Yusf}},\ }\href@noop {} {\bibinfo
  {title} {{Radiative processes on a quantum computer}}} (\bibinfo {year}
  {2022}),\ \Eprint {https://arxiv.org/abs/2209.09962} {arXiv:2209.09962
  [nucl-th]} \BibitemShut {NoStop}%
\bibitem [{\citenamefont {Ikeda}\ \emph {et~al.}(2023)\citenamefont {Ikeda},
  \citenamefont {Kharzeev}, \citenamefont {Meyer},\ and\ \citenamefont
  {Shi}}]{Ikeda:2023zil}%
  \BibitemOpen
  \bibfield  {author} {\bibinfo {author} {\bibfnamefont {K.}~\bibnamefont
  {Ikeda}}, \bibinfo {author} {\bibfnamefont {D.~E.}\ \bibnamefont {Kharzeev}},
  \bibinfo {author} {\bibfnamefont {R.}~\bibnamefont {Meyer}},\ and\ \bibinfo
  {author} {\bibfnamefont {S.}~\bibnamefont {Shi}},\ }\href@noop {} {\bibinfo
  {title} {{Detecting the critical point through entanglement in Schwinger
  model}}} (\bibinfo {year} {2023}),\ \Eprint
  {https://arxiv.org/abs/2305.00996} {arXiv:2305.00996 [hep-ph]} \BibitemShut
  {NoStop}%
\bibitem [{\citenamefont {Ciavarella}(2020)}]{Ciavarella:2020vqm}%
  \BibitemOpen
  \bibfield  {author} {\bibinfo {author} {\bibfnamefont {A.}~\bibnamefont
  {Ciavarella}},\ }\bibfield  {title} {\bibinfo {title} {{Algorithm for quantum
  computation of particle decays}},\ }\href
  {https://doi.org/10.1103/PhysRevD.102.094505} {\bibfield  {journal} {\bibinfo
   {journal} {Phys. Rev. D}\ }\textbf {\bibinfo {volume} {102}},\ \bibinfo
  {pages} {094505} (\bibinfo {year} {2020})},\ \Eprint
  {https://arxiv.org/abs/2007.04447} {arXiv:2007.04447 [hep-th]} \BibitemShut
  {NoStop}%
\bibitem [{\citenamefont {Huffman}\ \emph {et~al.}(2022)\citenamefont
  {Huffman}, \citenamefont {Garc\'\i{}a~Vera},\ and\ \citenamefont
  {Banerjee}}]{Huffman:2021gsi}%
  \BibitemOpen
  \bibfield  {author} {\bibinfo {author} {\bibfnamefont {E.}~\bibnamefont
  {Huffman}}, \bibinfo {author} {\bibfnamefont {M.}~\bibnamefont
  {Garc\'\i{}a~Vera}},\ and\ \bibinfo {author} {\bibfnamefont {D.}~\bibnamefont
  {Banerjee}},\ }\bibfield  {title} {\bibinfo {title} {{Toward the real-time
  evolution of gauge-invariant $\mathbb Z_2$ and $U(1)$ quantum link models on
  noisy intermediate-scale quantum hardware with error mitigation}},\ }\href
  {https://doi.org/10.1103/PhysRevD.106.094502} {\bibfield  {journal} {\bibinfo
   {journal} {Phys. Rev. D}\ }\textbf {\bibinfo {volume} {106}},\ \bibinfo
  {pages} {094502} (\bibinfo {year} {2022})},\ \Eprint
  {https://arxiv.org/abs/2109.15065} {arXiv:2109.15065 [quant-ph]} \BibitemShut
  {NoStop}%
\bibitem [{\citenamefont {Klco}\ and\ \citenamefont
  {Savage}(2021)}]{Klco:2021jxl}%
  \BibitemOpen
  \bibfield  {author} {\bibinfo {author} {\bibfnamefont {N.}~\bibnamefont
  {Klco}}\ and\ \bibinfo {author} {\bibfnamefont {M.~J.}\ \bibnamefont
  {Savage}},\ }\bibfield  {title} {\bibinfo {title} {{Hierarchical qubit maps
  and hierarchically implemented quantum error correction}},\ }\href
  {https://doi.org/10.1103/PhysRevA.104.062425} {\bibfield  {journal} {\bibinfo
   {journal} {Phys. Rev. A}\ }\textbf {\bibinfo {volume} {104}},\ \bibinfo
  {pages} {062425} (\bibinfo {year} {2021})},\ \Eprint
  {https://arxiv.org/abs/2109.01953} {arXiv:2109.01953 [quant-ph]} \BibitemShut
  {NoStop}%
\bibitem [{\citenamefont {Charles}\ \emph {et~al.}(2023)\citenamefont
  {Charles}, \citenamefont {Gustafson}, \citenamefont {Hardt}, \citenamefont
  {Herren}, \citenamefont {Hogan}, \citenamefont {Lamm}, \citenamefont
  {Starecheski}, \citenamefont {Van~de Water},\ and\ \citenamefont
  {Wagman}}]{Charles:2023zbl}%
  \BibitemOpen
  \bibfield  {author} {\bibinfo {author} {\bibfnamefont {C.}~\bibnamefont
  {Charles}}, \bibinfo {author} {\bibfnamefont {E.~J.}\ \bibnamefont
  {Gustafson}}, \bibinfo {author} {\bibfnamefont {E.}~\bibnamefont {Hardt}},
  \bibinfo {author} {\bibfnamefont {F.}~\bibnamefont {Herren}}, \bibinfo
  {author} {\bibfnamefont {N.}~\bibnamefont {Hogan}}, \bibinfo {author}
  {\bibfnamefont {H.}~\bibnamefont {Lamm}}, \bibinfo {author} {\bibfnamefont
  {S.}~\bibnamefont {Starecheski}}, \bibinfo {author} {\bibfnamefont {R.~S.}\
  \bibnamefont {Van~de Water}},\ and\ \bibinfo {author} {\bibfnamefont {M.~L.}\
  \bibnamefont {Wagman}},\ }\href@noop {} {\bibinfo {title} {{Simulating
  $\mathbb{Z}_2$ lattice gauge theory on a quantum computer}}} (\bibinfo {year}
  {2023}),\ \Eprint {https://arxiv.org/abs/2305.02361} {arXiv:2305.02361
  [hep-lat]} \BibitemShut {NoStop}%
\bibitem [{\citenamefont {Gustafson}(2022{\natexlab{a}})}]{Gustafson:2022xlj}%
  \BibitemOpen
  \bibfield  {author} {\bibinfo {author} {\bibfnamefont {E.}~\bibnamefont
  {Gustafson}},\ }\href@noop {} {\bibinfo {title} {{Noise Improvements in
  Quantum Simulations of sQED using Qutrits}}} (\bibinfo {year}
  {2022}{\natexlab{a}}),\ \Eprint {https://arxiv.org/abs/2201.04546}
  {arXiv:2201.04546 [quant-ph]} \BibitemShut {NoStop}%
\bibitem [{\citenamefont {Rajput}\ \emph {et~al.}(2021)\citenamefont {Rajput},
  \citenamefont {Roggero},\ and\ \citenamefont {Wiebe}}]{rajput2021quantum}%
  \BibitemOpen
  \bibfield  {author} {\bibinfo {author} {\bibfnamefont {A.}~\bibnamefont
  {Rajput}}, \bibinfo {author} {\bibfnamefont {A.}~\bibnamefont {Roggero}},\
  and\ \bibinfo {author} {\bibfnamefont {N.}~\bibnamefont {Wiebe}},\
  }\href@noop {} {\bibinfo {title} {Quantum error correction with gauge
  symmetries}} (\bibinfo {year} {2021}),\ \Eprint
  {https://arxiv.org/abs/2112.05186} {arXiv:2112.05186 [quant-ph]} \BibitemShut
  {NoStop}%
\bibitem [{\citenamefont {Gustafson}\ and\ \citenamefont
  {Lamm}(2023)}]{Gustafson:2023swx}%
  \BibitemOpen
  \bibfield  {author} {\bibinfo {author} {\bibfnamefont {E.~J.}\ \bibnamefont
  {Gustafson}}\ and\ \bibinfo {author} {\bibfnamefont {H.}~\bibnamefont
  {Lamm}},\ }\href@noop {} {\bibinfo {title} {{Robustness of Gauge Digitization
  to Quantum Noise}}} (\bibinfo {year} {2023}),\ \Eprint
  {https://arxiv.org/abs/2301.10207} {arXiv:2301.10207 [hep-lat]} \BibitemShut
  {NoStop}%
\bibitem [{\citenamefont {Halimeh}\ and\ \citenamefont
  {Hauke}(2019)}]{Halimeh:2019svu}%
  \BibitemOpen
  \bibfield  {author} {\bibinfo {author} {\bibfnamefont {J.~C.}\ \bibnamefont
  {Halimeh}}\ and\ \bibinfo {author} {\bibfnamefont {P.}~\bibnamefont
  {Hauke}},\ }\href@noop {} {\bibinfo {title} {{Reliability of lattice gauge
  theories}}} (\bibinfo {year} {2019}),\ \Eprint
  {https://arxiv.org/abs/2001.00024} {arXiv:2001.00024 [cond-mat.quant-gas]}
  \BibitemShut {NoStop}%
\bibitem [{\citenamefont {Lamm}\ \emph {et~al.}(2020)\citenamefont {Lamm},
  \citenamefont {Lawrence},\ and\ \citenamefont {Yamauchi}}]{Lamm:2020jwv}%
  \BibitemOpen
  \bibfield  {author} {\bibinfo {author} {\bibfnamefont {H.}~\bibnamefont
  {Lamm}}, \bibinfo {author} {\bibfnamefont {S.}~\bibnamefont {Lawrence}},\
  and\ \bibinfo {author} {\bibfnamefont {Y.}~\bibnamefont {Yamauchi}} (\bibinfo
  {collaboration} {NuQS}),\ }\href@noop {} {\bibinfo {title} {{Suppressing
  Coherent Gauge Drift in Quantum Simulations}}} (\bibinfo {year} {2020}),\
  \Eprint {https://arxiv.org/abs/2005.12688} {arXiv:2005.12688 [quant-ph]}
  \BibitemShut {NoStop}%
\bibitem [{\citenamefont {Tran}\ \emph {et~al.}(2020)\citenamefont {Tran},
  \citenamefont {Su}, \citenamefont {Carney},\ and\ \citenamefont
  {Taylor}}]{Tran:2020azk}%
  \BibitemOpen
  \bibfield  {author} {\bibinfo {author} {\bibfnamefont {M.~C.}\ \bibnamefont
  {Tran}}, \bibinfo {author} {\bibfnamefont {Y.}~\bibnamefont {Su}}, \bibinfo
  {author} {\bibfnamefont {D.}~\bibnamefont {Carney}},\ and\ \bibinfo {author}
  {\bibfnamefont {J.~M.}\ \bibnamefont {Taylor}},\ }\href@noop {} {\bibinfo
  {title} {{Faster Digital Quantum Simulation by Symmetry Protection}}}
  (\bibinfo {year} {2020}),\ \Eprint {https://arxiv.org/abs/2006.16248}
  {arXiv:2006.16248 [quant-ph]} \BibitemShut {NoStop}%
\bibitem [{\citenamefont {Kasper}\ \emph {et~al.}(2020)\citenamefont {Kasper},
  \citenamefont {Zache}, \citenamefont {Jendrzejewski}, \citenamefont
  {Lewenstein},\ and\ \citenamefont {Zohar}}]{Kasper:2020owz}%
  \BibitemOpen
  \bibfield  {author} {\bibinfo {author} {\bibfnamefont {V.}~\bibnamefont
  {Kasper}}, \bibinfo {author} {\bibfnamefont {T.~V.}\ \bibnamefont {Zache}},
  \bibinfo {author} {\bibfnamefont {F.}~\bibnamefont {Jendrzejewski}}, \bibinfo
  {author} {\bibfnamefont {M.}~\bibnamefont {Lewenstein}},\ and\ \bibinfo
  {author} {\bibfnamefont {E.}~\bibnamefont {Zohar}},\ }\href@noop {} {\bibinfo
  {title} {{Non-Abelian gauge invariance from dynamical decoupling}}} (\bibinfo
  {year} {2020}),\ \Eprint {https://arxiv.org/abs/2012.08620} {arXiv:2012.08620
  [quant-ph]} \BibitemShut {NoStop}%
\bibitem [{\citenamefont {Halimeh}\ \emph {et~al.}(2020)\citenamefont
  {Halimeh}, \citenamefont {Lang}, \citenamefont {Mildenberger}, \citenamefont
  {Jiang},\ and\ \citenamefont {Hauke}}]{Halimeh:2020ecg}%
  \BibitemOpen
  \bibfield  {author} {\bibinfo {author} {\bibfnamefont {J.~C.}\ \bibnamefont
  {Halimeh}}, \bibinfo {author} {\bibfnamefont {H.}~\bibnamefont {Lang}},
  \bibinfo {author} {\bibfnamefont {J.}~\bibnamefont {Mildenberger}}, \bibinfo
  {author} {\bibfnamefont {Z.}~\bibnamefont {Jiang}},\ and\ \bibinfo {author}
  {\bibfnamefont {P.}~\bibnamefont {Hauke}},\ }\href@noop {} {\bibinfo {title}
  {{Gauge-Symmetry Protection Using Single-Body Terms}}} (\bibinfo {year}
  {2020}),\ \Eprint {https://arxiv.org/abs/2007.00668} {arXiv:2007.00668
  [quant-ph]} \BibitemShut {NoStop}%
\bibitem [{\citenamefont {Van~Damme}\ \emph {et~al.}(2020)\citenamefont
  {Van~Damme}, \citenamefont {Halimeh},\ and\ \citenamefont
  {Hauke}}]{VanDamme:2020rur}%
  \BibitemOpen
  \bibfield  {author} {\bibinfo {author} {\bibfnamefont {M.}~\bibnamefont
  {Van~Damme}}, \bibinfo {author} {\bibfnamefont {J.~C.}\ \bibnamefont
  {Halimeh}},\ and\ \bibinfo {author} {\bibfnamefont {P.}~\bibnamefont
  {Hauke}},\ }\href@noop {} {\bibinfo {title} {{Gauge-Symmetry Violation
  Quantum Phase Transition in Lattice Gauge Theories}}} (\bibinfo {year}
  {2020}),\ \Eprint {https://arxiv.org/abs/2010.07338} {arXiv:2010.07338
  [cond-mat.quant-gas]} \BibitemShut {NoStop}%
\bibitem [{\citenamefont {Nguyen}\ \emph {et~al.}(2022)\citenamefont {Nguyen},
  \citenamefont {Tran}, \citenamefont {Zhu}, \citenamefont {Green},
  \citenamefont {Alderete}, \citenamefont {Davoudi},\ and\ \citenamefont
  {Linke}}]{Nguyen:2021hyk}%
  \BibitemOpen
  \bibfield  {author} {\bibinfo {author} {\bibfnamefont {N.~H.}\ \bibnamefont
  {Nguyen}}, \bibinfo {author} {\bibfnamefont {M.~C.}\ \bibnamefont {Tran}},
  \bibinfo {author} {\bibfnamefont {Y.}~\bibnamefont {Zhu}}, \bibinfo {author}
  {\bibfnamefont {A.~M.}\ \bibnamefont {Green}}, \bibinfo {author}
  {\bibfnamefont {C.~H.}\ \bibnamefont {Alderete}}, \bibinfo {author}
  {\bibfnamefont {Z.}~\bibnamefont {Davoudi}},\ and\ \bibinfo {author}
  {\bibfnamefont {N.~M.}\ \bibnamefont {Linke}},\ }\bibfield  {title} {\bibinfo
  {title} {{Digital Quantum Simulation of the Schwinger Model and Symmetry
  Protection with Trapped Ions}},\ }\href
  {https://doi.org/10.1103/PRXQuantum.3.020324} {\bibfield  {journal} {\bibinfo
   {journal} {PRX Quantum}\ }\textbf {\bibinfo {volume} {3}},\ \bibinfo {pages}
  {020324} (\bibinfo {year} {2022})},\ \Eprint
  {https://arxiv.org/abs/2112.14262} {arXiv:2112.14262 [quant-ph]} \BibitemShut
  {NoStop}%
\bibitem [{\citenamefont {Halimeh}\ \emph {et~al.}(2021)\citenamefont
  {Halimeh}, \citenamefont {Lang},\ and\ \citenamefont
  {Hauke}}]{Halimeh:2021vzf}%
  \BibitemOpen
  \bibfield  {author} {\bibinfo {author} {\bibfnamefont {J.~C.}\ \bibnamefont
  {Halimeh}}, \bibinfo {author} {\bibfnamefont {H.}~\bibnamefont {Lang}},\ and\
  \bibinfo {author} {\bibfnamefont {P.}~\bibnamefont {Hauke}},\ }\href@noop {}
  {\bibinfo {title} {{Gauge protection in non-Abelian lattice gauge theories}}}
  (\bibinfo {year} {2021}),\ \Eprint {https://arxiv.org/abs/2106.09032}
  {arXiv:2106.09032 [cond-mat.quant-gas]} \BibitemShut {NoStop}%
\bibitem [{\citenamefont {A~Rahman}\ \emph {et~al.}(2022)\citenamefont
  {A~Rahman}, \citenamefont {Lewis}, \citenamefont {Mendicelli},\ and\
  \citenamefont {Powell}}]{ARahman:2022tkr}%
  \BibitemOpen
  \bibfield  {author} {\bibinfo {author} {\bibfnamefont {S.}~\bibnamefont
  {A~Rahman}}, \bibinfo {author} {\bibfnamefont {R.}~\bibnamefont {Lewis}},
  \bibinfo {author} {\bibfnamefont {E.}~\bibnamefont {Mendicelli}},\ and\
  \bibinfo {author} {\bibfnamefont {S.}~\bibnamefont {Powell}},\ }\bibfield
  {title} {\bibinfo {title} {{Self-mitigating Trotter circuits for SU(2)
  lattice gauge theory on a quantum computer}},\ }\href
  {https://doi.org/10.1103/PhysRevD.106.074502} {\bibfield  {journal} {\bibinfo
   {journal} {Phys. Rev. D}\ }\textbf {\bibinfo {volume} {106}},\ \bibinfo
  {pages} {074502} (\bibinfo {year} {2022})},\ \Eprint
  {https://arxiv.org/abs/2205.09247} {arXiv:2205.09247 [hep-lat]} \BibitemShut
  {NoStop}%
\bibitem [{\citenamefont {Yeter-Aydeniz}\ \emph {et~al.}(2022)\citenamefont
  {Yeter-Aydeniz}, \citenamefont {Parks}, \citenamefont {Nair}, \citenamefont
  {Gustafson}, \citenamefont {Kemper}, \citenamefont {Pooser}, \citenamefont
  {Meurice},\ and\ \citenamefont {Dreher}}]{Yeter-Aydeniz:2022vuy}%
  \BibitemOpen
  \bibfield  {author} {\bibinfo {author} {\bibfnamefont {K.}~\bibnamefont
  {Yeter-Aydeniz}}, \bibinfo {author} {\bibfnamefont {Z.}~\bibnamefont
  {Parks}}, \bibinfo {author} {\bibfnamefont {A.}~\bibnamefont {Nair}},
  \bibinfo {author} {\bibfnamefont {E.}~\bibnamefont {Gustafson}}, \bibinfo
  {author} {\bibfnamefont {A.~F.}\ \bibnamefont {Kemper}}, \bibinfo {author}
  {\bibfnamefont {R.~C.}\ \bibnamefont {Pooser}}, \bibinfo {author}
  {\bibfnamefont {Y.}~\bibnamefont {Meurice}},\ and\ \bibinfo {author}
  {\bibfnamefont {P.}~\bibnamefont {Dreher}},\ }\href@noop {} {\bibinfo {title}
  {{Measuring NISQ Gate-Based Qubit Stability Using a 1+1 Field Theory and
  Cycle Benchmarking}}} (\bibinfo {year} {2022}),\ \Eprint
  {https://arxiv.org/abs/2201.02899} {arXiv:2201.02899 [quant-ph]} \BibitemShut
  {NoStop}%
\bibitem [{\citenamefont {Carena}\ \emph
  {et~al.}(2022{\natexlab{a}})\citenamefont {Carena}, \citenamefont {Lamm},
  \citenamefont {Li},\ and\ \citenamefont {Liu}}]{Carena:2022kpg}%
  \BibitemOpen
  \bibfield  {author} {\bibinfo {author} {\bibfnamefont {M.}~\bibnamefont
  {Carena}}, \bibinfo {author} {\bibfnamefont {H.}~\bibnamefont {Lamm}},
  \bibinfo {author} {\bibfnamefont {Y.-Y.}\ \bibnamefont {Li}},\ and\ \bibinfo
  {author} {\bibfnamefont {W.}~\bibnamefont {Liu}},\ }\href@noop {} {\bibinfo
  {title} {{Improved Hamiltonians for Quantum Simulations}}} (\bibinfo {year}
  {2022}{\natexlab{a}}),\ \Eprint {https://arxiv.org/abs/2203.02823}
  {arXiv:2203.02823 [hep-lat]} \BibitemShut {NoStop}%
\bibitem [{\citenamefont {Ciavarella}(2023)}]{Ciavarella:2023mfc}%
  \BibitemOpen
  \bibfield  {author} {\bibinfo {author} {\bibfnamefont {A.~N.}\ \bibnamefont
  {Ciavarella}},\ }\href@noop {} {\bibinfo {title} {{Quantum Simulation of
  Lattice QCD with Improved Hamiltonians}}} (\bibinfo {year} {2023}),\ \Eprint
  {https://arxiv.org/abs/2307.05593} {arXiv:2307.05593 [hep-lat]} \BibitemShut
  {NoStop}%
\bibitem [{\citenamefont {Gustafson}(2022{\natexlab{b}})}]{Gustafson:2022hjf}%
  \BibitemOpen
  \bibfield  {author} {\bibinfo {author} {\bibfnamefont {E.~J.}\ \bibnamefont
  {Gustafson}},\ }\href@noop {} {\bibinfo {title} {{Stout Smearing on a Quantum
  Computer}}} (\bibinfo {year} {2022}{\natexlab{b}}),\ \Eprint
  {https://arxiv.org/abs/2211.05607} {arXiv:2211.05607 [hep-lat]} \BibitemShut
  {NoStop}%
\bibitem [{\citenamefont {Temme}\ \emph {et~al.}(2011)\citenamefont {Temme},
  \citenamefont {Osborne}, \citenamefont {Vollbrecht}, \citenamefont {Poulin},\
  and\ \citenamefont {Verstraete}}]{Temme:2009wa}%
  \BibitemOpen
  \bibfield  {author} {\bibinfo {author} {\bibfnamefont {K.}~\bibnamefont
  {Temme}}, \bibinfo {author} {\bibfnamefont {T.}~\bibnamefont {Osborne}},
  \bibinfo {author} {\bibfnamefont {K.}~\bibnamefont {Vollbrecht}}, \bibinfo
  {author} {\bibfnamefont {D.}~\bibnamefont {Poulin}},\ and\ \bibinfo {author}
  {\bibfnamefont {F.}~\bibnamefont {Verstraete}},\ }\bibfield  {title}
  {\bibinfo {title} {{Quantum Metropolis Sampling}},\ }\href
  {https://doi.org/10.1038/nature09770} {\bibfield  {journal} {\bibinfo
  {journal} {Nature}\ }\textbf {\bibinfo {volume} {471}},\ \bibinfo {pages}
  {87} (\bibinfo {year} {2011})},\ \Eprint {https://arxiv.org/abs/0911.3635}
  {arXiv:0911.3635 [quant-ph]} \BibitemShut {NoStop}%
\bibitem [{\citenamefont {Clemente}\ \emph {et~al.}(2020)\citenamefont
  {Clemente} \emph {et~al.}}]{Clemente:2020lpr}%
  \BibitemOpen
  \bibfield  {author} {\bibinfo {author} {\bibfnamefont {G.}~\bibnamefont
  {Clemente}} \emph {et~al.} (\bibinfo {collaboration} {QuBiPF}),\ }\bibfield
  {title} {\bibinfo {title} {{Quantum computation of thermal averages in the
  presence of a sign problem}},\ }\href
  {https://doi.org/10.1103/PhysRevD.101.074510} {\bibfield  {journal} {\bibinfo
   {journal} {Phys. Rev. D}\ }\textbf {\bibinfo {volume} {101}},\ \bibinfo
  {pages} {074510} (\bibinfo {year} {2020})},\ \Eprint
  {https://arxiv.org/abs/2001.05328} {arXiv:2001.05328 [hep-lat]} \BibitemShut
  {NoStop}%
\bibitem [{\citenamefont {Yamamoto}(2022)}]{Yamamoto:2022jes}%
  \BibitemOpen
  \bibfield  {author} {\bibinfo {author} {\bibfnamefont {A.}~\bibnamefont
  {Yamamoto}},\ }\bibfield  {title} {\bibinfo {title} {{Quantum sampling for
  the Euclidean path integral of lattice gauge theory}},\ }\href
  {https://doi.org/10.1103/PhysRevD.105.094501} {\bibfield  {journal} {\bibinfo
   {journal} {Phys. Rev. D}\ }\textbf {\bibinfo {volume} {105}},\ \bibinfo
  {pages} {094501} (\bibinfo {year} {2022})},\ \Eprint
  {https://arxiv.org/abs/2201.12556} {arXiv:2201.12556 [quant-ph]} \BibitemShut
  {NoStop}%
\bibitem [{\citenamefont {Ballini}\ \emph {et~al.}(2023)\citenamefont
  {Ballini}, \citenamefont {Clemente}, \citenamefont {D'Elia}, \citenamefont
  {Maio},\ and\ \citenamefont {Zambello}}]{Ballini:2023ljs}%
  \BibitemOpen
  \bibfield  {author} {\bibinfo {author} {\bibfnamefont {E.}~\bibnamefont
  {Ballini}}, \bibinfo {author} {\bibfnamefont {G.}~\bibnamefont {Clemente}},
  \bibinfo {author} {\bibfnamefont {M.}~\bibnamefont {D'Elia}}, \bibinfo
  {author} {\bibfnamefont {L.}~\bibnamefont {Maio}},\ and\ \bibinfo {author}
  {\bibfnamefont {K.}~\bibnamefont {Zambello}},\ }\href@noop {} {\bibinfo
  {title} {{Quantum Computation of Thermal Averages for a Non-Abelian $D_4$
  Lattice Gauge Theory via Quantum Metropolis Sampling}}} (\bibinfo {year}
  {2023}),\ \Eprint {https://arxiv.org/abs/2309.07090} {arXiv:2309.07090
  [quant-ph]} \BibitemShut {NoStop}%
\bibitem [{\citenamefont {Avkhadiev}\ \emph {et~al.}(2020)\citenamefont
  {Avkhadiev}, \citenamefont {Shanahan},\ and\ \citenamefont
  {Young}}]{Avkhadiev:2019niu}%
  \BibitemOpen
  \bibfield  {author} {\bibinfo {author} {\bibfnamefont {A.}~\bibnamefont
  {Avkhadiev}}, \bibinfo {author} {\bibfnamefont {P.~E.}\ \bibnamefont
  {Shanahan}},\ and\ \bibinfo {author} {\bibfnamefont {R.~D.}\ \bibnamefont
  {Young}},\ }\bibfield  {title} {\bibinfo {title} {{Accelerating Lattice
  Quantum Field Theory Calculations via Interpolator Optimization Using Noisy
  Intermediate-Scale Quantum Computing}},\ }\href
  {https://doi.org/10.1103/PhysRevLett.124.080501} {\bibfield  {journal}
  {\bibinfo  {journal} {Phys. Rev. Lett.}\ }\textbf {\bibinfo {volume} {124}},\
  \bibinfo {pages} {080501} (\bibinfo {year} {2020})},\ \Eprint
  {https://arxiv.org/abs/1908.04194} {arXiv:1908.04194 [hep-lat]} \BibitemShut
  {NoStop}%
\bibitem [{\citenamefont {Avkhadiev}\ \emph {et~al.}(2023)\citenamefont
  {Avkhadiev}, \citenamefont {Shanahan},\ and\ \citenamefont
  {Young}}]{Avkhadiev:2022ttx}%
  \BibitemOpen
  \bibfield  {author} {\bibinfo {author} {\bibfnamefont {A.}~\bibnamefont
  {Avkhadiev}}, \bibinfo {author} {\bibfnamefont {P.~E.}\ \bibnamefont
  {Shanahan}},\ and\ \bibinfo {author} {\bibfnamefont {R.~D.}\ \bibnamefont
  {Young}},\ }\bibfield  {title} {\bibinfo {title} {{Strategies for
  quantum-optimized construction of interpolating operators in classical
  simulations of lattice quantum field theories}},\ }\href
  {https://doi.org/10.1103/PhysRevD.107.054507} {\bibfield  {journal} {\bibinfo
   {journal} {Phys. Rev. D}\ }\textbf {\bibinfo {volume} {107}},\ \bibinfo
  {pages} {054507} (\bibinfo {year} {2023})},\ \Eprint
  {https://arxiv.org/abs/2209.01209} {arXiv:2209.01209 [hep-lat]} \BibitemShut
  {NoStop}%
\bibitem [{\citenamefont {Alam}\ \emph {et~al.}(2022)\citenamefont {Alam},
  \citenamefont {Hadfield}, \citenamefont {Lamm},\ and\ \citenamefont
  {Li}}]{Alam:2021uuq}%
  \BibitemOpen
  \bibfield  {author} {\bibinfo {author} {\bibfnamefont {M.~S.}\ \bibnamefont
  {Alam}}, \bibinfo {author} {\bibfnamefont {S.}~\bibnamefont {Hadfield}},
  \bibinfo {author} {\bibfnamefont {H.}~\bibnamefont {Lamm}},\ and\ \bibinfo
  {author} {\bibfnamefont {A.~C.~Y.}\ \bibnamefont {Li}} (\bibinfo
  {collaboration} {SQMS}),\ }\bibfield  {title} {\bibinfo {title} {{Primitive
  quantum gates for dihedral gauge theories}},\ }\href
  {https://doi.org/10.1103/PhysRevD.105.114501} {\bibfield  {journal} {\bibinfo
   {journal} {Phys. Rev. D}\ }\textbf {\bibinfo {volume} {105}},\ \bibinfo
  {pages} {114501} (\bibinfo {year} {2022})},\ \Eprint
  {https://arxiv.org/abs/2108.13305} {arXiv:2108.13305 [quant-ph]} \BibitemShut
  {NoStop}%
\bibitem [{\citenamefont {Gustafson}\ \emph {et~al.}(2022)\citenamefont
  {Gustafson}, \citenamefont {Lamm}, \citenamefont {Lovelace},\ and\
  \citenamefont {Musk}}]{Gustafson:2022xdt}%
  \BibitemOpen
  \bibfield  {author} {\bibinfo {author} {\bibfnamefont {E.~J.}\ \bibnamefont
  {Gustafson}}, \bibinfo {author} {\bibfnamefont {H.}~\bibnamefont {Lamm}},
  \bibinfo {author} {\bibfnamefont {F.}~\bibnamefont {Lovelace}},\ and\
  \bibinfo {author} {\bibfnamefont {D.}~\bibnamefont {Musk}},\ }\bibfield
  {title} {\bibinfo {title} {{Primitive quantum gates for an SU(2) discrete
  subgroup: Binary tetrahedral}},\ }\href
  {https://doi.org/10.1103/PhysRevD.106.114501} {\bibfield  {journal} {\bibinfo
   {journal} {Phys. Rev. D}\ }\textbf {\bibinfo {volume} {106}},\ \bibinfo
  {pages} {114501} (\bibinfo {year} {2022})},\ \Eprint
  {https://arxiv.org/abs/2208.12309} {arXiv:2208.12309 [quant-ph]} \BibitemShut
  {NoStop}%
\bibitem [{\citenamefont {Zache}\ \emph
  {et~al.}(2023{\natexlab{a}})\citenamefont {Zache}, \citenamefont
  {Gonz\'alez-Cuadra},\ and\ \citenamefont {Zoller}}]{Zache:2023dko}%
  \BibitemOpen
  \bibfield  {author} {\bibinfo {author} {\bibfnamefont {T.~V.}\ \bibnamefont
  {Zache}}, \bibinfo {author} {\bibfnamefont {D.}~\bibnamefont
  {Gonz\'alez-Cuadra}},\ and\ \bibinfo {author} {\bibfnamefont
  {P.}~\bibnamefont {Zoller}},\ }\bibfield  {title} {\bibinfo {title} {{Quantum
  and Classical Spin-Network Algorithms for q-Deformed Kogut-Susskind Gauge
  Theories}},\ }\href {https://doi.org/10.1103/PhysRevLett.131.171902}
  {\bibfield  {journal} {\bibinfo  {journal} {Phys. Rev. Lett.}\ }\textbf
  {\bibinfo {volume} {131}},\ \bibinfo {pages} {171902} (\bibinfo {year}
  {2023}{\natexlab{a}})},\ \Eprint {https://arxiv.org/abs/2304.02527}
  {arXiv:2304.02527 [quant-ph]} \BibitemShut {NoStop}%
\bibitem [{\citenamefont {Zache}\ \emph
  {et~al.}(2023{\natexlab{b}})\citenamefont {Zache}, \citenamefont
  {Gonz\'alez-Cuadra},\ and\ \citenamefont {Zoller}}]{Zache:2023cfj}%
  \BibitemOpen
  \bibfield  {author} {\bibinfo {author} {\bibfnamefont {T.~V.}\ \bibnamefont
  {Zache}}, \bibinfo {author} {\bibfnamefont {D.}~\bibnamefont
  {Gonz\'alez-Cuadra}},\ and\ \bibinfo {author} {\bibfnamefont
  {P.}~\bibnamefont {Zoller}},\ }\bibfield  {title} {\bibinfo {title}
  {{Fermion-qudit quantum processors for simulating lattice gauge theories with
  matter}},\ }\href {https://doi.org/10.22331/q-2023-10-16-1140} {\bibfield
  {journal} {\bibinfo  {journal} {Quantum}\ }\textbf {\bibinfo {volume} {7}},\
  \bibinfo {pages} {1140} (\bibinfo {year} {2023}{\natexlab{b}})},\ \Eprint
  {https://arxiv.org/abs/2303.08683} {arXiv:2303.08683 [quant-ph]} \BibitemShut
  {NoStop}%
\bibitem [{\citenamefont {Zohar}\ \emph {et~al.}(2012)\citenamefont {Zohar},
  \citenamefont {Cirac},\ and\ \citenamefont {Reznik}}]{Zohar:2012ay}%
  \BibitemOpen
  \bibfield  {author} {\bibinfo {author} {\bibfnamefont {E.}~\bibnamefont
  {Zohar}}, \bibinfo {author} {\bibfnamefont {J.~I.}\ \bibnamefont {Cirac}},\
  and\ \bibinfo {author} {\bibfnamefont {B.}~\bibnamefont {Reznik}},\
  }\bibfield  {title} {\bibinfo {title} {{Simulating Compact Quantum
  Electrodynamics with ultracold atoms: Probing confinement and nonperturbative
  effects}},\ }\href {https://doi.org/10.1103/PhysRevLett.109.125302}
  {\bibfield  {journal} {\bibinfo  {journal} {Phys. Rev. Lett.}\ }\textbf
  {\bibinfo {volume} {109}},\ \bibinfo {pages} {125302} (\bibinfo {year}
  {2012})},\ \Eprint {https://arxiv.org/abs/1204.6574} {arXiv:1204.6574
  [quant-ph]} \BibitemShut {NoStop}%
\bibitem [{\citenamefont {Zohar}\ \emph
  {et~al.}(2013{\natexlab{a}})\citenamefont {Zohar}, \citenamefont {Cirac},\
  and\ \citenamefont {Reznik}}]{Zohar:2012xf}%
  \BibitemOpen
  \bibfield  {author} {\bibinfo {author} {\bibfnamefont {E.}~\bibnamefont
  {Zohar}}, \bibinfo {author} {\bibfnamefont {J.~I.}\ \bibnamefont {Cirac}},\
  and\ \bibinfo {author} {\bibfnamefont {B.}~\bibnamefont {Reznik}},\
  }\bibfield  {title} {\bibinfo {title} {{Cold-Atom Quantum Simulator for SU(2)
  Yang-Mills Lattice Gauge Theory}},\ }\href
  {https://doi.org/10.1103/PhysRevLett.110.125304} {\bibfield  {journal}
  {\bibinfo  {journal} {Phys. Rev. Lett.}\ }\textbf {\bibinfo {volume} {110}},\
  \bibinfo {pages} {125304} (\bibinfo {year} {2013}{\natexlab{a}})},\ \Eprint
  {https://arxiv.org/abs/1211.2241} {arXiv:1211.2241 [quant-ph]} \BibitemShut
  {NoStop}%
\bibitem [{\citenamefont {Zohar}\ \emph
  {et~al.}(2013{\natexlab{b}})\citenamefont {Zohar}, \citenamefont {Cirac},\
  and\ \citenamefont {Reznik}}]{Zohar:2013zla}%
  \BibitemOpen
  \bibfield  {author} {\bibinfo {author} {\bibfnamefont {E.}~\bibnamefont
  {Zohar}}, \bibinfo {author} {\bibfnamefont {J.~I.}\ \bibnamefont {Cirac}},\
  and\ \bibinfo {author} {\bibfnamefont {B.}~\bibnamefont {Reznik}},\
  }\bibfield  {title} {\bibinfo {title} {{Quantum simulations of gauge theories
  with ultracold atoms: local gauge invariance from angular momentum
  conservation}},\ }\href {https://doi.org/10.1103/PhysRevA.88.023617}
  {\bibfield  {journal} {\bibinfo  {journal} {Phys. Rev.}\ }\textbf {\bibinfo
  {volume} {A88}},\ \bibinfo {pages} {023617} (\bibinfo {year}
  {2013}{\natexlab{b}})},\ \Eprint {https://arxiv.org/abs/1303.5040}
  {arXiv:1303.5040 [quant-ph]} \BibitemShut {NoStop}%
\bibitem [{\citenamefont {Zohar}\ and\ \citenamefont
  {Burrello}(2015)}]{Zohar:2014qma}%
  \BibitemOpen
  \bibfield  {author} {\bibinfo {author} {\bibfnamefont {E.}~\bibnamefont
  {Zohar}}\ and\ \bibinfo {author} {\bibfnamefont {M.}~\bibnamefont
  {Burrello}},\ }\bibfield  {title} {\bibinfo {title} {{Formulation of lattice
  gauge theories for quantum simulations}},\ }\href
  {https://doi.org/10.1103/PhysRevD.91.054506} {\bibfield  {journal} {\bibinfo
  {journal} {Phys. Rev.}\ }\textbf {\bibinfo {volume} {D91}},\ \bibinfo {pages}
  {054506} (\bibinfo {year} {2015})},\ \Eprint
  {https://arxiv.org/abs/1409.3085} {arXiv:1409.3085 [quant-ph]} \BibitemShut
  {NoStop}%
\bibitem [{\citenamefont {Zohar}\ \emph {et~al.}(2016)\citenamefont {Zohar},
  \citenamefont {Cirac},\ and\ \citenamefont {Reznik}}]{Zohar:2015hwa}%
  \BibitemOpen
  \bibfield  {author} {\bibinfo {author} {\bibfnamefont {E.}~\bibnamefont
  {Zohar}}, \bibinfo {author} {\bibfnamefont {J.~I.}\ \bibnamefont {Cirac}},\
  and\ \bibinfo {author} {\bibfnamefont {B.}~\bibnamefont {Reznik}},\
  }\bibfield  {title} {\bibinfo {title} {{Quantum Simulations of Lattice Gauge
  Theories using Ultracold Atoms in Optical Lattices}},\ }\href
  {https://doi.org/10.1088/0034-4885/79/1/014401} {\bibfield  {journal}
  {\bibinfo  {journal} {Rept. Prog. Phys.}\ }\textbf {\bibinfo {volume} {79}},\
  \bibinfo {pages} {014401} (\bibinfo {year} {2016})},\ \Eprint
  {https://arxiv.org/abs/1503.02312} {arXiv:1503.02312 [quant-ph]} \BibitemShut
  {NoStop}%
\bibitem [{\citenamefont {Zohar}\ \emph {et~al.}(2017)\citenamefont {Zohar},
  \citenamefont {Farace}, \citenamefont {Reznik},\ and\ \citenamefont
  {Cirac}}]{Zohar:2016iic}%
  \BibitemOpen
  \bibfield  {author} {\bibinfo {author} {\bibfnamefont {E.}~\bibnamefont
  {Zohar}}, \bibinfo {author} {\bibfnamefont {A.}~\bibnamefont {Farace}},
  \bibinfo {author} {\bibfnamefont {B.}~\bibnamefont {Reznik}},\ and\ \bibinfo
  {author} {\bibfnamefont {J.~I.}\ \bibnamefont {Cirac}},\ }\bibfield  {title}
  {\bibinfo {title} {{Digital lattice gauge theories}},\ }\href
  {https://doi.org/10.1103/PhysRevA.95.023604} {\bibfield  {journal} {\bibinfo
  {journal} {Phys. Rev.}\ }\textbf {\bibinfo {volume} {A95}},\ \bibinfo {pages}
  {023604} (\bibinfo {year} {2017})},\ \Eprint
  {https://arxiv.org/abs/1607.08121} {arXiv:1607.08121 [quant-ph]} \BibitemShut
  {NoStop}%
\bibitem [{\citenamefont {Klco}\ \emph {et~al.}(2020)\citenamefont {Klco},
  \citenamefont {Stryker},\ and\ \citenamefont {Savage}}]{Klco:2019evd}%
  \BibitemOpen
  \bibfield  {author} {\bibinfo {author} {\bibfnamefont {N.}~\bibnamefont
  {Klco}}, \bibinfo {author} {\bibfnamefont {J.~R.}\ \bibnamefont {Stryker}},\
  and\ \bibinfo {author} {\bibfnamefont {M.~J.}\ \bibnamefont {Savage}},\
  }\bibfield  {title} {\bibinfo {title} {{SU(2) non-Abelian gauge field theory
  in one dimension on digital quantum computers}},\ }\href
  {https://doi.org/10.1103/PhysRevD.101.074512} {\bibfield  {journal} {\bibinfo
   {journal} {Phys. Rev. D}\ }\textbf {\bibinfo {volume} {101}},\ \bibinfo
  {pages} {074512} (\bibinfo {year} {2020})},\ \Eprint
  {https://arxiv.org/abs/1908.06935} {arXiv:1908.06935 [quant-ph]} \BibitemShut
  {NoStop}%
\bibitem [{\citenamefont {Ciavarella}\ \emph {et~al.}(2021)\citenamefont
  {Ciavarella}, \citenamefont {Klco},\ and\ \citenamefont
  {Savage}}]{Ciavarella:2021nmj}%
  \BibitemOpen
  \bibfield  {author} {\bibinfo {author} {\bibfnamefont {A.}~\bibnamefont
  {Ciavarella}}, \bibinfo {author} {\bibfnamefont {N.}~\bibnamefont {Klco}},\
  and\ \bibinfo {author} {\bibfnamefont {M.~J.}\ \bibnamefont {Savage}},\
  }\href@noop {} {\bibinfo {title} {{A Trailhead for Quantum Simulation of
  SU(3) Yang-Mills Lattice Gauge Theory in the Local Multiplet Basis}}}
  (\bibinfo {year} {2021}),\ \Eprint {https://arxiv.org/abs/2101.10227}
  {arXiv:2101.10227 [quant-ph]} \BibitemShut {NoStop}%
\bibitem [{\citenamefont {Bender}\ \emph {et~al.}(2018)\citenamefont {Bender},
  \citenamefont {Zohar}, \citenamefont {Farace},\ and\ \citenamefont
  {Cirac}}]{Bender:2018rdp}%
  \BibitemOpen
  \bibfield  {author} {\bibinfo {author} {\bibfnamefont {J.}~\bibnamefont
  {Bender}}, \bibinfo {author} {\bibfnamefont {E.}~\bibnamefont {Zohar}},
  \bibinfo {author} {\bibfnamefont {A.}~\bibnamefont {Farace}},\ and\ \bibinfo
  {author} {\bibfnamefont {J.~I.}\ \bibnamefont {Cirac}},\ }\bibfield  {title}
  {\bibinfo {title} {{Digital quantum simulation of lattice gauge theories in
  three spatial dimensions}},\ }\href
  {https://doi.org/10.1088/1367-2630/aadb71} {\bibfield  {journal} {\bibinfo
  {journal} {New J. Phys.}\ }\textbf {\bibinfo {volume} {20}},\ \bibinfo
  {pages} {093001} (\bibinfo {year} {2018})},\ \Eprint
  {https://arxiv.org/abs/1804.02082} {arXiv:1804.02082 [quant-ph]} \BibitemShut
  {NoStop}%
\bibitem [{\citenamefont {Liu}\ and\ \citenamefont {Xin}(2020)}]{Liu:2020eoa}%
  \BibitemOpen
  \bibfield  {author} {\bibinfo {author} {\bibfnamefont {J.}~\bibnamefont
  {Liu}}\ and\ \bibinfo {author} {\bibfnamefont {Y.}~\bibnamefont {Xin}},\
  }\href@noop {} {\bibinfo {title} {{Quantum simulation of quantum field
  theories as quantum chemistry}}} (\bibinfo {year} {2020}),\ \Eprint
  {https://arxiv.org/abs/2004.13234} {arXiv:2004.13234 [hep-th]} \BibitemShut
  {NoStop}%
\bibitem [{\citenamefont {Hackett}\ \emph {et~al.}(2019)\citenamefont
  {Hackett}, \citenamefont {Howe}, \citenamefont {Hughes}, \citenamefont {Jay},
  \citenamefont {Neil},\ and\ \citenamefont {Simone}}]{Hackett:2018cel}%
  \BibitemOpen
  \bibfield  {author} {\bibinfo {author} {\bibfnamefont {D.~C.}\ \bibnamefont
  {Hackett}}, \bibinfo {author} {\bibfnamefont {K.}~\bibnamefont {Howe}},
  \bibinfo {author} {\bibfnamefont {C.}~\bibnamefont {Hughes}}, \bibinfo
  {author} {\bibfnamefont {W.}~\bibnamefont {Jay}}, \bibinfo {author}
  {\bibfnamefont {E.~T.}\ \bibnamefont {Neil}},\ and\ \bibinfo {author}
  {\bibfnamefont {J.~N.}\ \bibnamefont {Simone}},\ }\bibfield  {title}
  {\bibinfo {title} {{Digitizing Gauge Fields: Lattice Monte Carlo Results for
  Future Quantum Computers}},\ }\href
  {https://doi.org/10.1103/PhysRevA.99.062341} {\bibfield  {journal} {\bibinfo
  {journal} {Phys.\ Rev.\ A}\ }\textbf {\bibinfo {volume} {99}},\ \bibinfo
  {pages} {062341} (\bibinfo {year} {2019})},\ \Eprint
  {https://arxiv.org/abs/1811.03629} {arXiv:1811.03629 [quant-ph]} \BibitemShut
  {NoStop}%
\bibitem [{\citenamefont {Alexandru}\ \emph {et~al.}(2019)\citenamefont
  {Alexandru}, \citenamefont {Bedaque}, \citenamefont {Harmalkar},
  \citenamefont {Lamm}, \citenamefont {Lawrence},\ and\ \citenamefont
  {Warrington}}]{Alexandru:2019nsa}%
  \BibitemOpen
  \bibfield  {author} {\bibinfo {author} {\bibfnamefont {A.}~\bibnamefont
  {Alexandru}}, \bibinfo {author} {\bibfnamefont {P.~F.}\ \bibnamefont
  {Bedaque}}, \bibinfo {author} {\bibfnamefont {S.}~\bibnamefont {Harmalkar}},
  \bibinfo {author} {\bibfnamefont {H.}~\bibnamefont {Lamm}}, \bibinfo {author}
  {\bibfnamefont {S.}~\bibnamefont {Lawrence}},\ and\ \bibinfo {author}
  {\bibfnamefont {N.~C.}\ \bibnamefont {Warrington}} (\bibinfo {collaboration}
  {NuQS}),\ }\bibfield  {title} {\bibinfo {title} {Gluon field digitization for
  quantum computers},\ }\href {https://doi.org/10.1103/PhysRevD.100.114501}
  {\bibfield  {journal} {\bibinfo  {journal} {Phys.Rev.D}\ }\textbf {\bibinfo
  {volume} {100}},\ \bibinfo {pages} {114501} (\bibinfo {year} {2019})},\
  \Eprint {https://arxiv.org/abs/1906.11213} {arXiv:1906.11213 [hep-lat]}
  \BibitemShut {NoStop}%
\bibitem [{\citenamefont {Yamamoto}(2021{\natexlab{b}})}]{Yamamoto:2020eqi}%
  \BibitemOpen
  \bibfield  {author} {\bibinfo {author} {\bibfnamefont {A.}~\bibnamefont
  {Yamamoto}},\ }\bibfield  {title} {\bibinfo {title} {{Real-time simulation of
  (2+1)-dimensional lattice gauge theory on qubits}},\ }\href
  {https://doi.org/10.1093/ptep/ptaa171} {\bibfield  {journal} {\bibinfo
  {journal} {PTEP}\ }\textbf {\bibinfo {volume} {2021}},\ \bibinfo {pages}
  {013B06} (\bibinfo {year} {2021}{\natexlab{b}})},\ \Eprint
  {https://arxiv.org/abs/2008.11395} {arXiv:2008.11395 [hep-lat]} \BibitemShut
  {NoStop}%
\bibitem [{\citenamefont {Haase}\ \emph {et~al.}(2021)\citenamefont {Haase},
  \citenamefont {Dellantonio}, \citenamefont {Celi}, \citenamefont {Paulson},
  \citenamefont {Kan}, \citenamefont {Jansen},\ and\ \citenamefont
  {Muschik}}]{Haase:2020kaj}%
  \BibitemOpen
  \bibfield  {author} {\bibinfo {author} {\bibfnamefont {J.~F.}\ \bibnamefont
  {Haase}}, \bibinfo {author} {\bibfnamefont {L.}~\bibnamefont {Dellantonio}},
  \bibinfo {author} {\bibfnamefont {A.}~\bibnamefont {Celi}}, \bibinfo {author}
  {\bibfnamefont {D.}~\bibnamefont {Paulson}}, \bibinfo {author} {\bibfnamefont
  {A.}~\bibnamefont {Kan}}, \bibinfo {author} {\bibfnamefont {K.}~\bibnamefont
  {Jansen}},\ and\ \bibinfo {author} {\bibfnamefont {C.~A.}\ \bibnamefont
  {Muschik}},\ }\bibfield  {title} {\bibinfo {title} {{A resource efficient
  approach for quantum and classical simulations of gauge theories in particle
  physics}},\ }\href {https://doi.org/10.22331/q-2021-02-04-393} {\bibfield
  {journal} {\bibinfo  {journal} {Quantum}\ }\textbf {\bibinfo {volume} {5}},\
  \bibinfo {pages} {393} (\bibinfo {year} {2021})},\ \Eprint
  {https://arxiv.org/abs/2006.14160} {arXiv:2006.14160 [quant-ph]} \BibitemShut
  {NoStop}%
\bibitem [{\citenamefont {Armon}\ \emph {et~al.}(2021)\citenamefont {Armon},
  \citenamefont {Ashkenazi}, \citenamefont {Garc\'\i{}a-Moreno}, \citenamefont
  {Gonz\'alez-Tudela},\ and\ \citenamefont {Zohar}}]{Armon:2021uqr}%
  \BibitemOpen
  \bibfield  {author} {\bibinfo {author} {\bibfnamefont {T.}~\bibnamefont
  {Armon}}, \bibinfo {author} {\bibfnamefont {S.}~\bibnamefont {Ashkenazi}},
  \bibinfo {author} {\bibfnamefont {G.}~\bibnamefont {Garc\'\i{}a-Moreno}},
  \bibinfo {author} {\bibfnamefont {A.}~\bibnamefont {Gonz\'alez-Tudela}},\
  and\ \bibinfo {author} {\bibfnamefont {E.}~\bibnamefont {Zohar}},\
  }\href@noop {} {\bibinfo {title} {{Photon-mediated Stroboscopic Quantum
  Simulation of a $\mathbb{Z}_{2}$ Lattice Gauge Theory}}} (\bibinfo {year}
  {2021}),\ \Eprint {https://arxiv.org/abs/2107.13024} {arXiv:2107.13024
  [quant-ph]} \BibitemShut {NoStop}%
\bibitem [{\citenamefont {Bazavov}\ \emph {et~al.}(2019)\citenamefont
  {Bazavov}, \citenamefont {Catterall}, \citenamefont {Jha},\ and\
  \citenamefont {Unmuth-Yockey}}]{PhysRevD.99.114507}%
  \BibitemOpen
  \bibfield  {author} {\bibinfo {author} {\bibfnamefont {A.}~\bibnamefont
  {Bazavov}}, \bibinfo {author} {\bibfnamefont {S.}~\bibnamefont {Catterall}},
  \bibinfo {author} {\bibfnamefont {R.~G.}\ \bibnamefont {Jha}},\ and\ \bibinfo
  {author} {\bibfnamefont {J.}~\bibnamefont {Unmuth-Yockey}},\ }\bibfield
  {title} {\bibinfo {title} {Tensor renormalization group study of the
  non-abelian higgs model in two dimensions},\ }\href
  {https://doi.org/10.1103/PhysRevD.99.114507} {\bibfield  {journal} {\bibinfo
  {journal} {Phys. Rev. D}\ }\textbf {\bibinfo {volume} {99}},\ \bibinfo
  {pages} {114507} (\bibinfo {year} {2019})}\BibitemShut {NoStop}%
\bibitem [{\citenamefont {Bazavov}\ \emph {et~al.}(2015)\citenamefont
  {Bazavov}, \citenamefont {Meurice}, \citenamefont {Tsai}, \citenamefont
  {Unmuth-Yockey},\ and\ \citenamefont {Zhang}}]{Bazavov:2015kka}%
  \BibitemOpen
  \bibfield  {author} {\bibinfo {author} {\bibfnamefont {A.}~\bibnamefont
  {Bazavov}}, \bibinfo {author} {\bibfnamefont {Y.}~\bibnamefont {Meurice}},
  \bibinfo {author} {\bibfnamefont {S.-W.}\ \bibnamefont {Tsai}}, \bibinfo
  {author} {\bibfnamefont {J.}~\bibnamefont {Unmuth-Yockey}},\ and\ \bibinfo
  {author} {\bibfnamefont {J.}~\bibnamefont {Zhang}},\ }\bibfield  {title}
  {\bibinfo {title} {{Gauge-invariant implementation of the Abelian Higgs model
  on optical lattices}},\ }\href {https://doi.org/10.1103/PhysRevD.92.076003}
  {\bibfield  {journal} {\bibinfo  {journal} {Phys. Rev.}\ }\textbf {\bibinfo
  {volume} {D92}},\ \bibinfo {pages} {076003} (\bibinfo {year} {2015})},\
  \Eprint {https://arxiv.org/abs/1503.08354} {arXiv:1503.08354 [hep-lat]}
  \BibitemShut {NoStop}%
\bibitem [{\citenamefont {Zhang}\ \emph {et~al.}(2018)\citenamefont {Zhang},
  \citenamefont {Unmuth-Yockey}, \citenamefont {Zeiher}, \citenamefont
  {Bazavov}, \citenamefont {Tsai},\ and\ \citenamefont
  {Meurice}}]{Zhang:2018ufj}%
  \BibitemOpen
  \bibfield  {author} {\bibinfo {author} {\bibfnamefont {J.}~\bibnamefont
  {Zhang}}, \bibinfo {author} {\bibfnamefont {J.}~\bibnamefont
  {Unmuth-Yockey}}, \bibinfo {author} {\bibfnamefont {J.}~\bibnamefont
  {Zeiher}}, \bibinfo {author} {\bibfnamefont {A.}~\bibnamefont {Bazavov}},
  \bibinfo {author} {\bibfnamefont {S.~W.}\ \bibnamefont {Tsai}},\ and\
  \bibinfo {author} {\bibfnamefont {Y.}~\bibnamefont {Meurice}},\ }\bibfield
  {title} {\bibinfo {title} {{Quantum simulation of the universal features of
  the Polyakov loop}},\ }\href {https://doi.org/10.1103/PhysRevLett.121.223201}
  {\bibfield  {journal} {\bibinfo  {journal} {Phys. Rev. Lett.}\ }\textbf
  {\bibinfo {volume} {121}},\ \bibinfo {pages} {223201} (\bibinfo {year}
  {2018})},\ \Eprint {https://arxiv.org/abs/1803.11166} {arXiv:1803.11166
  [hep-lat]} \BibitemShut {NoStop}%
\bibitem [{\citenamefont {Unmuth-Yockey}\ \emph {et~al.}(2018)\citenamefont
  {Unmuth-Yockey}, \citenamefont {Zhang}, \citenamefont {Bazavov},
  \citenamefont {Meurice},\ and\ \citenamefont {Tsai}}]{Unmuth-Yockey:2018ugm}%
  \BibitemOpen
  \bibfield  {author} {\bibinfo {author} {\bibfnamefont {J.}~\bibnamefont
  {Unmuth-Yockey}}, \bibinfo {author} {\bibfnamefont {J.}~\bibnamefont
  {Zhang}}, \bibinfo {author} {\bibfnamefont {A.}~\bibnamefont {Bazavov}},
  \bibinfo {author} {\bibfnamefont {Y.}~\bibnamefont {Meurice}},\ and\ \bibinfo
  {author} {\bibfnamefont {S.-W.}\ \bibnamefont {Tsai}},\ }\bibfield  {title}
  {\bibinfo {title} {{Universal features of the Abelian Polyakov loop in 1+1
  dimensions}},\ }\href {https://doi.org/10.1103/PhysRevD.98.094511} {\bibfield
   {journal} {\bibinfo  {journal} {Phys. Rev.}\ }\textbf {\bibinfo {volume}
  {D98}},\ \bibinfo {pages} {094511} (\bibinfo {year} {2018})},\ \Eprint
  {https://arxiv.org/abs/1807.09186} {arXiv:1807.09186 [hep-lat]} \BibitemShut
  {NoStop}%
\bibitem [{\citenamefont {Unmuth-Yockey}(2019)}]{Unmuth-Yockey:2018xak}%
  \BibitemOpen
  \bibfield  {author} {\bibinfo {author} {\bibfnamefont {J.~F.}\ \bibnamefont
  {Unmuth-Yockey}},\ }\bibfield  {title} {\bibinfo {title} {{Gauge-invariant
  rotor Hamiltonian from dual variables of 3D $U(1)$ gauge theory}},\ }\href
  {https://doi.org/10.1103/PhysRevD.99.074502} {\bibfield  {journal} {\bibinfo
  {journal} {Phys.\ Rev.\ D}\ }\textbf {\bibinfo {volume} {99}},\ \bibinfo
  {pages} {074502} (\bibinfo {year} {2019})},\ \Eprint
  {https://arxiv.org/abs/1811.05884} {arXiv:1811.05884 [hep-lat]} \BibitemShut
  {NoStop}%
\bibitem [{\citenamefont {Kreshchuk}\ \emph
  {et~al.}(2020{\natexlab{a}})\citenamefont {Kreshchuk}, \citenamefont {Kirby},
  \citenamefont {Goldstein}, \citenamefont {Beauchemin},\ and\ \citenamefont
  {Love}}]{Kreshchuk:2020dla}%
  \BibitemOpen
  \bibfield  {author} {\bibinfo {author} {\bibfnamefont {M.}~\bibnamefont
  {Kreshchuk}}, \bibinfo {author} {\bibfnamefont {W.~M.}\ \bibnamefont
  {Kirby}}, \bibinfo {author} {\bibfnamefont {G.}~\bibnamefont {Goldstein}},
  \bibinfo {author} {\bibfnamefont {H.}~\bibnamefont {Beauchemin}},\ and\
  \bibinfo {author} {\bibfnamefont {P.~J.}\ \bibnamefont {Love}},\ }\href@noop
  {} {\bibinfo {title} {{Quantum Simulation of Quantum Field Theory in the
  Light-Front Formulation}}} (\bibinfo {year} {2020}{\natexlab{a}}),\ \Eprint
  {https://arxiv.org/abs/2002.04016} {arXiv:2002.04016 [quant-ph]} \BibitemShut
  {NoStop}%
\bibitem [{\citenamefont {Kreshchuk}\ \emph
  {et~al.}(2020{\natexlab{b}})\citenamefont {Kreshchuk}, \citenamefont {Jia},
  \citenamefont {Kirby}, \citenamefont {Goldstein}, \citenamefont {Vary},\ and\
  \citenamefont {Love}}]{Kreshchuk:2020aiq}%
  \BibitemOpen
  \bibfield  {author} {\bibinfo {author} {\bibfnamefont {M.}~\bibnamefont
  {Kreshchuk}}, \bibinfo {author} {\bibfnamefont {S.}~\bibnamefont {Jia}},
  \bibinfo {author} {\bibfnamefont {W.~M.}\ \bibnamefont {Kirby}}, \bibinfo
  {author} {\bibfnamefont {G.}~\bibnamefont {Goldstein}}, \bibinfo {author}
  {\bibfnamefont {J.~P.}\ \bibnamefont {Vary}},\ and\ \bibinfo {author}
  {\bibfnamefont {P.~J.}\ \bibnamefont {Love}},\ }\href@noop {} {\bibinfo
  {title} {{Simulating Hadronic Physics on NISQ devices using Basis Light-Front
  Quantization}}} (\bibinfo {year} {2020}{\natexlab{b}}),\ \Eprint
  {https://arxiv.org/abs/2011.13443} {arXiv:2011.13443 [quant-ph]} \BibitemShut
  {NoStop}%
\bibitem [{\citenamefont {Raychowdhury}\ and\ \citenamefont
  {Stryker}(2018)}]{Raychowdhury:2018osk}%
  \BibitemOpen
  \bibfield  {author} {\bibinfo {author} {\bibfnamefont {I.}~\bibnamefont
  {Raychowdhury}}\ and\ \bibinfo {author} {\bibfnamefont {J.~R.}\ \bibnamefont
  {Stryker}},\ }\href@noop {} {\bibinfo {title} {{Solving Gauss's Law on
  Digital Quantum Computers with Loop-String-Hadron Digitization}}} (\bibinfo
  {year} {2018}),\ \Eprint {https://arxiv.org/abs/1812.07554} {arXiv:1812.07554
  [hep-lat]} \BibitemShut {NoStop}%
\bibitem [{\citenamefont {Raychowdhury}\ and\ \citenamefont
  {Stryker}(2020)}]{Raychowdhury:2019iki}%
  \BibitemOpen
  \bibfield  {author} {\bibinfo {author} {\bibfnamefont {I.}~\bibnamefont
  {Raychowdhury}}\ and\ \bibinfo {author} {\bibfnamefont {J.~R.}\ \bibnamefont
  {Stryker}},\ }\bibfield  {title} {\bibinfo {title} {{Loop, String, and Hadron
  Dynamics in SU(2) Hamiltonian Lattice Gauge Theories}},\ }\href
  {https://doi.org/10.1103/PhysRevD.101.114502} {\bibfield  {journal} {\bibinfo
   {journal} {Phys. Rev. D}\ }\textbf {\bibinfo {volume} {101}},\ \bibinfo
  {pages} {114502} (\bibinfo {year} {2020})},\ \Eprint
  {https://arxiv.org/abs/1912.06133} {arXiv:1912.06133 [hep-lat]} \BibitemShut
  {NoStop}%
\bibitem [{\citenamefont {Davoudi}\ \emph {et~al.}(2020)\citenamefont
  {Davoudi}, \citenamefont {Raychowdhury},\ and\ \citenamefont
  {Shaw}}]{Davoudi:2020yln}%
  \BibitemOpen
  \bibfield  {author} {\bibinfo {author} {\bibfnamefont {Z.}~\bibnamefont
  {Davoudi}}, \bibinfo {author} {\bibfnamefont {I.}~\bibnamefont
  {Raychowdhury}},\ and\ \bibinfo {author} {\bibfnamefont {A.}~\bibnamefont
  {Shaw}},\ }\href@noop {} {\bibinfo {title} {{Search for Efficient
  Formulations for Hamiltonian Simulation of non-Abelian Lattice Gauge
  Theories}}} (\bibinfo {year} {2020}),\ \Eprint
  {https://arxiv.org/abs/2009.11802} {arXiv:2009.11802 [hep-lat]} \BibitemShut
  {NoStop}%
\bibitem [{\citenamefont {Wiese}(2014)}]{Wiese:2014rla}%
  \BibitemOpen
  \bibfield  {author} {\bibinfo {author} {\bibfnamefont {U.-J.}\ \bibnamefont
  {Wiese}},\ }\bibfield  {title} {\bibinfo {title} {{Towards Quantum Simulating
  QCD}},\ }\bibfield  {booktitle} {\emph {\bibinfo {booktitle} {{Proceedings,
  24th International Conference on Ultra-Relativistic Nucleus-Nucleus
  Collisions (Quark Matter 2014): Darmstadt, Germany, May 19-24, 2014}}},\
  }\href {https://doi.org/10.1016/j.nuclphysa.2014.09.102} {\bibfield
  {journal} {\bibinfo  {journal} {Nucl. Phys.}\ }\textbf {\bibinfo {volume}
  {A931}},\ \bibinfo {pages} {246} (\bibinfo {year} {2014})},\ \Eprint
  {https://arxiv.org/abs/1409.7414} {arXiv:1409.7414 [hep-th]} \BibitemShut
  {NoStop}%
\bibitem [{\citenamefont {Luo}\ \emph {et~al.}(2019)\citenamefont {Luo},
  \citenamefont {Shen}, \citenamefont {Highman}, \citenamefont {Clark},
  \citenamefont {DeMarco}, \citenamefont {El-Khadra},\ and\ \citenamefont
  {Gadway}}]{Luo:2019vmi}%
  \BibitemOpen
  \bibfield  {author} {\bibinfo {author} {\bibfnamefont {D.}~\bibnamefont
  {Luo}}, \bibinfo {author} {\bibfnamefont {J.}~\bibnamefont {Shen}}, \bibinfo
  {author} {\bibfnamefont {M.}~\bibnamefont {Highman}}, \bibinfo {author}
  {\bibfnamefont {B.~K.}\ \bibnamefont {Clark}}, \bibinfo {author}
  {\bibfnamefont {B.}~\bibnamefont {DeMarco}}, \bibinfo {author} {\bibfnamefont
  {A.~X.}\ \bibnamefont {El-Khadra}},\ and\ \bibinfo {author} {\bibfnamefont
  {B.}~\bibnamefont {Gadway}},\ }\href@noop {} {\bibinfo {title} {{A Framework
  for Simulating Gauge Theories with Dipolar Spin Systems}}} (\bibinfo {year}
  {2019}),\ \Eprint {https://arxiv.org/abs/1912.11488} {arXiv:1912.11488
  [quant-ph]} \BibitemShut {NoStop}%
\bibitem [{\citenamefont {Brower}\ \emph {et~al.}(2019)\citenamefont {Brower},
  \citenamefont {Berenstein},\ and\ \citenamefont {Kawai}}]{Brower:2020huh}%
  \BibitemOpen
  \bibfield  {author} {\bibinfo {author} {\bibfnamefont {R.~C.}\ \bibnamefont
  {Brower}}, \bibinfo {author} {\bibfnamefont {D.}~\bibnamefont {Berenstein}},\
  and\ \bibinfo {author} {\bibfnamefont {H.}~\bibnamefont {Kawai}},\ }\bibfield
   {title} {\bibinfo {title} {{Lattice Gauge Theory for a Quantum Computer}},\
  }\href@noop {} {\bibfield  {journal} {\bibinfo  {journal} {PoS}\ }\textbf
  {\bibinfo {volume} {LATTICE2019}},\ \bibinfo {pages} {112} (\bibinfo {year}
  {2019})},\ \Eprint {https://arxiv.org/abs/2002.10028} {arXiv:2002.10028
  [hep-lat]} \BibitemShut {NoStop}%
\bibitem [{\citenamefont {Mathis}\ \emph {et~al.}(2020)\citenamefont {Mathis},
  \citenamefont {Mazzola},\ and\ \citenamefont {Tavernelli}}]{Mathis:2020fuo}%
  \BibitemOpen
  \bibfield  {author} {\bibinfo {author} {\bibfnamefont {S.~V.}\ \bibnamefont
  {Mathis}}, \bibinfo {author} {\bibfnamefont {G.}~\bibnamefont {Mazzola}},\
  and\ \bibinfo {author} {\bibfnamefont {I.}~\bibnamefont {Tavernelli}},\
  }\bibfield  {title} {\bibinfo {title} {{Toward scalable simulations of
  Lattice Gauge Theories on quantum computers}},\ }\href
  {https://doi.org/10.1103/PhysRevD.102.094501} {\bibfield  {journal} {\bibinfo
   {journal} {Phys. Rev. D}\ }\textbf {\bibinfo {volume} {102}},\ \bibinfo
  {pages} {094501} (\bibinfo {year} {2020})},\ \Eprint
  {https://arxiv.org/abs/2005.10271} {arXiv:2005.10271 [quant-ph]} \BibitemShut
  {NoStop}%
\bibitem [{\citenamefont {Singh}(2019)}]{Singh:2019jog}%
  \BibitemOpen
  \bibfield  {author} {\bibinfo {author} {\bibfnamefont {H.}~\bibnamefont
  {Singh}},\ }\href@noop {} {\bibinfo {title} {{Qubit $O(N)$ nonlinear sigma
  models}}} (\bibinfo {year} {2019}),\ \Eprint
  {https://arxiv.org/abs/1911.12353} {arXiv:1911.12353 [hep-lat]} \BibitemShut
  {NoStop}%
\bibitem [{\citenamefont {Singh}\ and\ \citenamefont
  {Chandrasekharan}(2019)}]{Singh:2019uwd}%
  \BibitemOpen
  \bibfield  {author} {\bibinfo {author} {\bibfnamefont {H.}~\bibnamefont
  {Singh}}\ and\ \bibinfo {author} {\bibfnamefont {S.}~\bibnamefont
  {Chandrasekharan}},\ }\bibfield  {title} {\bibinfo {title} {{Qubit
  regularization of the $O(3)$ sigma model}},\ }\href
  {https://doi.org/10.1103/PhysRevD.100.054505} {\bibfield  {journal} {\bibinfo
   {journal} {Phys. Rev. D}\ }\textbf {\bibinfo {volume} {100}},\ \bibinfo
  {pages} {054505} (\bibinfo {year} {2019})},\ \Eprint
  {https://arxiv.org/abs/1905.13204} {arXiv:1905.13204 [hep-lat]} \BibitemShut
  {NoStop}%
\bibitem [{\citenamefont {Buser}\ \emph {et~al.}(2020)\citenamefont {Buser},
  \citenamefont {Bhattacharya}, \citenamefont {Cincio},\ and\ \citenamefont
  {Gupta}}]{Buser:2020uzs}%
  \BibitemOpen
  \bibfield  {author} {\bibinfo {author} {\bibfnamefont {A.~J.}\ \bibnamefont
  {Buser}}, \bibinfo {author} {\bibfnamefont {T.}~\bibnamefont {Bhattacharya}},
  \bibinfo {author} {\bibfnamefont {L.}~\bibnamefont {Cincio}},\ and\ \bibinfo
  {author} {\bibfnamefont {R.}~\bibnamefont {Gupta}},\ }\bibfield  {title}
  {\bibinfo {title} {{State preparation and measurement in a quantum simulation
  of the $O$(3) sigma model}},\ }\href
  {https://doi.org/10.1103/PhysRevD.102.114514} {\bibfield  {journal} {\bibinfo
   {journal} {Phys. Rev. D}\ }\textbf {\bibinfo {volume} {102}},\ \bibinfo
  {pages} {114514} (\bibinfo {year} {2020})},\ \Eprint
  {https://arxiv.org/abs/2006.15746} {arXiv:2006.15746 [quant-ph]} \BibitemShut
  {NoStop}%
\bibitem [{\citenamefont {Bhattacharya}\ \emph {et~al.}(2021)\citenamefont
  {Bhattacharya}, \citenamefont {Buser}, \citenamefont {Chandrasekharan},
  \citenamefont {Gupta},\ and\ \citenamefont {Singh}}]{Bhattacharya:2020gpm}%
  \BibitemOpen
  \bibfield  {author} {\bibinfo {author} {\bibfnamefont {T.}~\bibnamefont
  {Bhattacharya}}, \bibinfo {author} {\bibfnamefont {A.~J.}\ \bibnamefont
  {Buser}}, \bibinfo {author} {\bibfnamefont {S.}~\bibnamefont
  {Chandrasekharan}}, \bibinfo {author} {\bibfnamefont {R.}~\bibnamefont
  {Gupta}},\ and\ \bibinfo {author} {\bibfnamefont {H.}~\bibnamefont {Singh}},\
  }\bibfield  {title} {\bibinfo {title} {{Qubit regularization of asymptotic
  freedom}},\ }\href {https://doi.org/10.1103/PhysRevLett.126.172001}
  {\bibfield  {journal} {\bibinfo  {journal} {Phys. Rev. Lett.}\ }\textbf
  {\bibinfo {volume} {126}},\ \bibinfo {pages} {172001} (\bibinfo {year}
  {2021})},\ \Eprint {https://arxiv.org/abs/2012.02153} {arXiv:2012.02153
  [hep-lat]} \BibitemShut {NoStop}%
\bibitem [{\citenamefont {Barata}\ \emph {et~al.}(2020)\citenamefont {Barata},
  \citenamefont {Mueller}, \citenamefont {Tarasov},\ and\ \citenamefont
  {Venugopalan}}]{Barata:2020jtq}%
  \BibitemOpen
  \bibfield  {author} {\bibinfo {author} {\bibfnamefont {J.~a.}\ \bibnamefont
  {Barata}}, \bibinfo {author} {\bibfnamefont {N.}~\bibnamefont {Mueller}},
  \bibinfo {author} {\bibfnamefont {A.}~\bibnamefont {Tarasov}},\ and\ \bibinfo
  {author} {\bibfnamefont {R.}~\bibnamefont {Venugopalan}},\ }\href@noop {}
  {\bibinfo {title} {{Single-particle digitization strategy for quantum
  computation of a $\phi^4$ scalar field theory}}} (\bibinfo {year} {2020}),\
  \Eprint {https://arxiv.org/abs/2012.00020} {arXiv:2012.00020 [hep-th]}
  \BibitemShut {NoStop}%
\bibitem [{\citenamefont {Kreshchuk}\ \emph
  {et~al.}(2020{\natexlab{c}})\citenamefont {Kreshchuk}, \citenamefont {Jia},
  \citenamefont {Kirby}, \citenamefont {Goldstein}, \citenamefont {Vary},\ and\
  \citenamefont {Love}}]{Kreshchuk:2020kcz}%
  \BibitemOpen
  \bibfield  {author} {\bibinfo {author} {\bibfnamefont {M.}~\bibnamefont
  {Kreshchuk}}, \bibinfo {author} {\bibfnamefont {S.}~\bibnamefont {Jia}},
  \bibinfo {author} {\bibfnamefont {W.~M.}\ \bibnamefont {Kirby}}, \bibinfo
  {author} {\bibfnamefont {G.}~\bibnamefont {Goldstein}}, \bibinfo {author}
  {\bibfnamefont {J.~P.}\ \bibnamefont {Vary}},\ and\ \bibinfo {author}
  {\bibfnamefont {P.~J.}\ \bibnamefont {Love}},\ }\href@noop {} {\bibinfo
  {title} {{Light-Front Field Theory on Current Quantum Computers}}} (\bibinfo
  {year} {2020}{\natexlab{c}}),\ \Eprint {https://arxiv.org/abs/2009.07885}
  {arXiv:2009.07885 [quant-ph]} \BibitemShut {NoStop}%
\bibitem [{\citenamefont {Ji}\ \emph {et~al.}(2020)\citenamefont {Ji},
  \citenamefont {Lamm},\ and\ \citenamefont {Zhu}}]{Ji:2020kjk}%
  \BibitemOpen
  \bibfield  {author} {\bibinfo {author} {\bibfnamefont {Y.}~\bibnamefont
  {Ji}}, \bibinfo {author} {\bibfnamefont {H.}~\bibnamefont {Lamm}},\ and\
  \bibinfo {author} {\bibfnamefont {S.}~\bibnamefont {Zhu}} (\bibinfo
  {collaboration} {NuQS}),\ }\bibfield  {title} {\bibinfo {title} {{Gluon Field
  Digitization via Group Space Decimation for Quantum Computers}},\ }\href
  {https://doi.org/10.1103/PhysRevD.102.114513} {\bibfield  {journal} {\bibinfo
   {journal} {Phys. Rev. D}\ }\textbf {\bibinfo {volume} {102}},\ \bibinfo
  {pages} {114513} (\bibinfo {year} {2020})},\ \Eprint
  {https://arxiv.org/abs/2005.14221} {arXiv:2005.14221 [hep-lat]} \BibitemShut
  {NoStop}%
\bibitem [{\citenamefont {Bauer}\ and\ \citenamefont
  {Grabowska}(2021)}]{Bauer:2021gek}%
  \BibitemOpen
  \bibfield  {author} {\bibinfo {author} {\bibfnamefont {C.~W.}\ \bibnamefont
  {Bauer}}\ and\ \bibinfo {author} {\bibfnamefont {D.~M.}\ \bibnamefont
  {Grabowska}},\ }\href@noop {} {\bibinfo {title} {{Efficient Representation
  for Simulating U(1) Gauge Theories on Digital Quantum Computers at All Values
  of the Coupling}}} (\bibinfo {year} {2021}),\ \Eprint
  {https://arxiv.org/abs/2111.08015} {arXiv:2111.08015 [hep-ph]} \BibitemShut
  {NoStop}%
\bibitem [{\citenamefont {Gustafson}(2021)}]{Gustafson:2021qbt}%
  \BibitemOpen
  \bibfield  {author} {\bibinfo {author} {\bibfnamefont {E.}~\bibnamefont
  {Gustafson}},\ }\bibfield  {title} {\bibinfo {title} {{Prospects for
  Simulating a Qudit Based Model of (1+1)d Scalar QED}},\ }\href
  {https://doi.org/10.1103/PhysRevD.103.114505} {\bibfield  {journal} {\bibinfo
   {journal} {Phys. Rev. D}\ }\textbf {\bibinfo {volume} {103}},\ \bibinfo
  {pages} {114505} (\bibinfo {year} {2021})},\ \Eprint
  {https://arxiv.org/abs/2104.10136} {arXiv:2104.10136 [quant-ph]} \BibitemShut
  {NoStop}%
\bibitem [{\citenamefont {Hartung}\ \emph {et~al.}(2022)\citenamefont
  {Hartung}, \citenamefont {Jakobs}, \citenamefont {Jansen}, \citenamefont
  {Ostmeyer},\ and\ \citenamefont {Urbach}}]{Hartung:2022hoz}%
  \BibitemOpen
  \bibfield  {author} {\bibinfo {author} {\bibfnamefont {T.}~\bibnamefont
  {Hartung}}, \bibinfo {author} {\bibfnamefont {T.}~\bibnamefont {Jakobs}},
  \bibinfo {author} {\bibfnamefont {K.}~\bibnamefont {Jansen}}, \bibinfo
  {author} {\bibfnamefont {J.}~\bibnamefont {Ostmeyer}},\ and\ \bibinfo
  {author} {\bibfnamefont {C.}~\bibnamefont {Urbach}},\ }\bibfield  {title}
  {\bibinfo {title} {{Digitising SU(2) gauge fields and the freezing
  transition}},\ }\href {https://doi.org/10.1140/epjc/s10052-022-10192-5}
  {\bibfield  {journal} {\bibinfo  {journal} {Eur. Phys. J. C}\ }\textbf
  {\bibinfo {volume} {82}},\ \bibinfo {pages} {237} (\bibinfo {year} {2022})},\
  \Eprint {https://arxiv.org/abs/2201.09625} {arXiv:2201.09625 [hep-lat]}
  \BibitemShut {NoStop}%
\bibitem [{\citenamefont {Grabowska}\ \emph {et~al.}(2022)\citenamefont
  {Grabowska}, \citenamefont {Kane}, \citenamefont {Nachman},\ and\
  \citenamefont {Bauer}}]{Grabowska:2022uos}%
  \BibitemOpen
  \bibfield  {author} {\bibinfo {author} {\bibfnamefont {D.~M.}\ \bibnamefont
  {Grabowska}}, \bibinfo {author} {\bibfnamefont {C.}~\bibnamefont {Kane}},
  \bibinfo {author} {\bibfnamefont {B.}~\bibnamefont {Nachman}},\ and\ \bibinfo
  {author} {\bibfnamefont {C.~W.}\ \bibnamefont {Bauer}},\ }\href@noop {}
  {\bibinfo {title} {{Overcoming exponential scaling with system size in
  Trotter-Suzuki implementations of constrained Hamiltonians: 2+1 U(1) lattice
  gauge theories}}} (\bibinfo {year} {2022}),\ \Eprint
  {https://arxiv.org/abs/2208.03333} {arXiv:2208.03333 [quant-ph]} \BibitemShut
  {NoStop}%
\bibitem [{\citenamefont {Murairi}\ \emph {et~al.}(2022)\citenamefont
  {Murairi}, \citenamefont {Cervia}, \citenamefont {Kumar}, \citenamefont
  {Bedaque},\ and\ \citenamefont {Alexandru}}]{Murairi:2022zdg}%
  \BibitemOpen
  \bibfield  {author} {\bibinfo {author} {\bibfnamefont {E.~M.}\ \bibnamefont
  {Murairi}}, \bibinfo {author} {\bibfnamefont {M.~J.}\ \bibnamefont {Cervia}},
  \bibinfo {author} {\bibfnamefont {H.}~\bibnamefont {Kumar}}, \bibinfo
  {author} {\bibfnamefont {P.~F.}\ \bibnamefont {Bedaque}},\ and\ \bibinfo
  {author} {\bibfnamefont {A.}~\bibnamefont {Alexandru}},\ }\href@noop {}
  {\bibinfo {title} {{How many quantum gates do gauge theories require?}}}
  (\bibinfo {year} {2022}),\ \Eprint {https://arxiv.org/abs/2208.11789}
  {arXiv:2208.11789 [hep-lat]} \BibitemShut {NoStop}%
\bibitem [{\citenamefont {Hasenfratz}\ and\ \citenamefont
  {Niedermayer}(2001{\natexlab{a}})}]{Hasenfratz:2001iz}%
  \BibitemOpen
  \bibfield  {author} {\bibinfo {author} {\bibfnamefont {P.}~\bibnamefont
  {Hasenfratz}}\ and\ \bibinfo {author} {\bibfnamefont {F.}~\bibnamefont
  {Niedermayer}},\ }\bibfield  {title} {\bibinfo {title} {{Asymptotic freedom
  with discrete spin variables?}},\ }\bibfield  {booktitle} {\emph {\bibinfo
  {booktitle} {{Proceedings, 2001 Europhysics Conference on High Energy Physics
  (EPS-HEP 2001): Budapest, Hungary, July 12-18, 2001}}},\ }\href
  {https://doi.org/10.22323/1.007.0229} {\bibfield  {journal} {\bibinfo
  {journal} {PoS}\ }\textbf {\bibinfo {volume} {HEP2001}},\ \bibinfo {pages}
  {229} (\bibinfo {year} {2001}{\natexlab{a}})},\ \Eprint
  {https://arxiv.org/abs/hep-lat/0112003} {arXiv:hep-lat/0112003 [hep-lat]}
  \BibitemShut {NoStop}%
\bibitem [{\citenamefont {Caracciolo}\ \emph
  {et~al.}(2001{\natexlab{a}})\citenamefont {Caracciolo}, \citenamefont
  {Montanari},\ and\ \citenamefont {Pelissetto}}]{Caracciolo:2001jd}%
  \BibitemOpen
  \bibfield  {author} {\bibinfo {author} {\bibfnamefont {S.}~\bibnamefont
  {Caracciolo}}, \bibinfo {author} {\bibfnamefont {A.}~\bibnamefont
  {Montanari}},\ and\ \bibinfo {author} {\bibfnamefont {A.}~\bibnamefont
  {Pelissetto}},\ }\bibfield  {title} {\bibinfo {title} {{Asymptotically free
  models and discrete nonAbelian groups}},\ }\href
  {https://doi.org/10.1016/S0370-2693(01)00674-8} {\bibfield  {journal}
  {\bibinfo  {journal} {Phys. Lett.}\ }\textbf {\bibinfo {volume} {B513}},\
  \bibinfo {pages} {223} (\bibinfo {year} {2001}{\natexlab{a}})},\ \Eprint
  {https://arxiv.org/abs/hep-lat/0103017} {arXiv:hep-lat/0103017 [hep-lat]}
  \BibitemShut {NoStop}%
\bibitem [{\citenamefont {Hasenfratz}\ and\ \citenamefont
  {Niedermayer}(2001{\natexlab{b}})}]{Hasenfratz:2000hd}%
  \BibitemOpen
  \bibfield  {author} {\bibinfo {author} {\bibfnamefont {P.}~\bibnamefont
  {Hasenfratz}}\ and\ \bibinfo {author} {\bibfnamefont {F.}~\bibnamefont
  {Niedermayer}},\ }\bibfield  {title} {\bibinfo {title} {{Asymptotically free
  theories based on discrete subgroups}},\ }\bibfield  {booktitle} {\emph
  {\bibinfo {booktitle} {{Lattice field theory. Proceedings, 18th International
  Symposium, Lattice 2000, Bangalore, India, August 17-22, 2000}}},\ }\href
  {https://doi.org/10.1016/S0920-5632(01)00870-2} {\bibfield  {journal}
  {\bibinfo  {journal} {Nucl. Phys. Proc. Suppl.}\ }\textbf {\bibinfo {volume}
  {94}},\ \bibinfo {pages} {575} (\bibinfo {year} {2001}{\natexlab{b}})},\
  \Eprint {https://arxiv.org/abs/hep-lat/0011056} {arXiv:hep-lat/0011056
  [hep-lat]} \BibitemShut {NoStop}%
\bibitem [{\citenamefont {Patrascioiu}\ and\ \citenamefont
  {Seiler}(1998)}]{PhysRevE.57.111}%
  \BibitemOpen
  \bibfield  {author} {\bibinfo {author} {\bibfnamefont {A.}~\bibnamefont
  {Patrascioiu}}\ and\ \bibinfo {author} {\bibfnamefont {E.}~\bibnamefont
  {Seiler}},\ }\bibfield  {title} {\bibinfo {title} {Continuum limit of
  two-dimensional spin models with continuous symmetry and conformal quantum
  field theory},\ }\href {https://doi.org/10.1103/PhysRevE.57.111} {\bibfield
  {journal} {\bibinfo  {journal} {Phys. Rev. E}\ }\textbf {\bibinfo {volume}
  {57}},\ \bibinfo {pages} {111} (\bibinfo {year} {1998})}\BibitemShut
  {NoStop}%
\bibitem [{\citenamefont {Krcmar}\ \emph {et~al.}(2016)\citenamefont {Krcmar},
  \citenamefont {Gendiar},\ and\ \citenamefont {Nishino}}]{PhysRevE.94.022134}%
  \BibitemOpen
  \bibfield  {author} {\bibinfo {author} {\bibfnamefont {R.}~\bibnamefont
  {Krcmar}}, \bibinfo {author} {\bibfnamefont {A.}~\bibnamefont {Gendiar}},\
  and\ \bibinfo {author} {\bibfnamefont {T.}~\bibnamefont {Nishino}},\
  }\bibfield  {title} {\bibinfo {title} {Phase diagram of a truncated
  tetrahedral model},\ }\href {https://doi.org/10.1103/PhysRevE.94.022134}
  {\bibfield  {journal} {\bibinfo  {journal} {Phys. Rev. E}\ }\textbf {\bibinfo
  {volume} {94}},\ \bibinfo {pages} {022134} (\bibinfo {year}
  {2016})}\BibitemShut {NoStop}%
\bibitem [{\citenamefont {Caracciolo}\ \emph
  {et~al.}(2001{\natexlab{b}})\citenamefont {Caracciolo}, \citenamefont
  {Montanari},\ and\ \citenamefont {Pelissetto}}]{car_article}%
  \BibitemOpen
  \bibfield  {author} {\bibinfo {author} {\bibfnamefont {S.}~\bibnamefont
  {Caracciolo}}, \bibinfo {author} {\bibfnamefont {A.}~\bibnamefont
  {Montanari}},\ and\ \bibinfo {author} {\bibfnamefont {A.}~\bibnamefont
  {Pelissetto}},\ }\bibfield  {title} {\bibinfo {title} {Asymptotically free
  models and discrete non-abelian groups},\ }\href
  {https://doi.org/10.1016/S0370-2693(01)00674-8} {\bibfield  {journal}
  {\bibinfo  {journal} {Physics Letters B}\ }\textbf {\bibinfo {volume}
  {513}},\ \bibinfo {pages} {223} (\bibinfo {year}
  {2001}{\natexlab{b}})}\BibitemShut {NoStop}%
\bibitem [{\citenamefont {Zhou}\ \emph {et~al.}(2022)\citenamefont {Zhou},
  \citenamefont {Singh}, \citenamefont {Bhattacharya}, \citenamefont
  {Chandrasekharan},\ and\ \citenamefont {Gupta}}]{Zhou:2021qpm}%
  \BibitemOpen
  \bibfield  {author} {\bibinfo {author} {\bibfnamefont {J.}~\bibnamefont
  {Zhou}}, \bibinfo {author} {\bibfnamefont {H.}~\bibnamefont {Singh}},
  \bibinfo {author} {\bibfnamefont {T.}~\bibnamefont {Bhattacharya}}, \bibinfo
  {author} {\bibfnamefont {S.}~\bibnamefont {Chandrasekharan}},\ and\ \bibinfo
  {author} {\bibfnamefont {R.}~\bibnamefont {Gupta}},\ }\bibfield  {title}
  {\bibinfo {title} {{Spacetime symmetric qubit regularization of the
  asymptotically free two-dimensional O(4) model}},\ }\href
  {https://doi.org/10.1103/PhysRevD.105.054510} {\bibfield  {journal} {\bibinfo
   {journal} {Phys. Rev. D}\ }\textbf {\bibinfo {volume} {105}},\ \bibinfo
  {pages} {054510} (\bibinfo {year} {2022})},\ \Eprint
  {https://arxiv.org/abs/2111.13780} {arXiv:2111.13780 [hep-lat]} \BibitemShut
  {NoStop}%
\bibitem [{\citenamefont {Caspar}\ and\ \citenamefont
  {Singh}(2022)}]{Caspar:2022llo}%
  \BibitemOpen
  \bibfield  {author} {\bibinfo {author} {\bibfnamefont {S.}~\bibnamefont
  {Caspar}}\ and\ \bibinfo {author} {\bibfnamefont {H.}~\bibnamefont {Singh}},\
  }\href@noop {} {\bibinfo {title} {{From asymptotic freedom to $\theta$ vacua:
  Qubit embeddings of the O(3) nonlinear $\sigma$ model}}} (\bibinfo {year}
  {2022}),\ \Eprint {https://arxiv.org/abs/2203.15766} {arXiv:2203.15766
  [hep-lat]} \BibitemShut {NoStop}%
\bibitem [{\citenamefont {Zohar}(2021)}]{Zohar:2021nyc}%
  \BibitemOpen
  \bibfield  {author} {\bibinfo {author} {\bibfnamefont {E.}~\bibnamefont
  {Zohar}},\ }\href@noop {} {\bibinfo {title} {{Quantum Simulation of Lattice
  Gauge Theories in more than One Space Dimension -- Requirements, Challenges,
  Methods}}} (\bibinfo {year} {2021}),\ \Eprint
  {https://arxiv.org/abs/2106.04609} {arXiv:2106.04609 [quant-ph]} \BibitemShut
  {NoStop}%
\bibitem [{\citenamefont {Kan}\ and\ \citenamefont {Nam}(2021)}]{Kan:2021xfc}%
  \BibitemOpen
  \bibfield  {author} {\bibinfo {author} {\bibfnamefont {A.}~\bibnamefont
  {Kan}}\ and\ \bibinfo {author} {\bibfnamefont {Y.}~\bibnamefont {Nam}},\
  }\href@noop {} {\bibinfo {title} {{Lattice Quantum Chromodynamics and
  Electrodynamics on a Universal Quantum Computer}}} (\bibinfo {year} {2021}),\
  \Eprint {https://arxiv.org/abs/2107.12769} {arXiv:2107.12769 [quant-ph]}
  \BibitemShut {NoStop}%
\bibitem [{\citenamefont {Carena}\ \emph {et~al.}(2021)\citenamefont {Carena},
  \citenamefont {Lamm}, \citenamefont {Li},\ and\ \citenamefont
  {Liu}}]{Carena:2021ltu}%
  \BibitemOpen
  \bibfield  {author} {\bibinfo {author} {\bibfnamefont {M.}~\bibnamefont
  {Carena}}, \bibinfo {author} {\bibfnamefont {H.}~\bibnamefont {Lamm}},
  \bibinfo {author} {\bibfnamefont {Y.-Y.}\ \bibnamefont {Li}},\ and\ \bibinfo
  {author} {\bibfnamefont {W.}~\bibnamefont {Liu}},\ }\bibfield  {title}
  {\bibinfo {title} {{Lattice renormalization of quantum simulations}},\ }\href
  {https://doi.org/10.1103/PhysRevD.104.094519} {\bibfield  {journal} {\bibinfo
   {journal} {Phys. Rev. D}\ }\textbf {\bibinfo {volume} {104}},\ \bibinfo
  {pages} {094519} (\bibinfo {year} {2021})},\ \Eprint
  {https://arxiv.org/abs/2107.01166} {arXiv:2107.01166 [hep-lat]} \BibitemShut
  {NoStop}%
\bibitem [{\citenamefont {Gonz\'alez-Cuadra}\ \emph {et~al.}(2022)\citenamefont
  {Gonz\'alez-Cuadra}, \citenamefont {Zache}, \citenamefont {Carrasco},
  \citenamefont {Kraus},\ and\ \citenamefont
  {Zoller}}]{Gonzalez-Cuadra:2022hxt}%
  \BibitemOpen
  \bibfield  {author} {\bibinfo {author} {\bibfnamefont {D.}~\bibnamefont
  {Gonz\'alez-Cuadra}}, \bibinfo {author} {\bibfnamefont {T.~V.}\ \bibnamefont
  {Zache}}, \bibinfo {author} {\bibfnamefont {J.}~\bibnamefont {Carrasco}},
  \bibinfo {author} {\bibfnamefont {B.}~\bibnamefont {Kraus}},\ and\ \bibinfo
  {author} {\bibfnamefont {P.}~\bibnamefont {Zoller}},\ }\href@noop {}
  {\bibinfo {title} {{Hardware efficient quantum simulation of non-abelian
  gauge theories with qudits on Rydberg platforms}}} (\bibinfo {year} {2022}),\
  \Eprint {https://arxiv.org/abs/2203.15541} {arXiv:2203.15541 [quant-ph]}
  \BibitemShut {NoStop}%
\bibitem [{\citenamefont {Creutz}\ \emph {et~al.}(1979)\citenamefont {Creutz},
  \citenamefont {Jacobs},\ and\ \citenamefont {Rebbi}}]{Creutz:1979zg}%
  \BibitemOpen
  \bibfield  {author} {\bibinfo {author} {\bibfnamefont {M.}~\bibnamefont
  {Creutz}}, \bibinfo {author} {\bibfnamefont {L.}~\bibnamefont {Jacobs}},\
  and\ \bibinfo {author} {\bibfnamefont {C.}~\bibnamefont {Rebbi}},\ }\bibfield
   {title} {\bibinfo {title} {{Monte Carlo Study of Abelian Lattice Gauge
  Theories}},\ }\href {https://doi.org/10.1103/PhysRevD.20.1915} {\bibfield
  {journal} {\bibinfo  {journal} {Phys. Rev.}\ }\textbf {\bibinfo {volume}
  {D20}},\ \bibinfo {pages} {1915} (\bibinfo {year} {1979})}\BibitemShut
  {NoStop}%
\bibitem [{\citenamefont {Creutz}\ and\ \citenamefont
  {Okawa}(1983)}]{Creutz:1982dn}%
  \BibitemOpen
  \bibfield  {author} {\bibinfo {author} {\bibfnamefont {M.}~\bibnamefont
  {Creutz}}\ and\ \bibinfo {author} {\bibfnamefont {M.}~\bibnamefont {Okawa}},\
  }\bibfield  {title} {\bibinfo {title} {{Generalized Actions in $Z(p$) Lattice
  Gauge Theory}},\ }\href {https://doi.org/10.1016/0550-3213(83)90220-1}
  {\bibfield  {journal} {\bibinfo  {journal} {Nucl. Phys.}\ }\textbf {\bibinfo
  {volume} {B220}},\ \bibinfo {pages} {149} (\bibinfo {year}
  {1983})}\BibitemShut {NoStop}%
\bibitem [{\citenamefont {Bhanot}\ and\ \citenamefont
  {Rebbi}(1981)}]{Bhanot:1981xp}%
  \BibitemOpen
  \bibfield  {author} {\bibinfo {author} {\bibfnamefont {G.}~\bibnamefont
  {Bhanot}}\ and\ \bibinfo {author} {\bibfnamefont {C.}~\bibnamefont {Rebbi}},\
  }\bibfield  {title} {\bibinfo {title} {{Monte Carlo Simulations of Lattice
  Models With Finite Subgroups of SU(3) as Gauge Groups}},\ }\href
  {https://doi.org/10.1103/PhysRevD.24.3319} {\bibfield  {journal} {\bibinfo
  {journal} {Phys. Rev.}\ }\textbf {\bibinfo {volume} {D24}},\ \bibinfo {pages}
  {3319} (\bibinfo {year} {1981})}\BibitemShut {NoStop}%
\bibitem [{\citenamefont {Petcher}\ and\ \citenamefont
  {Weingarten}(1980)}]{Petcher:1980cq}%
  \BibitemOpen
  \bibfield  {author} {\bibinfo {author} {\bibfnamefont {D.}~\bibnamefont
  {Petcher}}\ and\ \bibinfo {author} {\bibfnamefont {D.~H.}\ \bibnamefont
  {Weingarten}},\ }\bibfield  {title} {\bibinfo {title} {{Monte Carlo
  Calculations and a Model of the Phase Structure for Gauge Theories on
  Discrete Subgroups of SU(2)}},\ }\href
  {https://doi.org/10.1103/PhysRevD.22.2465} {\bibfield  {journal} {\bibinfo
  {journal} {Phys. Rev.}\ }\textbf {\bibinfo {volume} {D22}},\ \bibinfo {pages}
  {2465} (\bibinfo {year} {1980})}\BibitemShut {NoStop}%
\bibitem [{\citenamefont {Bhanot}(1982)}]{Bhanot:1981pj}%
  \BibitemOpen
  \bibfield  {author} {\bibinfo {author} {\bibfnamefont {G.}~\bibnamefont
  {Bhanot}},\ }\bibfield  {title} {\bibinfo {title} {{SU(3) Lattice Gauge
  Theory in Four-dimensions With a Modified Wilson Action}},\ }\href
  {https://doi.org/10.1016/0370-2693(82)91207-2} {\bibfield  {journal}
  {\bibinfo  {journal} {Phys. Lett.}\ }\textbf {\bibinfo {volume} {108B}},\
  \bibinfo {pages} {337} (\bibinfo {year} {1982})}\BibitemShut {NoStop}%
\bibitem [{\citenamefont {Ji}\ \emph {et~al.}(2022)\citenamefont {Ji},
  \citenamefont {Lamm},\ and\ \citenamefont {Zhu}}]{Ji:2022qvr}%
  \BibitemOpen
  \bibfield  {author} {\bibinfo {author} {\bibfnamefont {Y.}~\bibnamefont
  {Ji}}, \bibinfo {author} {\bibfnamefont {H.}~\bibnamefont {Lamm}},\ and\
  \bibinfo {author} {\bibfnamefont {S.}~\bibnamefont {Zhu}},\ }\href@noop {}
  {\bibinfo {title} {{Gluon Digitization via Character Expansion for Quantum
  Computers}}} (\bibinfo {year} {2022}),\ \Eprint
  {https://arxiv.org/abs/2203.02330} {arXiv:2203.02330 [hep-lat]} \BibitemShut
  {NoStop}%
\bibitem [{\citenamefont {Alexandru}\ \emph {et~al.}(2022)\citenamefont
  {Alexandru}, \citenamefont {Bedaque}, \citenamefont {Brett},\ and\
  \citenamefont {Lamm}}]{Alexandru:2021jpm}%
  \BibitemOpen
  \bibfield  {author} {\bibinfo {author} {\bibfnamefont {A.}~\bibnamefont
  {Alexandru}}, \bibinfo {author} {\bibfnamefont {P.~F.}\ \bibnamefont
  {Bedaque}}, \bibinfo {author} {\bibfnamefont {R.}~\bibnamefont {Brett}},\
  and\ \bibinfo {author} {\bibfnamefont {H.}~\bibnamefont {Lamm}},\ }\bibfield
  {title} {\bibinfo {title} {{Spectrum of digitized QCD: Glueballs in a S(1080)
  gauge theory}},\ }\href {https://doi.org/10.1103/PhysRevD.105.114508}
  {\bibfield  {journal} {\bibinfo  {journal} {Phys. Rev. D}\ }\textbf {\bibinfo
  {volume} {105}},\ \bibinfo {pages} {114508} (\bibinfo {year} {2022})},\
  \Eprint {https://arxiv.org/abs/2112.08482} {arXiv:2112.08482 [hep-lat]}
  \BibitemShut {NoStop}%
\bibitem [{\citenamefont {Carena}\ \emph
  {et~al.}(2022{\natexlab{b}})\citenamefont {Carena}, \citenamefont
  {Gustafson}, \citenamefont {Lamm}, \citenamefont {Li},\ and\ \citenamefont
  {Liu}}]{Carena:2022hpz}%
  \BibitemOpen
  \bibfield  {author} {\bibinfo {author} {\bibfnamefont {M.}~\bibnamefont
  {Carena}}, \bibinfo {author} {\bibfnamefont {E.~J.}\ \bibnamefont
  {Gustafson}}, \bibinfo {author} {\bibfnamefont {H.}~\bibnamefont {Lamm}},
  \bibinfo {author} {\bibfnamefont {Y.-Y.}\ \bibnamefont {Li}},\ and\ \bibinfo
  {author} {\bibfnamefont {W.}~\bibnamefont {Liu}},\ }\bibfield  {title}
  {\bibinfo {title} {{Gauge theory couplings on anisotropic lattices}},\ }\href
  {https://doi.org/10.1103/PhysRevD.106.114504} {\bibfield  {journal} {\bibinfo
   {journal} {Phys. Rev. D}\ }\textbf {\bibinfo {volume} {106}},\ \bibinfo
  {pages} {114504} (\bibinfo {year} {2022}{\natexlab{b}})},\ \Eprint
  {https://arxiv.org/abs/2208.10417} {arXiv:2208.10417 [hep-lat]} \BibitemShut
  {NoStop}%
\bibitem [{\citenamefont {Weingarten}\ and\ \citenamefont
  {Petcher}(1981)}]{Weingarten:1980hx}%
  \BibitemOpen
  \bibfield  {author} {\bibinfo {author} {\bibfnamefont {D.~H.}\ \bibnamefont
  {Weingarten}}\ and\ \bibinfo {author} {\bibfnamefont {D.~N.}\ \bibnamefont
  {Petcher}},\ }\bibfield  {title} {\bibinfo {title} {{Monte Carlo Integration
  for Lattice Gauge Theories with Fermions}},\ }\href
  {https://doi.org/10.1016/0370-2693(81)90112-X} {\bibfield  {journal}
  {\bibinfo  {journal} {Phys. Lett.}\ }\textbf {\bibinfo {volume} {99B}},\
  \bibinfo {pages} {333} (\bibinfo {year} {1981})}\BibitemShut {NoStop}%
\bibitem [{\citenamefont {Weingarten}(1982)}]{Weingarten:1981jy}%
  \BibitemOpen
  \bibfield  {author} {\bibinfo {author} {\bibfnamefont {D.}~\bibnamefont
  {Weingarten}},\ }\bibfield  {title} {\bibinfo {title} {{Monte Carlo
  Evaluation of Hadron Masses in Lattice Gauge Theories with Fermions}},\
  }\href {https://doi.org/10.1016/0370-2693(82)90463-4} {\bibfield  {journal}
  {\bibinfo  {journal} {Phys. Lett.}\ }\textbf {\bibinfo {volume} {109B}},\
  \bibinfo {pages} {57} (\bibinfo {year} {1982})},\ \bibinfo {note}
  {[,631(1981)]}\BibitemShut {NoStop}%
\bibitem [{\citenamefont {Kogut}(1980)}]{Kogut:1980qb}%
  \BibitemOpen
  \bibfield  {author} {\bibinfo {author} {\bibfnamefont {J.~B.}\ \bibnamefont
  {Kogut}},\ }\bibfield  {title} {\bibinfo {title} {{1/n Expansions and the
  Phase Diagram of Discrete Lattice Gauge Theories With Matter Fields}},\
  }\href {https://doi.org/10.1103/PhysRevD.21.2316} {\bibfield  {journal}
  {\bibinfo  {journal} {Phys. Rev. D}\ }\textbf {\bibinfo {volume} {21}},\
  \bibinfo {pages} {2316} (\bibinfo {year} {1980})}\BibitemShut {NoStop}%
\bibitem [{\citenamefont {Romers}(2007)}]{romers2007discrete}%
  \BibitemOpen
  \bibfield  {author} {\bibinfo {author} {\bibfnamefont {J.}~\bibnamefont
  {Romers}},\ }\emph {\bibinfo {title} {Discrete gauge theories in two spatial
  dimensions}},\ \href@noop {} {Ph.D. thesis},\ \bibinfo  {school} {Master’s
  thesis, Universiteit van Amsterdam} (\bibinfo {year} {2007})\BibitemShut
  {NoStop}%
\bibitem [{\citenamefont {Fradkin}\ and\ \citenamefont
  {Shenker}(1979)}]{Fradkin:1978dv}%
  \BibitemOpen
  \bibfield  {author} {\bibinfo {author} {\bibfnamefont {E.~H.}\ \bibnamefont
  {Fradkin}}\ and\ \bibinfo {author} {\bibfnamefont {S.~H.}\ \bibnamefont
  {Shenker}},\ }\bibfield  {title} {\bibinfo {title} {{Phase Diagrams of
  Lattice Gauge Theories with Higgs Fields}},\ }\href
  {https://doi.org/10.1103/PhysRevD.19.3682} {\bibfield  {journal} {\bibinfo
  {journal} {Phys. Rev. D}\ }\textbf {\bibinfo {volume} {19}},\ \bibinfo
  {pages} {3682} (\bibinfo {year} {1979})}\BibitemShut {NoStop}%
\bibitem [{\citenamefont {Harlow}\ and\ \citenamefont
  {Ooguri}(2018)}]{Harlow:2018tng}%
  \BibitemOpen
  \bibfield  {author} {\bibinfo {author} {\bibfnamefont {D.}~\bibnamefont
  {Harlow}}\ and\ \bibinfo {author} {\bibfnamefont {H.}~\bibnamefont
  {Ooguri}},\ }\href@noop {} {\bibinfo {title} {{Symmetries in quantum field
  theory and quantum gravity}}} (\bibinfo {year} {2018}),\ \Eprint
  {https://arxiv.org/abs/1810.05338} {arXiv:1810.05338 [hep-th]} \BibitemShut
  {NoStop}%
\bibitem [{\citenamefont {Horn}\ \emph {et~al.}(1979)\citenamefont {Horn},
  \citenamefont {Weinstein},\ and\ \citenamefont {Yankielowicz}}]{Horn:1979fy}%
  \BibitemOpen
  \bibfield  {author} {\bibinfo {author} {\bibfnamefont {D.}~\bibnamefont
  {Horn}}, \bibinfo {author} {\bibfnamefont {M.}~\bibnamefont {Weinstein}},\
  and\ \bibinfo {author} {\bibfnamefont {S.}~\bibnamefont {Yankielowicz}},\
  }\bibfield  {title} {\bibinfo {title} {{Hamiltonian Approach to Z(N) Lattice
  Gauge Theories}},\ }\href {https://doi.org/10.1103/PhysRevD.19.3715}
  {\bibfield  {journal} {\bibinfo  {journal} {Phys. Rev. D}\ }\textbf {\bibinfo
  {volume} {19}},\ \bibinfo {pages} {3715} (\bibinfo {year}
  {1979})}\BibitemShut {NoStop}%
\bibitem [{\citenamefont {Clemente}\ \emph {et~al.}(2022)\citenamefont
  {Clemente}, \citenamefont {Crippa},\ and\ \citenamefont
  {Jansen}}]{Clemente:2022cka}%
  \BibitemOpen
  \bibfield  {author} {\bibinfo {author} {\bibfnamefont {G.}~\bibnamefont
  {Clemente}}, \bibinfo {author} {\bibfnamefont {A.}~\bibnamefont {Crippa}},\
  and\ \bibinfo {author} {\bibfnamefont {K.}~\bibnamefont {Jansen}},\
  }\bibfield  {title} {\bibinfo {title} {{Strategies for the determination of
  the running coupling of (2+1)-dimensional QED with quantum computing}},\
  }\href {https://doi.org/10.1103/PhysRevD.106.114511} {\bibfield  {journal}
  {\bibinfo  {journal} {Phys. Rev. D}\ }\textbf {\bibinfo {volume} {106}},\
  \bibinfo {pages} {114511} (\bibinfo {year} {2022})},\ \Eprint
  {https://arxiv.org/abs/2206.12454} {arXiv:2206.12454 [hep-lat]} \BibitemShut
  {NoStop}%
\bibitem [{\citenamefont {Creutz}(1985)}]{Creutz:1984mg}%
  \BibitemOpen
  \bibfield  {author} {\bibinfo {author} {\bibfnamefont {M.}~\bibnamefont
  {Creutz}},\ }\href@noop {} {\emph {\bibinfo {title} {{Quarks, gluons and
  lattices}}}},\ Cambridge Monographs on Mathematical Physics\ (\bibinfo
  {publisher} {Cambridge Univ. Press},\ \bibinfo {address} {Cambridge, UK},\
  \bibinfo {year} {1985})\BibitemShut {NoStop}%
\bibitem [{\citenamefont {Fromm}\ \emph {et~al.}(2022)\citenamefont {Fromm},
  \citenamefont {Philipsen},\ and\ \citenamefont {Winterowd}}]{Fromm:2022vaj}%
  \BibitemOpen
  \bibfield  {author} {\bibinfo {author} {\bibfnamefont {M.}~\bibnamefont
  {Fromm}}, \bibinfo {author} {\bibfnamefont {O.}~\bibnamefont {Philipsen}},\
  and\ \bibinfo {author} {\bibfnamefont {C.}~\bibnamefont {Winterowd}},\
  }\href@noop {} {\bibinfo {title} {{Dihedral Lattice Gauge Theories on a
  Quantum Annealer}}} (\bibinfo {year} {2022}),\ \Eprint
  {https://arxiv.org/abs/2206.14679} {arXiv:2206.14679 [hep-lat]} \BibitemShut
  {NoStop}%
\bibitem [{\citenamefont {Kogut}\ and\ \citenamefont
  {Susskind}(1975)}]{PhysRevD.11.395}%
  \BibitemOpen
  \bibfield  {author} {\bibinfo {author} {\bibfnamefont {J.}~\bibnamefont
  {Kogut}}\ and\ \bibinfo {author} {\bibfnamefont {L.}~\bibnamefont
  {Susskind}},\ }\bibfield  {title} {\bibinfo {title} {Hamiltonian formulation
  of {W}ilson's lattice gauge theories},\ }\href
  {https://doi.org/10.1103/PhysRevD.11.395} {\bibfield  {journal} {\bibinfo
  {journal} {Phys. Rev. D}\ }\textbf {\bibinfo {volume} {11}},\ \bibinfo
  {pages} {395} (\bibinfo {year} {1975})}\BibitemShut {NoStop}%
\bibitem [{\citenamefont {Grimus}\ and\ \citenamefont
  {Ludl}(2012)}]{Grimus:2011fk}%
  \BibitemOpen
  \bibfield  {author} {\bibinfo {author} {\bibfnamefont {W.}~\bibnamefont
  {Grimus}}\ and\ \bibinfo {author} {\bibfnamefont {P.~O.}\ \bibnamefont
  {Ludl}},\ }\bibfield  {title} {\bibinfo {title} {{Finite flavour groups of
  fermions}},\ }\href {https://doi.org/10.1088/1751-8113/45/23/233001}
  {\bibfield  {journal} {\bibinfo  {journal} {J. Phys. A}\ }\textbf {\bibinfo
  {volume} {45}},\ \bibinfo {pages} {233001} (\bibinfo {year} {2012})},\
  \Eprint {https://arxiv.org/abs/1110.6376} {arXiv:1110.6376 [hep-ph]}
  \BibitemShut {NoStop}%
\bibitem [{\citenamefont {Nielsen}\ and\ \citenamefont
  {Chuang}(2010)}]{nielsen_chuang_2010}%
  \BibitemOpen
  \bibfield  {author} {\bibinfo {author} {\bibfnamefont {M.~A.}\ \bibnamefont
  {Nielsen}}\ and\ \bibinfo {author} {\bibfnamefont {I.~L.}\ \bibnamefont
  {Chuang}},\ }\href {https://doi.org/10.1017/CBO9780511976667} {\emph
  {\bibinfo {title} {Quantum Computation and Quantum Information: 10th
  Anniversary Edition}}}\ (\bibinfo  {publisher} {Cambridge University Press},\
  \bibinfo {year} {2010})\BibitemShut {NoStop}%
\bibitem [{\citenamefont {Hoyer}(1997)}]{hoyer1997efficient}%
  \BibitemOpen
  \bibfield  {author} {\bibinfo {author} {\bibfnamefont {P.}~\bibnamefont
  {Hoyer}},\ }\href@noop {} {\bibinfo {title} {Efficient quantum transforms}}
  (\bibinfo {year} {1997}),\ \Eprint {https://arxiv.org/abs/quant-ph/9702028}
  {arXiv:quant-ph/9702028 [quant-ph]} \BibitemShut {NoStop}%
\bibitem [{\citenamefont {Beals}(1997)}]{beals1997quantum}%
  \BibitemOpen
  \bibfield  {author} {\bibinfo {author} {\bibfnamefont {R.}~\bibnamefont
  {Beals}},\ }\bibfield  {title} {\bibinfo {title} {Quantum computation of
  {F}ourier transforms over symmetric groups},\ }in\ \href@noop {} {\emph
  {\bibinfo {booktitle} {Proceedings of the twenty-ninth annual ACM symposium
  on Theory of computing}}}\ (\bibinfo {organization} {Citeseer},\ \bibinfo
  {year} {1997})\ pp.\ \bibinfo {pages} {48--53}\BibitemShut {NoStop}%
\bibitem [{\citenamefont {P{\"u}schel}\ \emph {et~al.}(1999)\citenamefont
  {P{\"u}schel}, \citenamefont {R{\"o}tteler},\ and\ \citenamefont
  {Beth}}]{puschel1999fast}%
  \BibitemOpen
  \bibfield  {author} {\bibinfo {author} {\bibfnamefont {M.}~\bibnamefont
  {P{\"u}schel}}, \bibinfo {author} {\bibfnamefont {M.}~\bibnamefont
  {R{\"o}tteler}},\ and\ \bibinfo {author} {\bibfnamefont {T.}~\bibnamefont
  {Beth}},\ }\bibfield  {title} {\bibinfo {title} {Fast quantum {F}ourier
  transforms for a class of non-abelian groups},\ }in\ \href@noop {} {\emph
  {\bibinfo {booktitle} {International Symposium on Applied Algebra, Algebraic
  Algorithms, and Error-Correcting Codes}}}\ (\bibinfo {organization}
  {Springer},\ \bibinfo {year} {1999})\ pp.\ \bibinfo {pages}
  {148--159}\BibitemShut {NoStop}%
\bibitem [{\citenamefont {Moore}\ \emph {et~al.}(2006)\citenamefont {Moore},
  \citenamefont {Rockmore},\ and\ \citenamefont {Russell}}]{moore2006generic}%
  \BibitemOpen
  \bibfield  {author} {\bibinfo {author} {\bibfnamefont {C.}~\bibnamefont
  {Moore}}, \bibinfo {author} {\bibfnamefont {D.}~\bibnamefont {Rockmore}},\
  and\ \bibinfo {author} {\bibfnamefont {A.}~\bibnamefont {Russell}},\
  }\bibfield  {title} {\bibinfo {title} {Generic quantum {F}ourier
  transforms},\ }\href@noop {} {\bibfield  {journal} {\bibinfo  {journal} {ACM
  Transactions on Algorithms (TALG)}\ }\textbf {\bibinfo {volume} {2}},\
  \bibinfo {pages} {707} (\bibinfo {year} {2006})}\BibitemShut {NoStop}%
\bibitem [{\citenamefont {Childs}\ and\ \citenamefont {van
  Dam}(2010)}]{childs2010quantum}%
  \BibitemOpen
  \bibfield  {author} {\bibinfo {author} {\bibfnamefont {A.~M.}\ \bibnamefont
  {Childs}}\ and\ \bibinfo {author} {\bibfnamefont {W.}~\bibnamefont {van
  Dam}},\ }\bibfield  {title} {\bibinfo {title} {Quantum algorithms for
  algebraic problems},\ }\href {https://doi.org/10.1103/RevModPhys.82.1}
  {\bibfield  {journal} {\bibinfo  {journal} {Rev. Mod. Phys.}\ }\textbf
  {\bibinfo {volume} {82}},\ \bibinfo {pages} {1} (\bibinfo {year}
  {2010})}\BibitemShut {NoStop}%
\bibitem [{\citenamefont {Eastin}\ and\ \citenamefont
  {Knill}(2009)}]{PhysRevLett.102.110502}%
  \BibitemOpen
  \bibfield  {author} {\bibinfo {author} {\bibfnamefont {B.}~\bibnamefont
  {Eastin}}\ and\ \bibinfo {author} {\bibfnamefont {E.}~\bibnamefont {Knill}},\
  }\bibfield  {title} {\bibinfo {title} {Restrictions on transversal encoded
  quantum gate sets},\ }\href {https://doi.org/10.1103/PhysRevLett.102.110502}
  {\bibfield  {journal} {\bibinfo  {journal} {Phys. Rev. Lett.}\ }\textbf
  {\bibinfo {volume} {102}},\ \bibinfo {pages} {110502} (\bibinfo {year}
  {2009})}\BibitemShut {NoStop}%
\bibitem [{\citenamefont {Chuang}\ and\ \citenamefont
  {Nielsen}(1997)}]{Chuang:1996hw}%
  \BibitemOpen
  \bibfield  {author} {\bibinfo {author} {\bibfnamefont {I.~L.}\ \bibnamefont
  {Chuang}}\ and\ \bibinfo {author} {\bibfnamefont {M.~A.}\ \bibnamefont
  {Nielsen}},\ }\bibfield  {title} {\bibinfo {title} {{Prescription for
  experimental determination of the dynamics of a quantum black box}},\ }\href
  {https://doi.org/10.1080/09500349708231894} {\bibfield  {journal} {\bibinfo
  {journal} {J. Mod. Opt.}\ }\textbf {\bibinfo {volume} {44}},\ \bibinfo
  {pages} {2455} (\bibinfo {year} {1997})},\ \Eprint
  {https://arxiv.org/abs/quant-ph/9610001} {arXiv:quant-ph/9610001}
  \BibitemShut {NoStop}%
\bibitem [{\citenamefont {{Calderbank}}\ and\ \citenamefont
  {{Shor}}(1996)}]{1996PhRvA..54.1098C}%
  \BibitemOpen
  \bibfield  {author} {\bibinfo {author} {\bibfnamefont {A.~R.}\ \bibnamefont
  {{Calderbank}}}\ and\ \bibinfo {author} {\bibfnamefont {P.~W.}\ \bibnamefont
  {{Shor}}},\ }\bibfield  {title} {\bibinfo {title} {{Good quantum
  error-correcting codes exist}},\ }\href
  {https://doi.org/10.1103/PhysRevA.54.1098} {\bibfield  {journal} {\bibinfo
  {journal} {\pra}\ }\textbf {\bibinfo {volume} {54}},\ \bibinfo {pages} {1098}
  (\bibinfo {year} {1996})},\ \Eprint {https://arxiv.org/abs/quant-ph/9512032}
  {arXiv:quant-ph/9512032 [quant-ph]} \BibitemShut {NoStop}%
\bibitem [{\citenamefont {Steane}(1996)}]{PhysRevLett.77.793}%
  \BibitemOpen
  \bibfield  {author} {\bibinfo {author} {\bibfnamefont {A.~M.}\ \bibnamefont
  {Steane}},\ }\bibfield  {title} {\bibinfo {title} {Error correcting codes in
  quantum theory},\ }\href {https://doi.org/10.1103/PhysRevLett.77.793}
  {\bibfield  {journal} {\bibinfo  {journal} {Phys. Rev. Lett.}\ }\textbf
  {\bibinfo {volume} {77}},\ \bibinfo {pages} {793} (\bibinfo {year}
  {1996})}\BibitemShut {NoStop}%
\bibitem [{\citenamefont {{Steane}}(1996)}]{1996RSPSA.452.2551S}%
  \BibitemOpen
  \bibfield  {author} {\bibinfo {author} {\bibfnamefont {A.}~\bibnamefont
  {{Steane}}},\ }\bibfield  {title} {\bibinfo {title} {{Multiple-Particle
  Interference and Quantum Error Correction}},\ }\href
  {https://doi.org/10.1098/rspa.1996.0136} {\bibfield  {journal} {\bibinfo
  {journal} {Proceedings of the Royal Society of London Series A}\ }\textbf
  {\bibinfo {volume} {452}},\ \bibinfo {pages} {2551} (\bibinfo {year}
  {1996})},\ \Eprint {https://arxiv.org/abs/quant-ph/9601029}
  {arXiv:quant-ph/9601029 [quant-ph]} \BibitemShut {NoStop}%
\bibitem [{\citenamefont {Steane}(1996)}]{PhysRevA.54.4741}%
  \BibitemOpen
  \bibfield  {author} {\bibinfo {author} {\bibfnamefont {A.~M.}\ \bibnamefont
  {Steane}},\ }\bibfield  {title} {\bibinfo {title} {Simple quantum
  error-correcting codes},\ }\href {https://doi.org/10.1103/PhysRevA.54.4741}
  {\bibfield  {journal} {\bibinfo  {journal} {Phys. Rev. A}\ }\textbf {\bibinfo
  {volume} {54}},\ \bibinfo {pages} {4741} (\bibinfo {year}
  {1996})}\BibitemShut {NoStop}%
\bibitem [{\citenamefont {{Kitaev}}(1997)}]{1997RuMaS..52.1191K}%
  \BibitemOpen
  \bibfield  {author} {\bibinfo {author} {\bibfnamefont {A.~Y.}\ \bibnamefont
  {{Kitaev}}},\ }\bibfield  {title} {\bibinfo {title} {{Quantum computations:
  algorithms and error correction}},\ }\href
  {https://doi.org/10.1070/RM1997v052n06ABEH002155} {\bibfield  {journal}
  {\bibinfo  {journal} {Russian Mathematical Surveys}\ }\textbf {\bibinfo
  {volume} {52}},\ \bibinfo {pages} {1191} (\bibinfo {year}
  {1997})}\BibitemShut {NoStop}%
\bibitem [{\citenamefont {Kubischta}\ and\ \citenamefont
  {Teixeira}(2023)}]{Kubischta:2023nlb}%
  \BibitemOpen
  \bibfield  {author} {\bibinfo {author} {\bibfnamefont {E.}~\bibnamefont
  {Kubischta}}\ and\ \bibinfo {author} {\bibfnamefont {I.}~\bibnamefont
  {Teixeira}},\ }\href@noop {} {\bibinfo {title} {{A Family of Quantum Codes
  with Exotic Transversal Gates}}} (\bibinfo {year} {2023}),\ \Eprint
  {https://arxiv.org/abs/2305.07023} {arXiv:2305.07023 [quant-ph]} \BibitemShut
  {NoStop}%
\bibitem [{\citenamefont {Denys}\ and\ \citenamefont
  {Leverrier}(2023{\natexlab{a}})}]{Denys:2023syu}%
  \BibitemOpen
  \bibfield  {author} {\bibinfo {author} {\bibfnamefont {A.}~\bibnamefont
  {Denys}}\ and\ \bibinfo {author} {\bibfnamefont {A.}~\bibnamefont
  {Leverrier}},\ }\href@noop {} {\bibinfo {title} {{Multimode bosonic cat codes
  with an easily implementable universal gate set}}} (\bibinfo {year}
  {2023}{\natexlab{a}}),\ \Eprint {https://arxiv.org/abs/2306.11621}
  {arXiv:2306.11621 [quant-ph]} \BibitemShut {NoStop}%
\bibitem [{\citenamefont {Jain}\ \emph {et~al.}(2023)\citenamefont {Jain},
  \citenamefont {Iosue}, \citenamefont {Barg},\ and\ \citenamefont
  {Albert}}]{Jain:2023deu}%
  \BibitemOpen
  \bibfield  {author} {\bibinfo {author} {\bibfnamefont {S.~P.}\ \bibnamefont
  {Jain}}, \bibinfo {author} {\bibfnamefont {J.~T.}\ \bibnamefont {Iosue}},
  \bibinfo {author} {\bibfnamefont {A.}~\bibnamefont {Barg}},\ and\ \bibinfo
  {author} {\bibfnamefont {V.~V.}\ \bibnamefont {Albert}},\ }\href@noop {}
  {\bibinfo {title} {{Quantum spherical codes}}} (\bibinfo {year} {2023}),\
  \Eprint {https://arxiv.org/abs/2302.11593} {arXiv:2302.11593 [quant-ph]}
  \BibitemShut {NoStop}%
\bibitem [{\citenamefont {Denys}\ and\ \citenamefont
  {Leverrier}(2023{\natexlab{b}})}]{Denys:2022iyj}%
  \BibitemOpen
  \bibfield  {author} {\bibinfo {author} {\bibfnamefont {A.}~\bibnamefont
  {Denys}}\ and\ \bibinfo {author} {\bibfnamefont {A.}~\bibnamefont
  {Leverrier}},\ }\bibfield  {title} {\bibinfo {title} {{The $2T$-qutrit, a
  two-mode bosonic qutrit}},\ }\href
  {https://doi.org/10.22331/q-2023-06-05-1032} {\bibfield  {journal} {\bibinfo
  {journal} {Quantum}\ }\textbf {\bibinfo {volume} {7}},\ \bibinfo {pages}
  {1032} (\bibinfo {year} {2023}{\natexlab{b}})},\ \Eprint
  {https://arxiv.org/abs/2210.16188} {arXiv:2210.16188 [quant-ph]} \BibitemShut
  {NoStop}%
\bibitem [{\citenamefont {{Baker}}\ \emph {et~al.}(2019)\citenamefont
  {{Baker}}, \citenamefont {{Duckering}}, \citenamefont {{Hoover}},\ and\
  \citenamefont {{Chong}}}]{2019arXiv190401671B}%
  \BibitemOpen
  \bibfield  {author} {\bibinfo {author} {\bibfnamefont {J.~M.}\ \bibnamefont
  {{Baker}}}, \bibinfo {author} {\bibfnamefont {C.}~\bibnamefont
  {{Duckering}}}, \bibinfo {author} {\bibfnamefont {A.}~\bibnamefont
  {{Hoover}}},\ and\ \bibinfo {author} {\bibfnamefont {F.~T.}\ \bibnamefont
  {{Chong}}},\ }\href@noop {} {\bibinfo {title} {{Decomposing Quantum
  Generalized Toffoli with an Arbitrary Number of Ancilla}}} (\bibinfo {year}
  {2019}),\ \Eprint {https://arxiv.org/abs/1904.01671} {arXiv:1904.01671
  [quant-ph]} \BibitemShut {NoStop}%
\bibitem [{\citenamefont {Barenco}\ \emph {et~al.}(1995)\citenamefont
  {Barenco}, \citenamefont {Bennett}, \citenamefont {Cleve}, \citenamefont
  {DiVincenzo}, \citenamefont {Margolus}, \citenamefont {Shor}, \citenamefont
  {Sleator}, \citenamefont {Smolin},\ and\ \citenamefont
  {Weinfurter}}]{PhysRevA.52.3457}%
  \BibitemOpen
  \bibfield  {author} {\bibinfo {author} {\bibfnamefont {A.}~\bibnamefont
  {Barenco}}, \bibinfo {author} {\bibfnamefont {C.~H.}\ \bibnamefont
  {Bennett}}, \bibinfo {author} {\bibfnamefont {R.}~\bibnamefont {Cleve}},
  \bibinfo {author} {\bibfnamefont {D.~P.}\ \bibnamefont {DiVincenzo}},
  \bibinfo {author} {\bibfnamefont {N.}~\bibnamefont {Margolus}}, \bibinfo
  {author} {\bibfnamefont {P.}~\bibnamefont {Shor}}, \bibinfo {author}
  {\bibfnamefont {T.}~\bibnamefont {Sleator}}, \bibinfo {author} {\bibfnamefont
  {J.~A.}\ \bibnamefont {Smolin}},\ and\ \bibinfo {author} {\bibfnamefont
  {H.}~\bibnamefont {Weinfurter}},\ }\bibfield  {title} {\bibinfo {title}
  {Elementary gates for quantum computation},\ }\href
  {https://doi.org/10.1103/PhysRevA.52.3457} {\bibfield  {journal} {\bibinfo
  {journal} {Phys. Rev. A}\ }\textbf {\bibinfo {volume} {52}},\ \bibinfo
  {pages} {3457} (\bibinfo {year} {1995})}\BibitemShut {NoStop}%
\bibitem [{\citenamefont {Bocharov}\ \emph {et~al.}(2015)\citenamefont
  {Bocharov}, \citenamefont {Roetteler},\ and\ \citenamefont
  {Svore}}]{PhysRevLett.114.080502}%
  \BibitemOpen
  \bibfield  {author} {\bibinfo {author} {\bibfnamefont {A.}~\bibnamefont
  {Bocharov}}, \bibinfo {author} {\bibfnamefont {M.}~\bibnamefont
  {Roetteler}},\ and\ \bibinfo {author} {\bibfnamefont {K.~M.}\ \bibnamefont
  {Svore}},\ }\bibfield  {title} {\bibinfo {title} {Efficient synthesis of
  universal repeat-until-success quantum circuits},\ }\href
  {https://doi.org/10.1103/PhysRevLett.114.080502} {\bibfield  {journal}
  {\bibinfo  {journal} {Phys. Rev. Lett.}\ }\textbf {\bibinfo {volume} {114}},\
  \bibinfo {pages} {080502} (\bibinfo {year} {2015})}\BibitemShut {NoStop}%
\bibitem [{\citenamefont {Selinger}(2015)}]{10.5555/2685188.2685198}%
  \BibitemOpen
  \bibfield  {author} {\bibinfo {author} {\bibfnamefont {P.}~\bibnamefont
  {Selinger}},\ }\bibfield  {title} {\bibinfo {title} {Efficient clifford+t
  approximation of single-qubit operators},\ }\href@noop {} {\bibfield
  {journal} {\bibinfo  {journal} {Quantum Info. Comput.}\ }\textbf {\bibinfo
  {volume} {15}},\ \bibinfo {pages} {159–180} (\bibinfo {year}
  {2015})}\BibitemShut {NoStop}%
\bibitem [{\citenamefont {Peruzzo}\ \emph {et~al.}(2014)\citenamefont
  {Peruzzo}, \citenamefont {McClean}, \citenamefont {Shadbolt}, \citenamefont
  {Yung}, \citenamefont {Zhou}, \citenamefont {Love}, \citenamefont
  {Aspuru-Guzik},\ and\ \citenamefont {O’brien}}]{peruzzo2014variational}%
  \BibitemOpen
  \bibfield  {author} {\bibinfo {author} {\bibfnamefont {A.}~\bibnamefont
  {Peruzzo}}, \bibinfo {author} {\bibfnamefont {J.}~\bibnamefont {McClean}},
  \bibinfo {author} {\bibfnamefont {P.}~\bibnamefont {Shadbolt}}, \bibinfo
  {author} {\bibfnamefont {M.-H.}\ \bibnamefont {Yung}}, \bibinfo {author}
  {\bibfnamefont {X.-Q.}\ \bibnamefont {Zhou}}, \bibinfo {author}
  {\bibfnamefont {P.~J.}\ \bibnamefont {Love}}, \bibinfo {author}
  {\bibfnamefont {A.}~\bibnamefont {Aspuru-Guzik}},\ and\ \bibinfo {author}
  {\bibfnamefont {J.~L.}\ \bibnamefont {O’brien}},\ }\bibfield  {title}
  {\bibinfo {title} {A variational eigenvalue solver on a photonic quantum
  processor},\ }\href@noop {} {\bibfield  {journal} {\bibinfo  {journal}
  {Nature communications}\ }\textbf {\bibinfo {volume} {5}},\ \bibinfo {pages}
  {4213} (\bibinfo {year} {2014})}\BibitemShut {NoStop}%
\end{thebibliography}%

\end{document}